\documentclass[12pt,oneside,a4paper,onecolumn]{elsarticle}
\usepackage{hyperref}

\journal{International Journal of Multiphase Flow}

\usepackage{fullpage}
\usepackage{siunitx}
\usepackage{cancel}
\usepackage[usenames,dvipsnames]{xcolor}
\usepackage{multirow}

\usepackage{gensymb}

\usepackage{alltt}
\usepackage{titlesec}
\usepackage{upquote}
\usepackage{float}
\usepackage{array}
\usepackage{pgfplots}
\usetikzlibrary{positioning}
\usepackage{graphicx}
\usepackage[toc,page]{appendix}
\usepackage{amsmath,mathtools}
\usetikzlibrary{calc,decorations.markings,fit,shapes,chains,arrows}
\usepackage{amsthm}
\usepackage{amsfonts}
\usepackage{enumitem}
\usepackage{subcaption}
\usepackage{caption}
\usepackage[format=hang,indention=-1cm,labelfont={bf},margin=1em]{caption}
\usepackage{booktabs}
\usepackage{xcolor}
\usepackage{scalerel}
\usepackage{algorithm}
\usepackage{multicol}
\usepackage{mdframed}
\usepackage{mathtools}
\theoremstyle{remark}

\usepackage{tikzit}
\usetikzlibrary{patterns}

\tikzstyle{black dot}=[fill=black, draw=black, shape=circle]
\tikzstyle{green dot}=[fill=green, draw=black, shape=circle]

\tikzstyle{red solid line}=[-, fill=none, draw=red]
\tikzstyle{black dashed line}=[-, fill=none, draw=black, dashed]
\tikzstyle{black solid line}=[-, fill=none, draw=black]
\tikzstyle{black arrow solid}=[->, fill=none]
\tikzstyle{black arrow dashed}=[fill=none, ->, dashed, draw=black]
\tikzstyle{black dashed double sided arrow}=[<->, dashed]
\tikzstyle{red dashed}=[-, draw=red, fill=none, dashed]
\tikzstyle{red dashed arrow}=[draw=red, fill=none, dashed, ->]
\tikzstyle{blue solid}=[-, fill=none, draw=blue]
\tikzstyle{blue dashed}=[-, fill=none, draw=blue, dashed]

\usepackage{natbib}
\setcitestyle{square, numbers}

\pgfdeclarepatternformonly{crossHatchDots}{\pgfqpoint{-1pt}{-1pt}}{\pgfqpoint{5pt}{5pt}}{\pgfqpoint{6pt}{6pt}}%
{
    \pgfpathcircle{\pgfqpoint{0pt}{0pt}}{.5pt}
    \pgfpathcircle{\pgfqpoint{3pt}{3pt}}{.5pt}
    \pgfusepath{fill}
}

\tikzset{
  block/.style={rectangle, draw, rounded corners, text centered,text width = 16em, minimum height = 2em},
  line/.style={draw, -latex'}
  }
\tikzset{
  block2/.style={text centered,text width = 22em, minimum height = 2em},
  line/.style={draw, -latex'}
  }
\tikzset{
  block3/.style={rectangle, draw, rounded corners, text centered,text width = 19em, minimum height = 1em},
  line/.style={draw, -latex'}
  }
\tikzset{
  block4/.style={rectangle, draw, rounded corners, text centered,text width = 15em, minimum height = 1em},
  line/.style={draw, -latex'}
  }
\tikzset{
  blockNS/.style={rectangle, draw, fill=black!20, rounded corners, text centered,text width = 20em, minimum height = 2em, label={center:Navier-Stokes solver}},
  line/.style={draw, -latex'}
  }
\tikzset{
  blockAC/.style={rectangle, draw, fill=black!20, rounded corners, text centered,text width = 20em, minimum height = 2em, label={center:Allen-Cahn solver}},
  line/.style={draw, -latex'}
  }
\tikzset{  
  decision/.style = {diamond, draw, minimum width=4cm, minimum height=0.2cm},
  line/.style={draw, -latex'}
  }
  
\def\@author#1{\g@addto@macro\elsauthors{\normalsize%
    \def\baselinestretch{1}%
    \upshape\authorsep#1\unskip\textsuperscript{%
      \ifx\@fnmark\@empty\else\unskip\sep\@fnmark\let\sep=,\fi
      \ifx\@corref\@empty\else\unskip\sep\@corref\let\sep=,\fi
      }%
    \def\authorsep{\unskip,\space}%
    \global\let\@fnmark\@empty
    \global\let\@corref\@empty
    \global\let\sep\@empty}%
    \@eadauthor={#1}
}

\graphicspath{{figures/}}

\begin{document}
\begin{frontmatter}
\title{Unsteady cavitation dynamics and frequency lock-in of a freely vibrating hydrofoil at high Reynolds number}
\author[ubc]{Suraj R. Kashyap}
\ead{suraj.kashyap@ubc.ca}

\author[ubc]{Rajeev K. Jaiman\corref{cor1}}
\ead{rjaiman@mech.ubc.ca}
\cortext[cor1]{Corresponding author}
\address[ubc]{Department of Mechanical Engineering, The University of British Columbia, Vancouver, BC V6T 1Z4}

\begin{abstract}
In the current work, we investigate the influence of unsteady partial cavitation on the fluid-structure interaction of a freely vibrating hydrofoil section at high Reynolds numbers. We consider an elastically-mounted NACA66 hydrofoil section that is free to vibrate in the transverse flow direction. Cavitating flow dynamics coupled with the transverse vibration are studied at low angles of attack.
For this numerical study, we employ a recently developed unified variational fluid-structure interaction framework based on homogeneous mixture-based cavitation with a hybrid URANS-LES turbulence modeling. We first validate the numerical implementation against the experimental data for turbulent cavitating flow at high Reynolds numbers.
%
For the freely oscillating hydrofoil, we observe large-amplitude vibrations during unsteady partial cavitating conditions that are absent in the non-cavitating flow configuration. We identify a frequency lock-in phenomenon as the main source of sustained large-amplitude vibration whereby the unsteady lift forces lock into a sub-harmonic of the hydrofoil natural frequency.  During the cavity collapse and shedding, we find a periodic generation of clockwise vorticity, leading to the unsteady lift generation. We determine the origin of this flow unsteadiness at the vicinity of the trailing edge of the hydrofoil through the interplay between the growing cavity with the adverse pressure gradient. The flow-induced structural vibration is also observed to have a consequent impact on the cavity dynamics. In the frequency lock-in regime, large coherent cavitating structures are seen over the hydrofoil suction surface undergoing a full cavity growth-detachment-collapse cycle. For the post-lock-in regime, the cavity length is shorter and the attached cavity length is observed to undergo high frequency spatially localized oscillations. In this regime, cavity shedding is primarily limited to the cavity trailing end and the frequency of a complete cavity detachment and shedding event is reduced.
This work has a practical relevance to the cavitation-induced vibration of marine propellers with the target of noise mitigation by active or passive control mechanisms.
\end{abstract}


\end{frontmatter}

	
\section{Introduction}\label{sec:introduction}
Cavitating flows are widespread in nature and numerous engineering applications.  The phenomenon of cavitation in liquids has been adapted for aiding industrial processes such as homogenization \cite{guo2021effect}, machining \cite{guo2018effect}, metrology \cite{saint2020bubble, bruning2021soft}, surface cleaning \cite{song2004cleaning, chahine2016cleaning}, and biomedical procedures of lithotripsy \cite{bailey2003cavitation} and drug delivery \cite{stride2019nucleation}. On the other hand, detrimental effects of cavitation abound in the form of noise, vibration, material erosion and drop in hydrodynamic efficiency \cite{carlton2018marine,kerr1940problems, brennen_2013, arndt2015singing}. The phenomenon of cavitation involves the phase change of a liquid into vapor and a highly complex interaction between the vapor and the liquid phases. When the flowing liquid encounters a region of low pressure, cavities filled with the entrained gas or vapor due to evaporation are generated at any inherent points of weaknesses such as  microscopic crevices on the solid surface and ephemeral voids created by the thermal motions of the liquid \cite{brennen_2013}. Subsequently, these cavities can be convected by the flow until they encounter a region of high pressure which may accompany by a violent cavity collapse. Of particular interest to this study is marine propellers wherein cavitation is often encountered over the propeller blades  \cite{carlton2018marine,arndt2015singing}. Depending on the extent of cavitation, which can be represented by the dimensionless cavitation number,  several distinct cavitating flow structures of varying energy content can be observed \cite{ross1989mechanics, arndt2012some}. Development of cavities impacts the coupled fluid-structure dynamics of propeller operation, which can result in performance loss, vibration, material erosion, and noise emission \cite{brennen_2013}.

 A continuous formation and rapid collapse of bubbles during propeller-induced cavitation is the dominant source of underwater noise produced by marine vessels. In the event of cavitation as studied in \cite{van1974calculation, ross1989mechanics}, propeller noise dominates all other sources of self-noise from ships, including electrical noise, machinery noise and boundary layer noise.
 Cavitation can further reduce propulsion efficiency, as well as introduce the unwelcome risks of longer-term propeller damage. The reduction of noise and vibration in marine vessels is of interest both from an industrial and a marine-environmental perspective. For example, increased underwater noise has been shown to have a serious impact on all marine species \cite{duarte2021soundscape}. In particular, marine mammals are severely affected both in chronic behavioral and physical aspects as well as in their vital life activities such as communication, foraging, mating and migration \cite{duarte2021soundscape, erbe2019effects, marley2017effects}. 
 In a classical work, \citet{kerr1940problems} identified that tonal noise emission in propellers was a result of blade vibration due to irregular cavitation and vortex-shedding dynamics. Recently, \citet{carlton2018marine} provided an excellent review of noise from cavitating propellers and identified two broad categories: (i) a broadband noise component resulting from the sudden collapse of cavities and vortices, and (ii) tonal noise components from periodic fluctuations in the cavity volumes.   

\subsection{Cavitating flow over hydrofoils}
\begin{figure}[!h]
    \centering
    \begin{subfigure}{\textwidth}
      \centering
      \includegraphics[width=\linewidth]{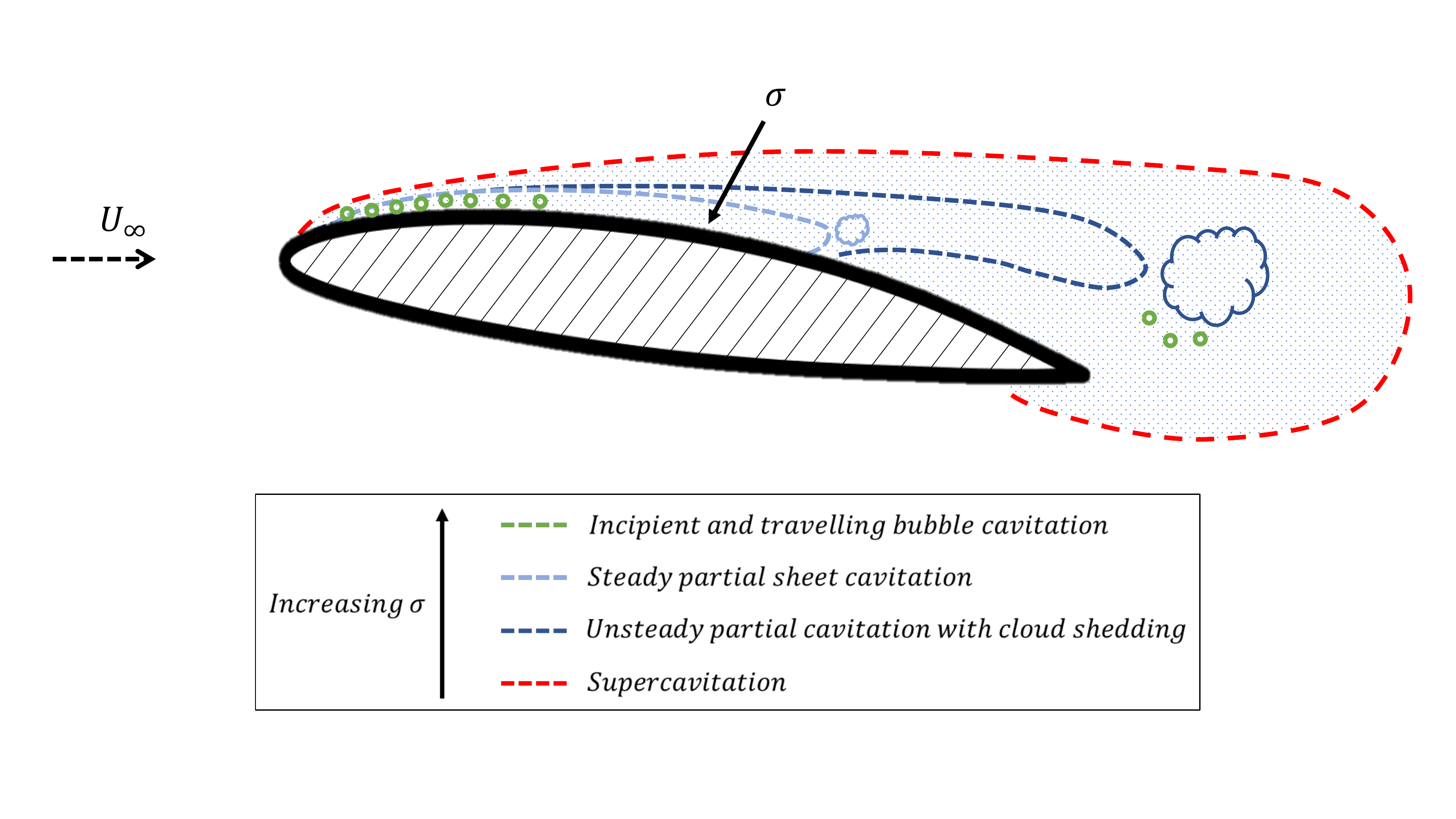}
      \caption{\label{fig:cavRegimes}}
    \end{subfigure}
    \begin{subfigure}{\textwidth}
      \centering
      \ctikzfig{Introduction/noiseSchematic}
      \caption{\label{fig:noiseSchematic}}
    \end{subfigure}
    \caption{(a) Illustration of some prominent cavitation regimes over hydrofoils and relation to cavitation number $\sigma$, (b) Sketch to demonstrate some of the prominent noise sources from cavitating flow around a deformable blade. $\Omega^{\mathrm{f}}$ and $\Omega^{\mathrm{s}}$ are the fluid and solid domains, $\Gamma^{\mathrm{fs}}$ represents the interface between them.} 
    \end{figure}

 A section of the propeller blade can be represented as a hydrofoil, and the interactions between the blade and the surrounding fluid can be simplified to an elastically mounted rigid body.  This arrangement can serve a prototypical problem for the fundamental understanding of the rich and complex  coupled physics of cavitating flows with fluid-structure interaction. 
Hydrofoil cavitation demonstrates the salient cavitation regimes observed in propeller blades away from root and tip effects. Cavitation over hydrofoils can exist in several forms and temporal-spatial scales \cite{arndt2015singing, brennen_2013}. While the specific regime of cavitation depends on various physical and geometric factors, the cavitation number $\sigma$ and the angle of attack $\alpha$ can be considered two important parameters for the hydrofoil study \cite{arndt2012some, watanabe1998linear, akcabay2014cavity}. Figure~\ref{fig:cavRegimes} shows a few of the prominent cavitation regimes observed on hydrofoils away from root and tip effects \cite{brennen_2013, franc2006fundamentals, arndt2015singing}.
 A representative fluid-structure system with a cavitating flow is shown in Fig.~\ref{fig:noiseSchematic}, where $\Omega^{\mathrm{f}}$ represents the fluid domain, $\Omega^{\mathrm{s}}$ the solid structural domain and $\Gamma^{\mathrm{fs}}$ the fluid-structure interface.
 %
 %
 Noise emission from propellers can be attributed to a complex multiphase fluid-structure interaction (FSI) between three key components namely the cavitating flow dynamics, the vortex shedding and the blade structural dynamics. Each of these components has its own fundamental frequency. In addition to the coupled multiphase FSI problem, resonance in cavity-filled vortices shed from the blade tip can also emit intense tonal noise \cite{arakeri1988model,maines1997case,arndt2015singing}. The shed cavity clouds and cavity-filled vortices possess bubbles of a wide distribution of radii. The implosion of these bubbles contributes to a range of broadband noise emission~\cite{brennen_2013}.
  Tonal noise emission from propellers, popularly known as propeller singing, is phenomenologically similar to hydrofoil singing \cite{blake2017mechanicsv2}.  Partial sheet cavitation is the regime where the cavities close on the cavity generating hydrofoil surface. The partial cavities can display unsteady periodic growth-collapse cycles, with cavity cloud shedding from the trailing end of the cavity \cite{brennen_2013, franc2006fundamentals}. The mechanism of collapse can vary depending on the cavitation number, with re-entrant jets and bubbly shock waves as two identified methods \cite{franc1988unsteady, arndt2000instability, bhatt2020numerical, smith2020influence}.

%
Cavity behavior over hydrofoils has been studied in several experimental and numerical works.  In two classical works, \citet{franc1985attached, franc1988unsteady} used flow visualizations to investigate the interaction between the boundary layer and attached cavities and related the cavity detachment with the separation of the boundary layer. \citet{arndt2012some, arndt2000instability} presented water-tunnel experiments on a stationary (force-balanced) spanwise two-dimensional (2D) hydrofoil section and observed high oscillations in lift measurements when unsteady partial cavitation occurs. This was largely associated with the regime where the ratio of the cavitation number $\sigma$ to the angle of attack $\alpha$ is in the range ${\sigma}/{2\alpha}\in\left(3, 5\right)$. Further, unsteadiness in the lift coefficient was found to exist primarily in the cavitation regime where the relative cavity length to the chord length $L/C\in \left({3}/{4}, {4}/{3}\right)$. These observations were consistent with previous theoretical predictions by \citet{watanabe1998linear} and \citet{ brennen_2013}. A relation between the shedding cavities and the lift oscillations was indicated. Prominent frequencies associated with different cavitation regimes over a stationary hydrofoil were highlighted. \citet{arndt2006experimental} studied large scale structures in the wakes of cavitating flows and highlighted the contribution of a `re-entrant jet' to negative (clockwise) vortex structures during cavity shedding. Recently, \citet{ji2015large} numerically studied the transient cavitating flow around a stationary NACA66 hydrofoil section and related pressure fluctuations with the cavity shedding process. The interaction of the cavity with vorticity was also discussed using the vorticity transport equation.

\subsection{Cavitation with flow-induced vibrations (FIV)}
A handful of studies has been carried out to understand the impact of cavitation on fluid-structure interaction and vice versa. In the case of coupled cavitation and fluid-structure interaction, cavitating flow induces structural vibration that in turn moves the fluid-structure interface and modifies the flow locally. 
Similar to a flexible or elastically-mounted bluff body, there exists a strong coupling between the hydrofoil and the vortices forming in its wake. As the natural frequency of the hydrofoil approaches the frequency of the unsteady vortex shedding, a frequency lock-in with a relatively large transverse vibration can be expected.
Frequency lock-in is a general nonlinear physical phenomenon in fluid-structure systems whereby the coupled system has an intrinsic ability to lock at a preferred frequency. Due to large structural vibration, the vortex strength is enhanced as well as the unsteady periodic loading. The phenomenon of frequency lock-in and flow-induced vibrations are extensively reviewed for bluff bodies in \cite{sarpkaya2004,williamson2004,bearman2011}.
\citet{ausoni2007cavitation} experimentally studied the influence of cavitation on the wake vortex shedding and trailing edge vibrations of a blunt truncated hydrofoil. Increased trailing edge displacements during cavitation were in turn observed to amplify the vortex strength as well as promote cavitation inception at higher cavitation numbers.  \citet{smith2020influence} experimentally compared the influence of FSI on cloud cavitation about a flexible and rigid hydrofoil. The transition between cavitation regimes was observed to be accelerated for the flexible hydrofoil indicating a significant influence of the structural dynamics on the cavity. An increase in the cavity length was reported to accompany the twist deformations, resulting in lower shedding frequencies. The flexible hydrofoil was also reported to attenuate relatively high-frequency oscillations compared to the rigid counterpart. 
 
 Most numerical studies of cavitating flows around hydrofoils have focused on configurations that are stationary (e.g., \cite{huang2014large}) or are prescribed motions (e.g., \cite{huang2013physical}). Among recent numerical works that consider fluid-structure interaction, \citet{akcabay2014cavity} used a loosely coupled framework between a 2D URANS solver with a 2DOF hydrofoil model to study the cavitation-induced vibration of flexible hydrofoils. Focusing of the vibration frequency content to the closest sub-harmonics of the hydrofoil's wetted natural frequencies was observed. In \citet{wu2018transient}, the authors employed a similar numerical approach to study the cavity shedding dynamics and flow-induced vibration over a hydrofoil section. The transient cavity behaviors were shown to lead to periodic pressure fluctuations on the hydrofoil. 

Theoretically, \citet{benaouicha2012analysis} studied the effect of added mass in cavitating flows developing over a vibrating body. Strong space-time variations in fluid density at the fluid-structure interface can influence the added mass significantly in cavitating flows (e.g., sheet cavitation) through large-scale pulsating changes from the vapor density to the liquid density. In contrast to the homogeneous flow, the cavitating non-homogeneous flow has an asymmetrical added mass operator during cavitation-structure interaction and it strongly depends on the geometry and the flow conditions. Oscillations in the cavity length were shown to be strongly correlated with the oscillating added mass coefficients. In general, the modal frequencies of the structure were found to increase as the cavity length increased over the hydrofoil surface.

While hydrofoil cavitation has been studied for several decades, most of the research has focused on the study of stationary hydrofoils, with only a few studies considered the fluid-structure interaction effects. There is a need for further investigations to understand the complex interplay among the unsteady cavitation, the vortex dynamics and the structural vibration characteristics. While there exists some computational modeling of cavitation in the literature, there has been relatively little work on the  flow-induced vibration of a hydrofoil with cavitating flow. More specifically, the impact of cavitation on flow-induced vibration and the frequency lock-in phenomenon is not fully explored via fully-coupled Navier-Stokes equations with turbulence and cavitation modeling. Hence, this is the focus of the current work.

\subsection{Current work and contributions}
%
The current work employs a 3D high-fidelity computational framework reported recently in \citet{KASHYAP202119}  based on a unified variational finite element formulation for fluid-structure interaction and cavitating flows at high Reynolds numbers. The unsteady viscous flow equations with an arbitrary Lagrangian-Eulerian (ALE) frame are discretized using the stabilized Petrov–Galerkin variational formulation \cite{jaimancomputational}.
We represent the cavitating flow as a homogeneous mixture of liquid and vapor via semi-empirical transport-equation-based modeling.
A fully-implicit residual-based stabilization and consistent linearizations have been incorporated that address numerical challenges normally encountered in the state-of-the-art cavitation solvers for unstructured meshes. A hybrid unsteady Reynolds-averaged Navier–Stokes (URANS) and large eddy simulation (LES) model based on the delayed detached eddy simulation treatment is employed to simulate the separated turbulent flow. A variationally consistent and robust hybrid URANS-LES developed by \citet{joshi2017variationally} is used for the modeling of turbulence with moving body-fitted fluid-structure interfaces. A nonlinear interface force correction algorithm is employed to correct and stabilize the fluid forces at each iterative step  \cite{jaiman2016partitioned,jaiman2016stable}. 
%
%
To begin, the body-conforming FSI framework is applied to the study of a freely oscillating hydrofoil subjected to unsteady cavitating conditions.  This validation work attempts to quantify the hydrodynamic interaction of the cavitating flow with the hydrofoil undergoing flow-induced vibration.


The central intent of this work is to perform a numerical investigation of the cavitating flow and the vibrational characteristics of an elastically-mounted hydrofoil subjected to the vorticity/cavity interactions. Coupled dynamics of unsteady cavitation and oscillating hydrofoil interaction, the force and amplitude characteristics and the vorticity and pressure distributions are investigated during the oscillation. The study seeks to answer two key questions related to cavitation-induced vibration of an elastically-mounted hydrofoil: (i) How the hydrofoil sustains the increased amplitude during the flow-induced vibration with cavity shedding? (ii) What is the underlying mechanism behind increased frequency of vortex shedding in the presence of cavitation? 
We primarily focus on the flow regime where unsteady partial
cavitation occurs, while an attention is paid to the corresponding evolution of lift force and the vorticity field. We restrain ourselves to low angles of attack which is aligned with a practical operating range of marine propeller blades.
 We explore a periodic generation of clockwise vorticity and its connection to the unsteady forces at the vicinity of the trailing edge of the hydrofoil. In the frequency lock-in regime, we examine coherent cavitating structures over the hydrofoil suction surface during a full cavity growth-detachment-collapse cycle. By quantifying the vorticity generation, we analyze the cavity dynamics and its impact on the unsteady lift generation and the frequency lock-in with large-amplitude vibrations.
 Such analysis and physical insight on the frequency lock-in and cavitation interactions may guide the development of effective active or passive suppression devices.

The paper is structured as follows. In Section 2, the mathematical model and computational framework are first presented. Section 3 discusses the problem set-up and the validation of turbulent cavitating and non-cavitating flow over a stationary and an elastically-mounted hydrofoil. In Section 4, we then employ the validated framework to study the flow-induced vibrations of a freely oscillating hydrofoil subjected to unsteady cavitating flows. The influence of the unsteady cavitation dynamics on the lift forces and structural response is systematically investigated. Concluding remarks and the
key findings of the current work are provided in Section 5.

\section{Mathematical model and computational framework}
\label{sec:methodology}
For the sake of completeness, we briefly review the salient features of the computational framework and interested readers are directed to the original work by \citet{KASHYAP202119} for further details of the numerical implementation.

\subsection{Governing fluid equations}
\label{sec:GE}
\noindent We consider the fluid physical domain $\Omega^{\mathrm{f}}(\boldsymbol{x}^{\mathrm{f}},t)$ with an associated fluid boundary $\Gamma^{\mathrm{f}}(t)$, where $\boldsymbol{x}^{\mathrm{f}}$ and $t$ represent the spatial and temporal coordinates. The working fluid, consisting of the liquid and vapor phases, is assumed to be present in the form of a continuous homogeneous mixture. The phase indicator $\phi^{\mathrm{f}}(\boldsymbol{x}^{\mathrm{f}},t)$ is used to represent the phase fraction of the liquid phase at any coordinate $(\boldsymbol{x}^{\mathrm{f}},t)$ in the homogeneous two-phase liquid-vapor mixture. The fluid density ($\rho^{\mathrm{f}}$) and dynamic viscosity ($\mu^{\mathrm{f}}$) are taken as linear combinations of $\phi^{\mathrm{f}}$
\begin{align} 
	\rho^{\mathrm{f}} 
	&=\rho_l\boldsymbol{\phi}^{\mathrm{f}} + \rho_v\left( 1-\boldsymbol{\phi}^{\mathrm{f}} \right),\label{densInterp}\\
	\mu^{\mathrm{f}}
	&=\mu_l\boldsymbol{\phi}^{\mathrm{f}} + \mu_v\left( 1-\boldsymbol{\phi}^{\mathrm{f}} \right),\label{viscInterp}
\end{align} 
\noindent where $\rho_l$ and $\rho_v$ are the densities of the pure liquid and vapor phases, respectively. $\mu_l$ and $\mu_v$ are the dynamic viscosities of the liquid and the vapor phases. 
\subsubsection{Cavitation modeling}
\noindent The phase indicator $\phi^{\mathrm{f}}$ is obtained as the solution of a scalar transport equation, which can be written in the conservative form in the Arbitrary Lagrangian-Eulerian (ALE) framework as: 
\begin{align} \label{TEM}
  \left. \frac{\partial \boldsymbol{\phi}^{\mathrm{f}}}{\partial t}\right|_{\boldsymbol{\chi}}
	+ \boldsymbol{\phi}^{\mathrm{f}}\nabla \cdot \boldsymbol{u}^{\mathrm{f}} 
	+ \left(\boldsymbol{u}^{\mathrm{f}}-\boldsymbol{u}^{\mathrm{m}}\right)\cdot\nabla\boldsymbol{\phi}^{\mathrm{f}} 
    = \dfrac{\dot{m}}{\rho_{l}},&&\mathrm{on}\ (\boldsymbol{x}^{\mathrm{f}},t)\in \Omega^{\mathrm{f}}
\end{align}
\noindent where $\boldsymbol{\chi}$ is the referential coordinate system, ${\boldsymbol{u}^{\mathrm{f}}} = {\boldsymbol{u}^{\mathrm{f}}}(\boldsymbol{x}^{\mathrm{f}},t)$ is the fluid velocity at each spatial point $\boldsymbol{x}^{\mathrm{f}} \in \Omega^{\mathrm{f}}$ and $\boldsymbol{u}^{\mathrm{m}}$ is the relative velocity of the spatial coordinates $\boldsymbol{x}^{\mathrm{f}}$ with respect to the referential coordinate system $\boldsymbol{\chi}$. The source term $\dot{m}$ in the transport equation is representative of a finite mass transfer rate that governs the rates of destruction and production of liquid by the process of cavitation. \cite{schnerr2001physical} proposed the source term $\dot{m}$ to be a non-linear function of $\boldsymbol{\phi}^{\mathrm{f}}$ and $p^{\mathrm{f}}$
\begin{align}
        \dot{m}  
        =  \frac{3 \rho_{l} \rho_{v}}{\rho^{\mathrm{f}} R_{B}} \sqrt{\frac{2}{3 \rho_{l}\left|p^{\mathrm{f}}-p_{v}\right|}}
         \bigg[ C_{c} \boldsymbol{\phi}^{\mathrm{f}}&(1-\boldsymbol{\phi}^{\mathrm{f}}) \operatorname{max}\left(p^{\mathrm{f}}-p_{v}, 0\right) \nonumber \\
        &+ C_{v} \boldsymbol{\phi}^{\mathrm{f}}(1 + \phi_{nuc} -\boldsymbol{\phi}^{\mathrm{f}}) \operatorname{min}\left(p^{\mathrm{f}}-p_{v}, 0\right)
        \bigg] \label{eq:schnerrSauer}
\end{align}
\noindent This model attempts to relate the finite mass transfer rate to the rate of growth/collapse of an equivalent spherical bubble under an external pressure field. Cavitation is assumed to initiate from nucleation sites present in the flow by a heterogeneous nucleation process \cite{brennen_2013}. The initial concentration of nuclei per unit volume ($n_0$) with an associated nuclei diameter($d_{nuc}$) is assumed to be a constant. It is also assumed that only vaporous cavitation occurs, and the effect of non-condensable gases is not considered. $R_B(\boldsymbol{x}^{\mathrm{f}},t)$ in Eq.~(\ref{eq:schnerrSauer}) is representative of the equivalent radius of the vapor volume at the coordinates $(\boldsymbol{x}^{\mathrm{f}},t)$, while $\phi_{nuc}$ is the phase fraction of the initial nucleation sites in an unit volume. These are calculated as
\begin{equation}
    R_B = \left( \frac{3}{4\pi n_0} \frac{1+\phi_{nuc}-\phi^{\mathrm{f}}}{\phi^{\mathrm{f}}} \right)^{1/3} \quad\mathrm{and}\quad 
    \phi_{nuc} = \frac{\dfrac{\pi n_0 d^3_{nuc}}{6}}{1+\dfrac{\pi n_0 d^3_{nuc}}{6}}
\end{equation}
  The vapor phase at any spatial location is assumed to be present in the form of a concentration of bubbles with identical radii. The model requires as input the condensation coefficient $C_c$ and the evaporation coefficient $C_v$. $C_c$ and $C_v$ are not part of the original model but are introduced in numerical implementations\cite{ghahramani2019comparative, cazzoli2016assessment} for enhancing the condensation and evaporation effect in the study of specific flow configurations. The cavitation model has been applied to the study of different cavitating flow configurations, including the collapse of vaporous bubbles \cite{KASHYAP202119, ghahramani2019comparative} and cavitating flow over hydrofoils \cite{ji2015large}. 

\subsubsection{Fluid momentum and mass conservation}
\noindent The unsteady Navier-Stokes equations for the fluid momentum and mass conservation can be written in an ALE framework as
\begin{align} 
	\left.\rho^{\mathrm{f}} \frac{\partial \boldsymbol{u}^{\mathrm{f}}}{\partial t}\right|_{\boldsymbol{\chi}}
	+\rho^{\mathrm{f}}\left(\boldsymbol{u}^{\mathrm{f}}-\boldsymbol{u}^{\mathrm{m}}\right) \cdot \nabla \boldsymbol{u}^{\mathrm{f}}
	-\nabla \cdot \boldsymbol{\sigma}
	=\boldsymbol{f}^{\mathrm{f}},&&\mathrm{on}\ (\boldsymbol{x}^{\mathrm{f}},t)\in \Omega^{\mathrm{f}}, \label{NS_mom}\\
	\left. \frac{\partial \rho^{\mathrm{f}}}{\partial t}\right|_{\boldsymbol{\chi}}+ \rho^{\mathrm{f}}\nabla \cdot \boldsymbol{u}^{\mathrm{f}} + \left(\boldsymbol{u}^{\mathrm{f}}-\boldsymbol{u}^{\mathrm{m}}\right)\cdot\nabla\rho^{\mathrm{f}} = 0,&&\mathrm{on}\ (\boldsymbol{x}^{\mathrm{f}},t)\in \Omega^{\mathrm{f}}, \label{NS_mass}
\end{align} 
\noindent \noindent where $\boldsymbol{f}^{\mathrm{f}}$ is the body force applied on the fluid and
\begin{equation}
    \boldsymbol{\sigma} = \boldsymbol{\sigma}^{\mathrm{f}} + \boldsymbol{\sigma}^{\mathrm{des}}
\end{equation}
where ${\boldsymbol{\sigma}^{\mathrm{f}}}$ and $\boldsymbol{\sigma}^{\mathrm{des}}$ are the Cauchy stress tensor for a Newtonian fluid and the turbulent stress tensor respectively, given by
\begin{align}
	{\boldsymbol{\sigma}^{\mathrm{f}}} 
	&= -{p^{\mathrm{f}}}\boldsymbol{I} + \mu^{\mathrm{f}}( \nabla{\boldsymbol{u}^{\mathrm{f}}}+ (\nabla{\boldsymbol{u}^{\mathrm{f}}})^T),\\
	{\boldsymbol{\sigma}^{\mathrm{des}}} 
	&= \mu_T( \nabla{\boldsymbol{u}^{\mathrm{f}}}+ (\nabla{\boldsymbol{u}^{\mathrm{f}}})^T),
\end{align}
where ${p^{\mathrm{f}}}$ denotes the fluid pressure and $\mu_T$ is the turbulent viscosity. $\boldsymbol{\sigma}^{\mathrm{des}}$ is modeled using the Boussinesq approximation and in the current work a hybrid URANS-LES turbulence model is applied. The details of the turbulence model implementation can be found in \citet{joshi2017variationally}. 

\subsubsection{Convective form of cavitation transport equation and local fluid compressibility}
\noindent In the present work, the conservative form of the transport equation is re-arranged in the form a convection-reaction equation. Taking the material derivative of Eq.~(\ref{densInterp}) in the ALE framework, we obtain
\begin{equation}\label{densMaterialDerivative}
  \left. \frac{\partial \rho^{\mathrm{f}}}{\partial t}\right|_{\boldsymbol{\chi}}+\left(\boldsymbol{u}^{\mathrm{f}}-\boldsymbol{u}^{\mathrm{m}}\right) \cdot \nabla \rho^{\mathrm{f}} = \left( \rho_l - \rho_v \right) \left( \left. \frac{\partial \boldsymbol{\phi}^{\mathrm{f}}}{\partial t}\right|_{\boldsymbol{\chi}}+\left(\boldsymbol{u}^{\mathrm{f}}-\boldsymbol{u}^{\mathrm{m}}\right) \cdot \nabla \boldsymbol{\phi}^{\mathrm{f}} \right)
\end{equation} 
\noindent Combining equations (\ref{TEM}), (\ref{NS_mass}) and (\ref{densMaterialDerivative}), the following forms of the mass continuity equation and the phase indicator transport equation are obtained, which are used in the current implementation.
\begin{align} 
	\nabla \cdot \boldsymbol{u}^{\mathrm{f}} 
	= \left(\frac{1}{\rho_{l}} - \frac{1}{\rho_{v}} \right) \dot{m},
	&&\mathrm{on}\ (\boldsymbol{x}^{\mathrm{f}},t)\in \Omega^{\mathrm{f}}, \label{eq:NS_massMod}\\
	\left. \frac{\partial \boldsymbol{\phi}^{\mathrm{f}}}{\partial t}\right|_{\boldsymbol{\chi}}
	+\left(\boldsymbol{u}^{\mathrm{f}}-\boldsymbol{u}^{\mathrm{m}}\right) \cdot \nabla \boldsymbol{\phi}^{\mathrm{f}} 
	= \frac{\rho^{\mathrm{f}}}{\rho_{l}\rho_{v}}\dot{m},
	&&\mathrm{on}\ (\boldsymbol{x}^{\mathrm{f}},t)\in \Omega^{\mathrm{f}}, \label{phiMod}
\end{align} 
It is observed that the divergence of the velocity field is no longer zero, and local dilation effects are introduced that are governed by the finite mass transfer rate. This local compressibility exists only within the two-phase mixture and the pure phases are incompressible, since for the cavitation model no mass transfer occurs when $\boldsymbol{\phi}^{\mathrm{f}}$ equals $0$ or $1$.

\subsubsection{Fluid-structure boundary conditions and fluid mesh deformation}
In the current work, we study cavitating flow over a freely oscillating hydrofoil. We briefly review the fluid-structure interaction boundary conditions and the ALE mesh motion in the continuum setting. The modeling of FSI requires the satisfaction of the velocity continuity and traction equilibrium at the fluid-structure boundary $\Gamma^{\mathrm{fs}}$. 
Let us consider a structural domain $\Omega^{s} \subset \mathbb{R}^{d}$ with an associated structural boundary $\Gamma^{\mathrm{s}}(0)$ at time $t=0$. Let the function $\boldsymbol{\varphi}^{\mathrm{s}}\left(\boldsymbol{x}^{\mathrm{s}}, t\right)$ maps the deformation of the structure from its initial configuration $\Omega^{s}$ to a deformed configuration $\Omega^{s}(t)$ at time $t$, where $\boldsymbol{x}^{\mathrm{s}}$ denote the material coordinates. We denote the initial fluid-structure interface at $t=0$ by $\Gamma^{\mathrm{fs}}(0)=\Gamma^{\mathrm{f}}(0) \cap \Gamma^{\mathrm{s}}(0)$. At time $t$ the interface will then be deformed as $\Gamma^{\mathrm{fs}}(t)=\boldsymbol{\varphi}^{\mathrm{s}}\left(\Gamma^{\mathrm{fs}}, t\right)$. The following kinematic and dynamic conditions are satisfied on $\Gamma^{\mathrm{fs}}$
\begin{align}
\boldsymbol{u}^{\mathrm{f}}\left(\boldsymbol{\varphi}^{\mathrm{s}}\left(\boldsymbol{x}^{\mathrm{s}}, t\right), t\right) &=\boldsymbol{u}^{\mathrm{s}}\left(\boldsymbol{x}^{\mathrm{s}}, t\right), & & \forall \boldsymbol{x}^{\mathrm{s}} \in \Gamma^{\mathrm{fs}} \\
\int_{\boldsymbol{\varphi}^{\mathrm{s}}(\gamma, t)} \boldsymbol{\sigma}^{\mathrm{f}} \cdot \mathbf{n}^{\mathrm{f}} d \Gamma+\int_{\gamma} \boldsymbol{\sigma}^{\mathrm{s}} \cdot \mathbf{n}^{\mathrm{s}} d \Gamma &=0, && \forall \gamma \subset \Gamma^{\mathrm{fs}}
\end{align}
where $u^{s}$ is the velocity of the structural domain, $\mathbf{n}^{\mathrm{f}}$ and $\mathbf{n}^{\mathrm{s}}$ are the unit normals to the deformed fluid elements $\boldsymbol{\varphi}^{\mathrm{s}}(\gamma, t)$ and their corresponding structural elements $\gamma$ on the interface $\Gamma^{\mathrm{fs}}$ respectively. The structural stress tensor $\sigma^{\mathrm{s}}$ is modeled depending on the type of material.

Away from the interface $\Gamma^{\mathrm{fs}}$ and any Dirichlet conditions on $\boldsymbol{\eta}^{\mathrm{f}}$ on the Dirichlet boundary $\Gamma_{D}^{\mathrm{m}}$, the fluid spatial coordinates are updated to conform to the structural deformation. The motion of the coordinates which are not at $\Gamma^{\mathrm{fs}}$ is modeled as an elastic material in equilibrium and the mesh equation is solved as
\begin{align}
\nabla \cdot \boldsymbol{\sigma}^{\mathrm{m}} &=\mathbf{0}, && \text { on } \Omega^{\mathrm{f}}, \label{eq:meshUpdate}\\
\boldsymbol{\eta}^{\mathrm{f}} &=\boldsymbol{\eta}_{D}^{\mathrm{f}}, && \forall \boldsymbol{x}^{\mathrm{f}} \in \Gamma_{D}^{\mathrm{m}}
\end{align}
where $\boldsymbol{\sigma}^{\mathrm{m}}=\left(1+k_{m}\right)\left[\nabla \boldsymbol{\eta}^{\mathrm{f}}+\left(\nabla \boldsymbol{\eta}^{\mathrm{f}}\right)^{T}+\left(\nabla \cdot \boldsymbol{\eta}^{\mathrm{f}}\right) \boldsymbol{I}\right]$ is the stress experienced at the fluid spatial coordinates due to the
strain induced by the deformation of the interface, $\boldsymbol{\eta}^{\mathrm{f}}$ is the displacement of the fluid spatial coordinates. The amount of deformation of the spatial coordinates is controlled using the local stiffness parameter $k_{m}$. Dirichlet conditions for the fluid mesh displacement $\boldsymbol{\eta}_{D}^{\mathrm{f}}$ are satisfied on the boundary $\Gamma_{D}^{\mathrm{m}}$.

All the aforementioned continuum equations are solved implicitly via a nonlinear partitioned iterative manner \cite{Jaiman-2021-computational}. For our stabilized Petrov-Galerkin discretization of the variables $\boldsymbol{u}^{\mathrm{f}}$, $p^{\mathrm{f}}$ and $\phi^{\mathrm{f}}$, we consider equal-order interpolations and the generalized-$\alpha$ method for performing the time integration \cite{jansen2000generalized}. The equations are linearized via the Newton-Raphson technique and are then solved in a predictor-corrector format. The left-hand side matrix is not constructed explicitly and the Harwell-Boeing sparse matrix format is used to store the matrices for the linear system of equations. A Generalized Minimal RESidual (GMRES)\cite{saad1986gmres} algorithm is used to solve the linear system. The solver uses communication protocols based on standard message passing interface \cite{MPI} for parallel computing on distributed memory clusters. 
The adopted variational solver has been extensively validated for a wide range of single and two-phase FSI problems \cite{jaiman2016stable,law2017wake,joshi2019hybrid,chizfahm2021transverse,miyanawala2018square}.

\section{Simulation setup}
\begin{figure}[!h]
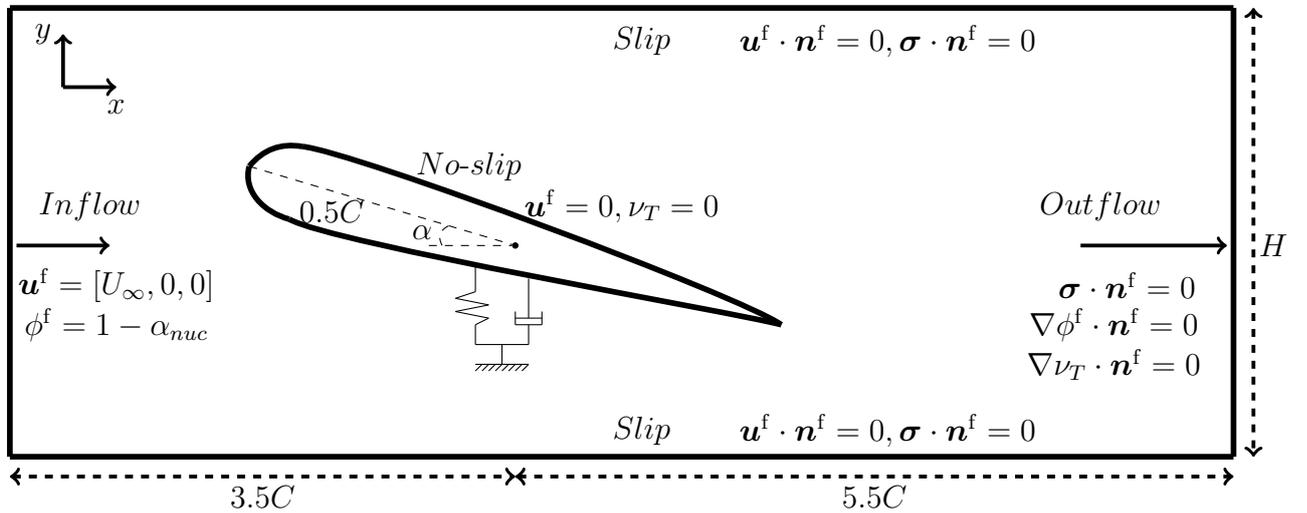

\centering
\ctikzfig{Methods/pitchingSchematic}
\caption{Representative computational domain and associated boundary conditions for cavitating flow over hydrofoil}
\label{fig:pitchingSchematic}
\end{figure}

For our numerical study, we consider a NACA66 hydrofoil section. Turbulent cavitating flow over a NACA66 hydrofoil has been extensively studied using laboratory experiments \cite{leroux2004experimental, ji2015large}. This is an often-encountered scenario in marine propellers where fluid acceleration over the hydrofoil surface can result in very low pressures and cavity inception near the blade leading edge. Figure~\ref{fig:pitchingSchematic} shows the general schematic of the computational domain used in the sections to follow. Here $C$ denotes the hydrofoil chord length, $\alpha$ is the angle of attack of the incoming flow, $H$ is the channel height and $\nu_{T}$ is the kinematic turbulence viscosity.  

\subsection{Stationary hydrofoil validation}
\begin{figure}[!h]
\centering
\includegraphics[width=\columnwidth]{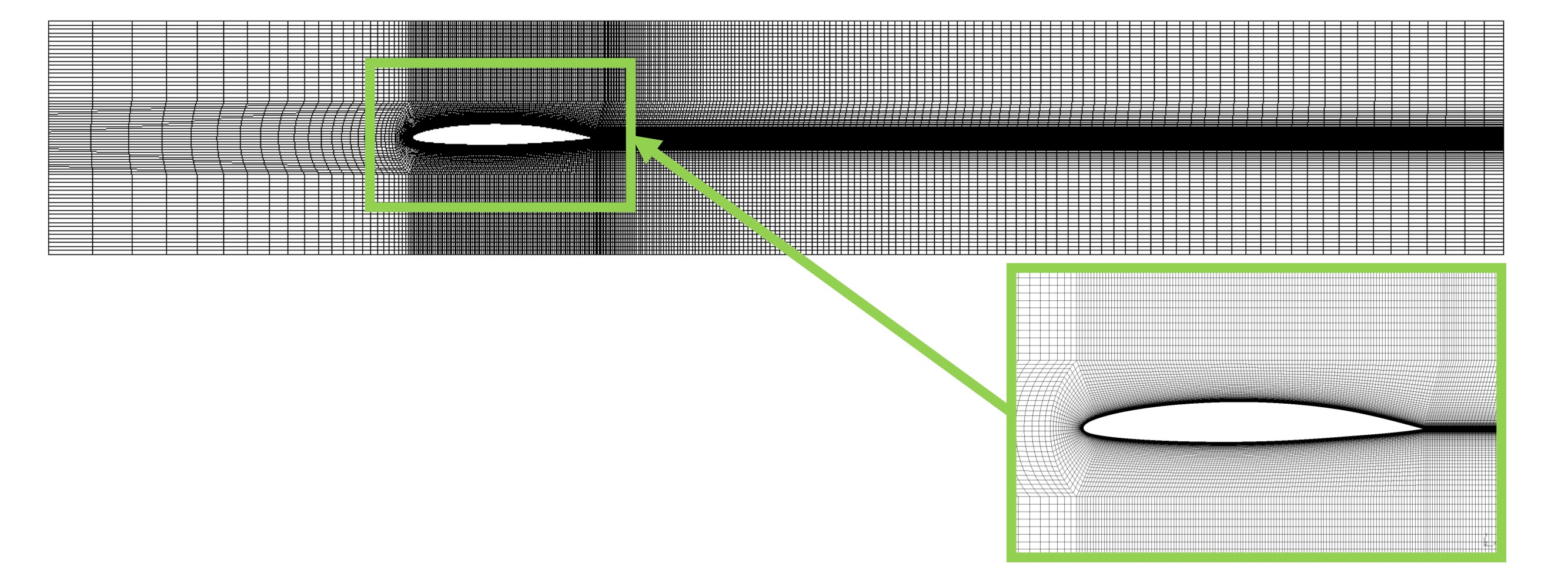}
\caption{Computational mesh for NACA66 hydrofoil. Inlay showing mesh in the vicinity of the hydrofoil}
\label{fig:NACA66grid}        
\end{figure}

Using our computational framework, we attempt to numerically replicate the experimental setup employed by \citet{leroux2004experimental} turbulent cavitating flows over a stationary hydrofoil. Before proceeding to the numerical study, we first determine the appropriate finite element mesh to be utilized. A grid-sensitivity test is performed using three computational grids. A target $y^+=y u_{\tau}/\nu=1$ is maintained in the discretization of the hydrofoil boundary layer for all three grids, where $y$ is the height of the first node from the wall, $u_{\tau}$ is the friction velocity and $\nu$ is the kinematic viscosity of the single-phase liquid. The fine grid used in the validation study is shown in Fig.~\ref{fig:NACA66grid}.
The domain is discretized with 93296 hexahedral and prism elements and a 2D periodic boundary condition is applied in the spanwise direction. The fluid domain is initialized with a liquid phase fraction of $\phi^{\mathrm{f}}=1$. A freestream velocity $U_\infty$ is applied at the inlet as a Dirichlet boundary condition. A traction-free outflow boundary condition is used, weakly setting $p_\infty=0$.

\begin{figure}[!h]
\centering
\includegraphics[width=0.95\columnwidth]{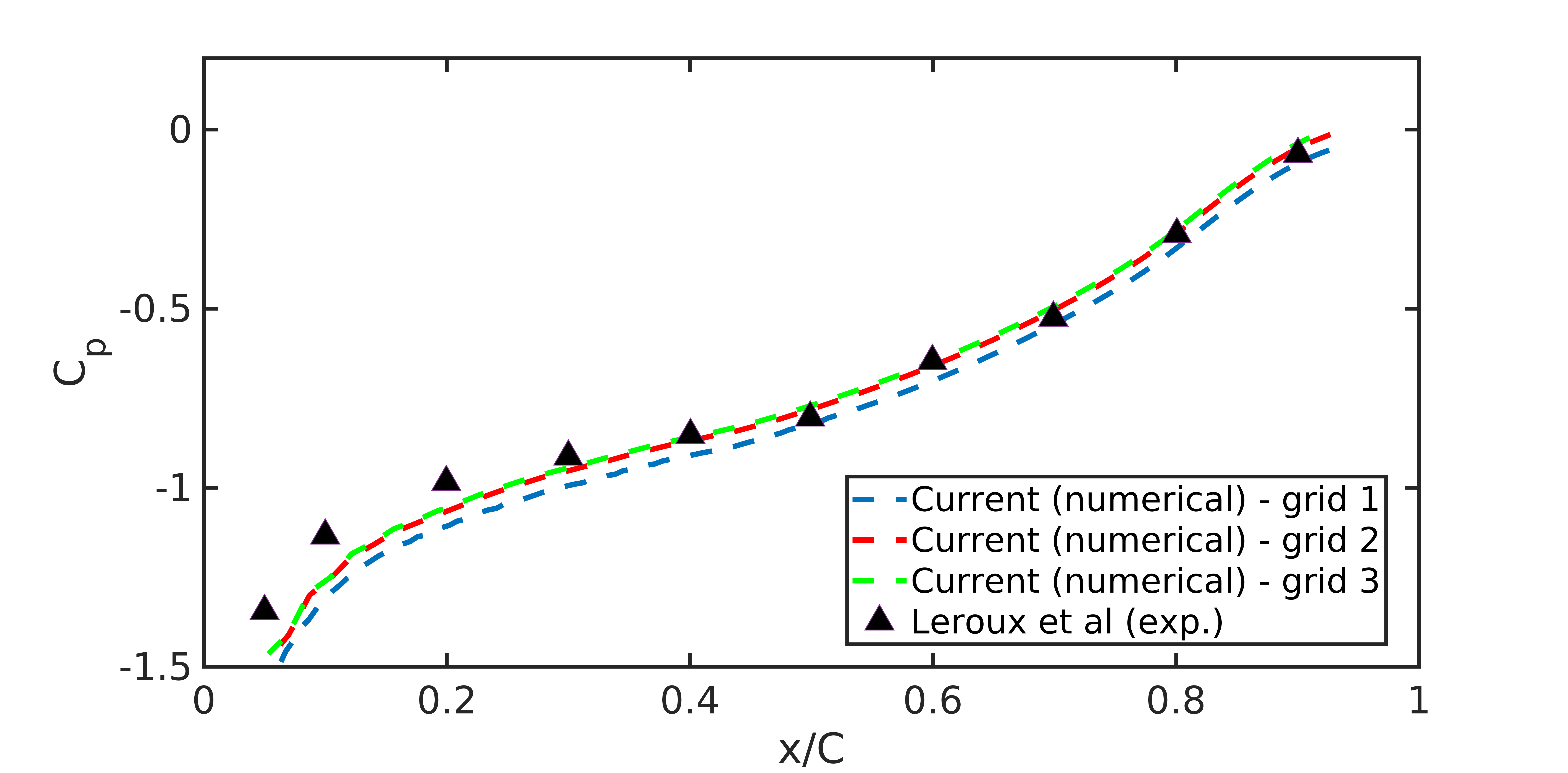}
\caption{Comparison of the predicted pressure coefficient with experimental values for the three grid sizes. The physical conditions correspond to $\alpha = 6.5^{\circ}, Re = 0.8\times10^6, U_\infty = 5.33~\si{m.s^{-1}}, \sigma = 8$.}
\label{fig:gridSensitivity}        
\end{figure}

Figure~\ref{fig:gridSensitivity} shows the comparison of the numerically predicted pressure coefficient with the experimental data of \cite{leroux2004experimental} on the suction surface of the hydrofoil.   The cavitation number of the flow is defined as $\sigma = \dfrac{p_\infty-p_v}{0.5\rho_l U_{\infty}^2}$, where $p_{\infty}$ is the free-steam hydrostatic pressure. For non-cavitating flow configurations we set the cavitation number to a value of $\sigma = 8$. The vapor pressure is prescribed as an input parameter to define the cavitation number of the flow. Liquid water at $25^{\circ}C$ is taken as the working fluid, with a pure liquid density of $\rho_l=998~\si{kg.m^{-3}}$, pure vapor density of $\rho_l=0.023~\si{kg.m^{-3}}$, liquid dynamic viscosity $\mu_l=1.1\times10^{-3}~\si{Pa.s}$ and a vapor phase dynamic viscosity $\mu_v=9.95\times10^{-6}~\si{Pa.s}$. For the cavitation model, the numerical parameters $n_0$, $d_{nuc}$, $C_c$ and $C_v$ are required as inputs. Consistent with \cite{schnerr2001physical, ji2015large},  the value of $n_0$ is set to $10^{13}$. Numerical experiments are then performed to determine the value of the parameter $d_{nuc}$ as $2.5\times10^{-6}$. In this work, $C_c$ and $C_v$ are set to $10^{-3}$ and $5\times10^{-3}$, respectively. A solver time-step of $\Delta t=t_{\infty}/200$ \cite{coutier2003numerical} is used for all the numerical studies, where $t_\infty = C/U_\infty$.

\begin{figure*}[t!]
    \centering
    \begin{subfigure}[b]{0.5\columnwidth}
        \centering
        \centerline{\includegraphics[height=7.5cm,clip=false]{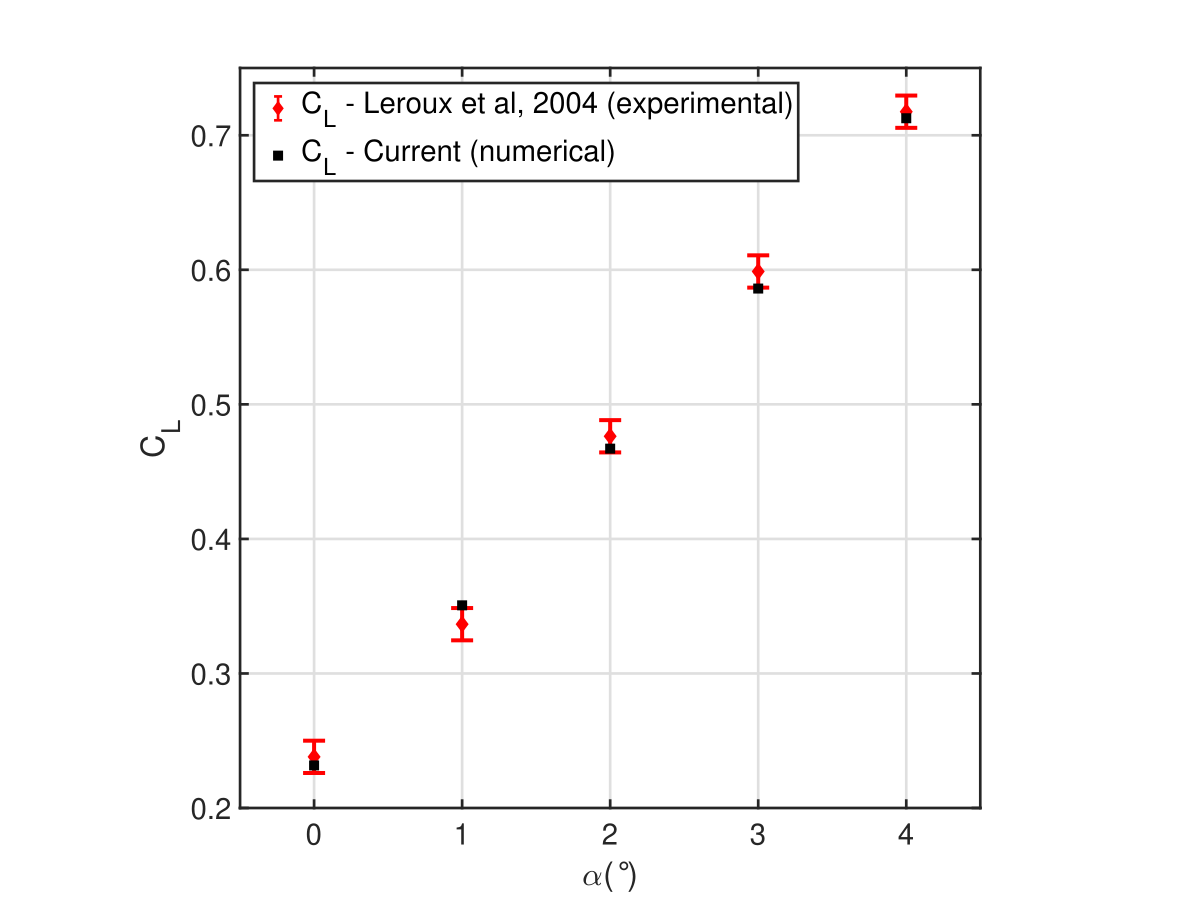}}
        \subcaption{}
    \end{subfigure}%
    ~ 
    \begin{subfigure}[b]{0.5\columnwidth}
        \centering
        \centerline{\includegraphics[height=7.5cm,clip=false]{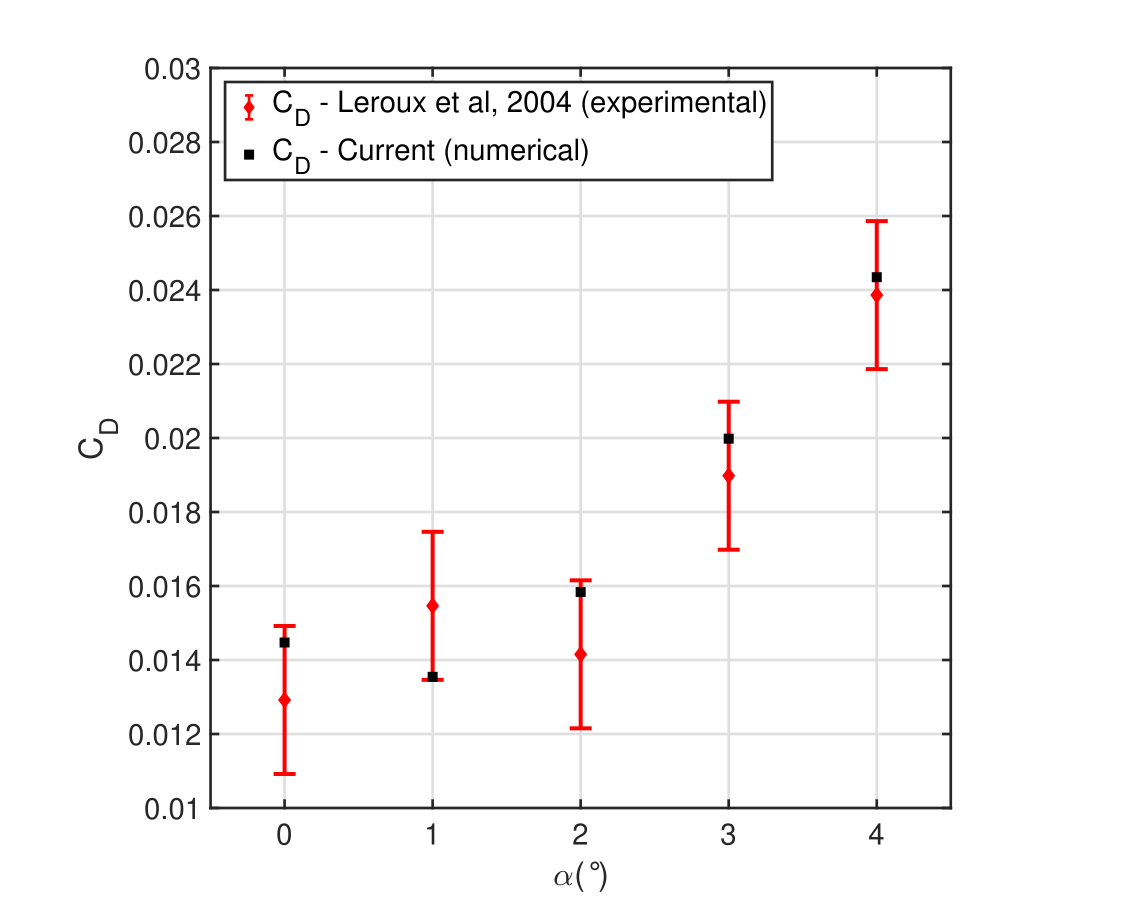}}
        \subcaption{}
    \end{subfigure}
    \caption{Comparisons of predicted time-averaged lift ($C_L$) and drag ($C_D$) coefficients  with the experimental values from \cite{leroux2004experimental} in the non-cavitating regime.}\label{fig:turbValidation}
\end{figure*}

\begin{figure}[!h]
\centering
\includegraphics[width=0.7\columnwidth]{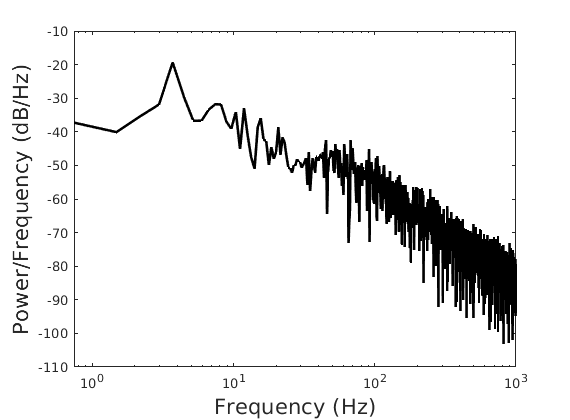}
\caption{Power spectrum density of the predicted pressure coefficient on the suction surface of the hydrofoil at $x/c = 0.7$. Spectrum peak at 3.78Hz.}
\label{fig:PSD_Cd}        
\end{figure}

\begin{figure}
\begin{subfigure}{\textwidth}
  \centering
  \includegraphics[width=0.5\linewidth]{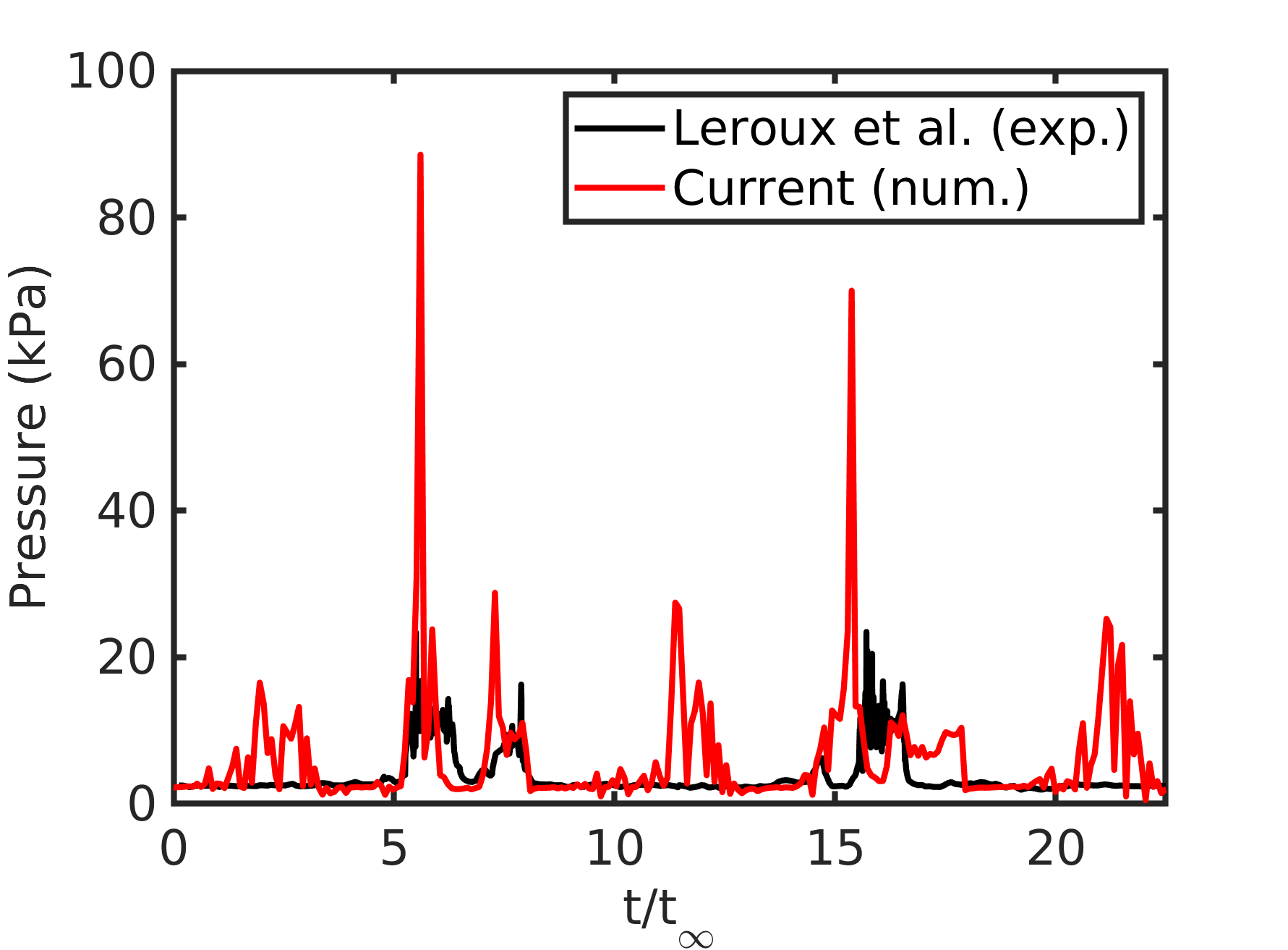}
  \caption{$x/C = 0.3$}
  \label{fig:presValidationD3}
\end{subfigure}
\begin{subfigure}{\textwidth}
  \centering
  \includegraphics[width=0.5\linewidth]{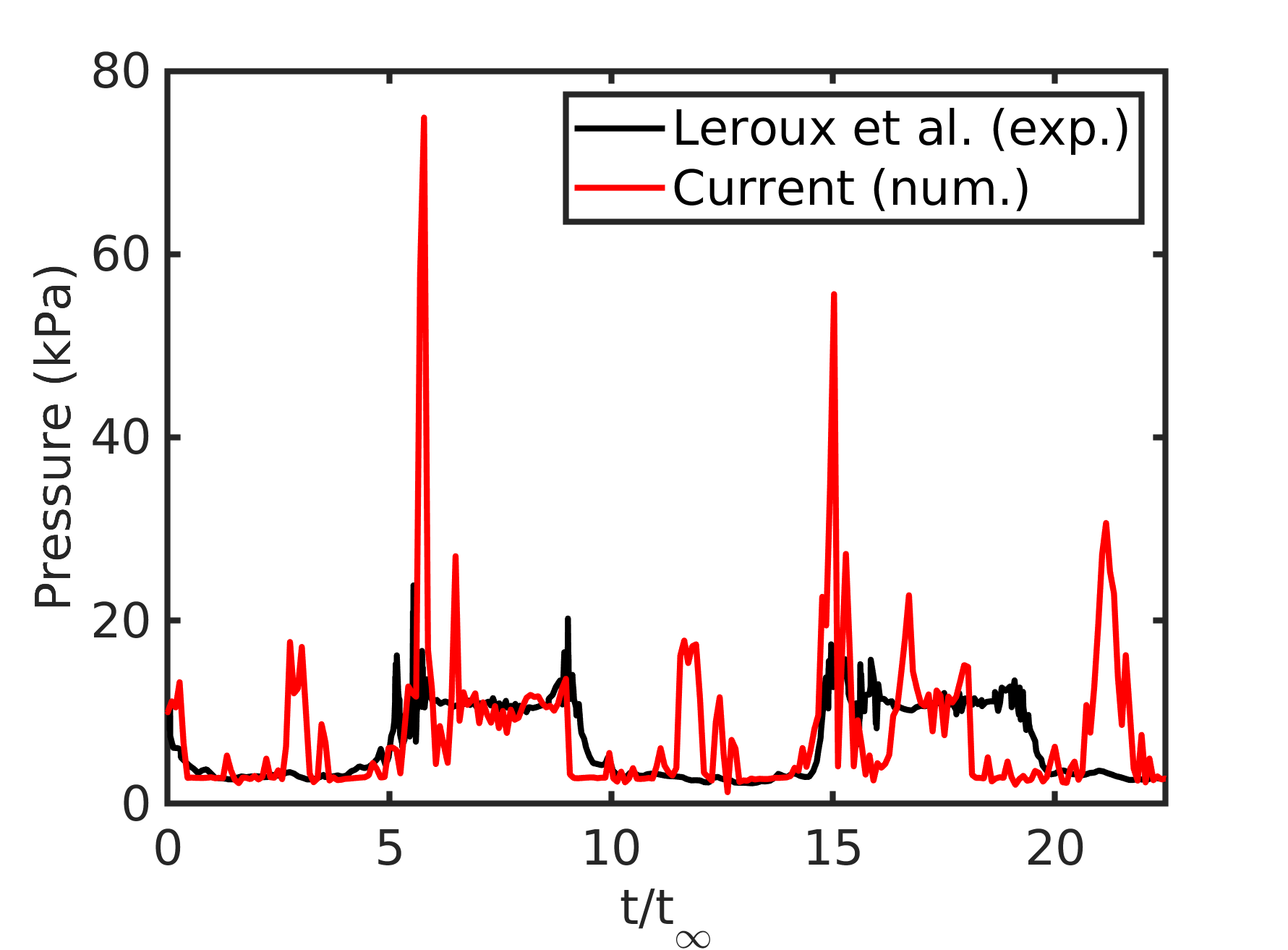}
  \caption{$x/C = 0.5$}
  \label{fig:presValidationD5}
\end{subfigure}
\begin{subfigure}{\textwidth}
  \centering
  \includegraphics[width=0.5\linewidth]{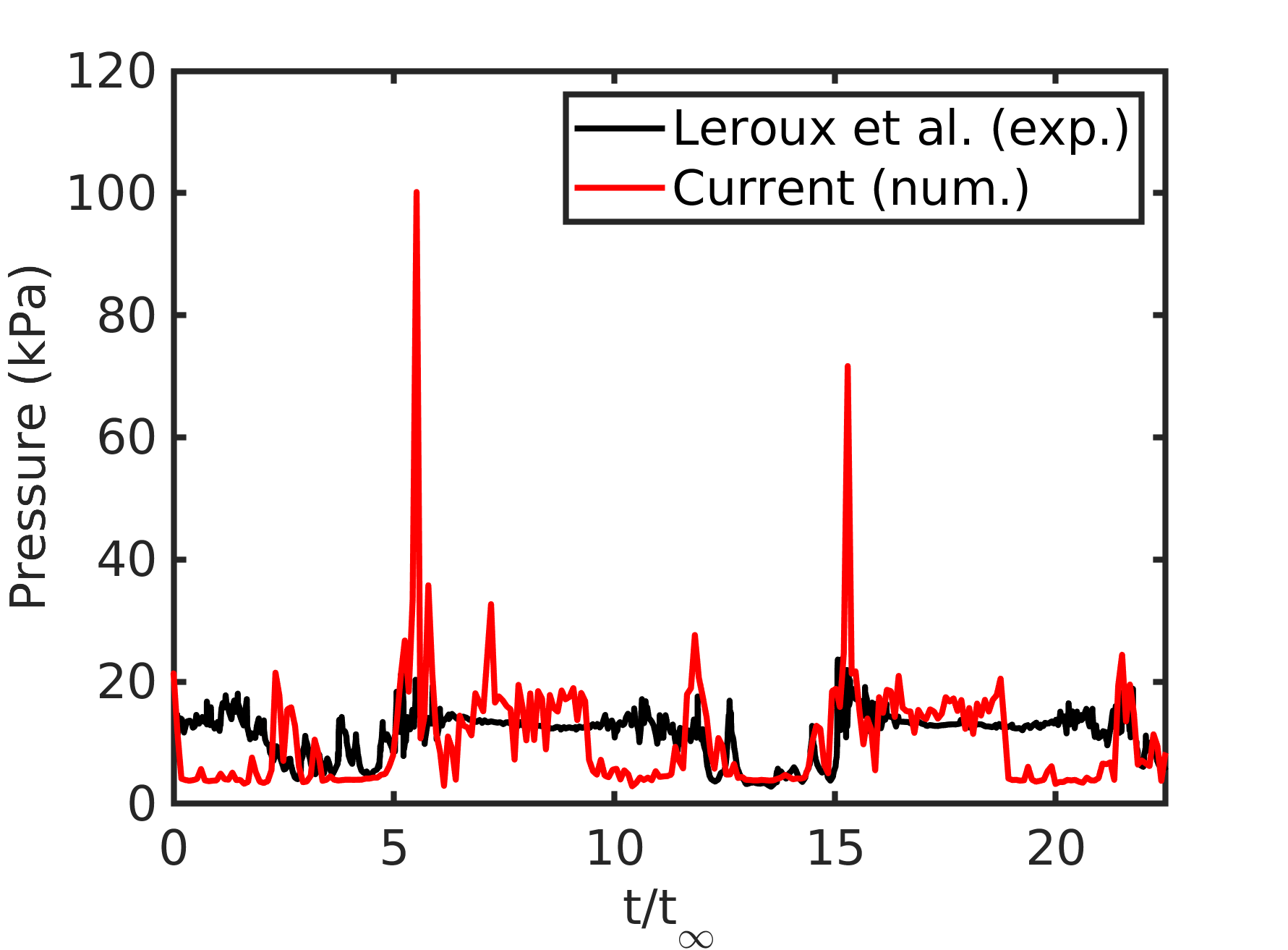} 
  \caption{$x/C = 0.7$}
  \label{fig:presValidationD7}
\end{subfigure}
\caption{Pressure predicted on the suction surface of the hydrofoil compared to experimental values \cite{leroux2004experimental}. Data from three pressure probes were used for comparison - aligned along the chord at distances $x/C=0.3$, $x/C=0.5$ and $x/C=0.7$ from the leading edge.}
\label{fig:presProbeValidation}
\end{figure}

For our validation, the angle of attack $\alpha$ of the hydrofoil is varied between $0^{\circ}-4^{\circ}$ for the non-cavitating flow condition, and the time-averaged lift $C_L$ and drag $C_D$ coefficients are monitored.
Figure~\ref{fig:turbValidation} shows the comparison of $C_L$ and $C_D$ obtained from the numerical simulation with the experimental values. The predicted numerical results agree well with the experimental values in the non-cavitating regime and are within the uncertainties for $C_L (\Delta C_L = 0.012)$ and $C_D (\Delta C_D = 0.002)$ reported in \cite{leroux2004experimental}.
Next, we decrease the cavitation number of the flow to $\sigma = 1.25$. The angle of attack $\alpha$ of the hydrofoil is set to $6^{\circ}$. This is a condition for which leading-edge cavitation develops over the hydrofoil suction surface. The cavity follows a quasi-periodic growth and shedding cycle. The resulting periodic pressure oscillations over the hydrofoil surface are of particular interest due to their potential contribution to noise and blade vibration. To determine the ability of the current cavitation model to capture these oscillating pressures, we compare the numerically predicted pressures at specific locations on the suction surface and compare them with experimental data from \cite{leroux2004experimental}. Figure~\ref{fig:presProbeValidation} shows the numerically predicted and experimental pressure at three probe locations along the suction surface characterized by the chord length. The numerical simulations can recover the pressure inside the cavity, although the peak pressures during cavity collapse are overestimated. In particular, the periodicity of pressure oscillations is reasonably captured.  Figure~\ref{fig:PSD_Cd} shows the power spectrum density of the time-varying pressure coefficient at a relative distance $x/C = 0.7$ on hydrofoil suction surface. The dominant frequency is driven by the cavity shedding cycles at $3.78$~Hz and agrees well with the experimentally obtained shedding frequency of $3.67$~Hz. 

\noindent 

\subsection{Freely vibrating hydrofoil}
After validating our numerical framework, we study the effect of cavitation on flow-induced vibrations of the hydrofoil. The hydrofoil section is modeled as an elastically mounted rigid body and allowed to freely oscillate in the transverse direction, as shown in Fig.~\ref{fig:pitchingSchematic}.
We first identify the key parameters influencing the motion characteristics of the hydrofoil. The translational flow-induced vibration of a cavitating hydrofoil is strongly influenced by the four key non-dimensional parameters, namely mass-ratio $\left(m^{*}\right)$, Reynolds number $(R e)$, reduced velocity $\left(U_{r}\right)$, and critical damping ratio $(\zeta)$ defined as:
$$
m^{*}=\frac{M}{m_{f}}, \quad R e=\frac{\rho^{\mathrm{f}} U_\infty C}{\mu^{\mathrm{f}}}, \quad U_{r}=\frac{U_\infty}{f_{N} C}, \quad \zeta=\frac{C_\zeta}{2 \sqrt{K M}}
$$
where $M$ is the mass per unit length of the body, $C_\zeta$ and $K$ are the damping and stiffness coefficients, respectively for an equivalent spring-mass-damper system of a vibrating structure, $U_\infty$ and $C$ the free-stream speed and the hydrofoil chord length, respectively. The wetted natural frequency of the body is given by $f_{N}=(1 / 2 \pi) \sqrt{K / M}$. The mass of displaced fluid by the structure is $m_{f}=\rho^{\mathrm{f}} A C S$, where $A$ is the cross-sectional area and $S$ denotes the span of the hydrofoil section. 

The hydrofoil section submerged in the flow stream experiences transient flow-induced forces and consequently may undergo rigid body motion if mounted elastically. In the absence of external body forces, the rigid-body motion of the hydrofoil along the Cartesian axes, is governed by the following equation:
\begin{equation}
    {M} \frac{\partial \boldsymbol{u}^{\mathrm{s}}}{\partial t} + C_\zeta  \boldsymbol{u}^{\mathrm{s}} + K \left(\boldsymbol{\varphi}^{\mathrm{s}}\left({y}_{0}, t\right)-{y}_{0}\right)=\boldsymbol{F}^{\mathrm{s}} 
\end{equation}
where ${M}, {C_\zeta}$ and ${K}$ denote the mass, damping coefficient and stiffness.  $\Omega^{\mathrm{s}}$ denotes the rigid body, $\boldsymbol{u}^{\mathrm{s}}(t)$ represents the rigid-body velocity at time $t$, and $\boldsymbol{F}^{\mathrm{s}}$ is the fluid traction acting on the rigid body, respectively. Here $\boldsymbol{\varphi}^{s}$ denotes the position vector mapping the initial position $y_{0}$ of the rigid body to its position at time $t$. The hydrofoil transverse displacement $y_{disp}(t)$ is given by 
\begin{equation}
    y_{disp}(t) = \boldsymbol{\varphi}^{\mathrm{s}}\left({y}_{0}, t\right)-{y}_{0}
\end{equation}
The spatial and temporal coordinates are denoted by $\boldsymbol{x}$ and $t$, respectively. 
In the presence of cavitation, the angle of attack $\alpha$ of the hydrofoil and the cavitation number $\sigma$ are additional important parameters. In addition, we define the frequencies $f_{cav}$, $f_{disp}$ and $f_{C_L}$ as the dominant frequencies in the unsteady cavity length, the structural displacement and the lift coefficient respectively. 
The fluid loading is computed by integrating the surface traction considering the first layer of elements located on the hydrofoil surface. The instantaneous force coefficients are defined as
\begin{equation}
    \begin{aligned}
        C_{L} &=\frac{1}{\frac{1}{2} \rho^{\mathrm{f}} U_\infty^{2} S C} \int_{\Gamma}\left(\boldsymbol{\sigma}^{f} \cdot \boldsymbol{n}\right) e_{y} d {\Gamma} \\
        C_{D} &=\frac{1}{\frac{1}{2} \rho^{\mathrm{f}} U_\infty^{2} S C} \int_{\Gamma}\left(\boldsymbol{\sigma}^{f} \cdot \boldsymbol{n}\right) e_{x} d {\Gamma}
    \end{aligned}
    \label{eq:forceCoefficients}
\end{equation}
Here $e_{x}$ and $e_{y}$ are the streamwise and cross-flow components of the unit normal $\boldsymbol{e}$.
We limit ourselves to low angles of attack $\alpha$ keeping in mind the general operating ranges of marine propellers and hydrofoils. In this study we are particularly interested in the flow configuration where unsteady partial cavitation occurs. For this, a target value of $\dfrac{\sigma}{2\alpha}=3$ is taken for all the angles of attack \cite{arndt2012some}. The Reynolds number $Re$ of the flow is set to $10^5$, which corresponds to the freestream velocity of $U_\infty = 0.735\si ~\mathrm{m. s^{-1}}$. A mass ratio $m^* = 5$ is taken for all the cases.

\section{Results and Discussion}
It is well-known that cavitation can cause unsteady vibrations of underwater structures. The complexity of the coupled fluid-structure dynamics of a freely vibrating hydrofoil is significantly enhanced by the cavity shedding. 
In the current study, we allow the hydrofoil to freely oscillate and vary the reduced velocity $U_r$. The reduced velocity $U_r \propto 1/\sqrt{K}$ and thus an increase in $U_r$ is indicative of a reduction of the stiffness of the hydrofoil while keeping a fixed flow speed and the mass of the hydrofoil.

\subsection{Vibration response}
\begin{figure}[!h]
\centering
\includegraphics[width=0.85\columnwidth]{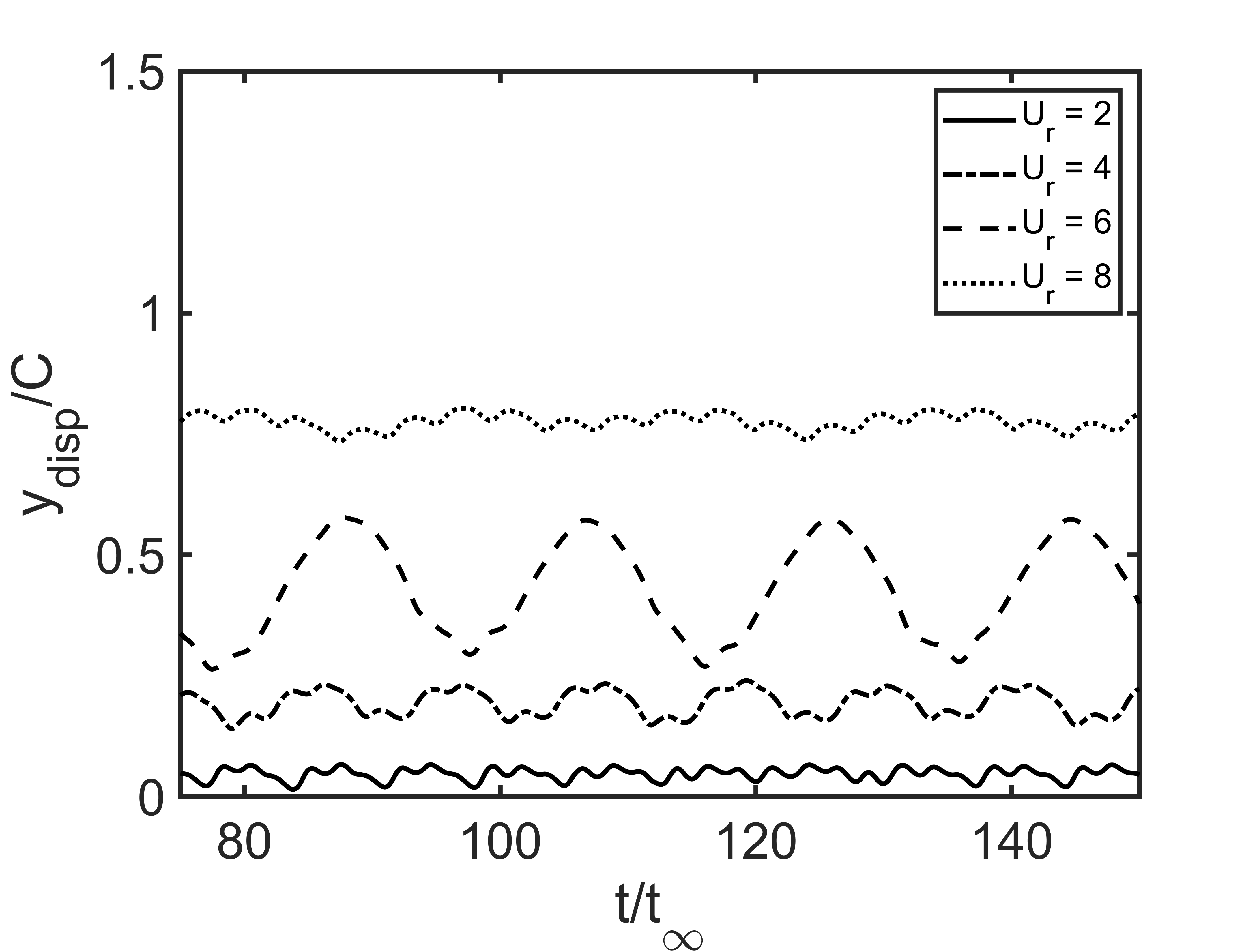}
\caption{Time evolution of hydrofoil transverse displacement $y_{disp}$  for varying $U_r$ at representative system parameters $\alpha = 5^{\circ}, \sigma = 0.52$.}
\label{fig:a5_yDispSampled}        
\end{figure}

\begin{figure}[!h]
\centering
\includegraphics[width=0.95\textwidth]{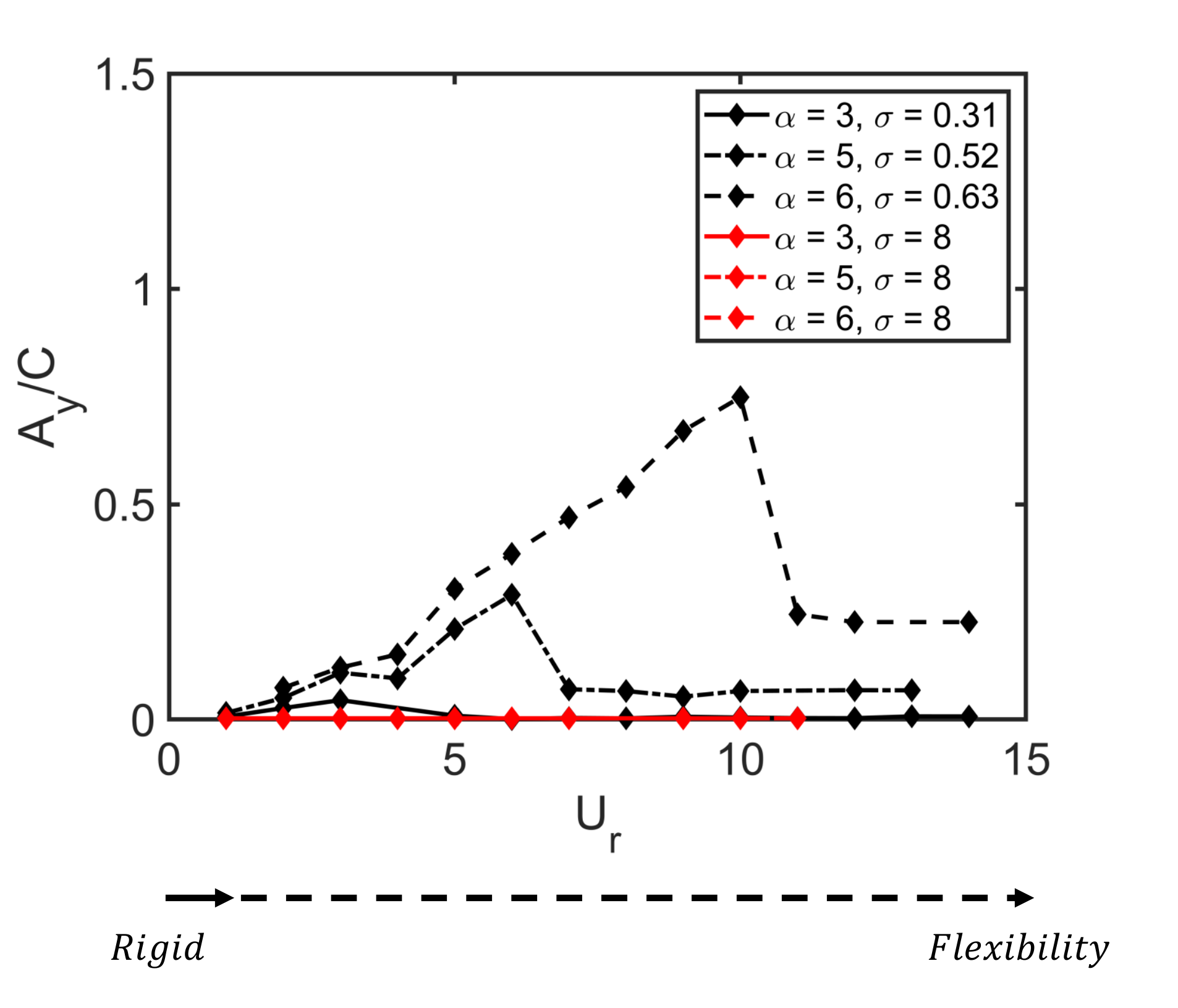}
\caption{Peak-to-peak amplitude $A_y$ of oscillations in transverse motion $y_{disp}$, normalized by the hydrofoil chord length $C$. Non-cavitating flow configurations ($\sigma = 8$) are shown in red, while the cavitating configurations are plotted in black.}
\label{fig:ampCompAll}        
\end{figure}

 As the reduced velocity $U_r$ increases and the stiffness decreases, the hydrofoil transverse displacement $y_{disp}$ follows a general increasing trend. At certain values of $U_r$  periodic oscillations in $y_{disp}$ are observed. Figure \ref{fig:a5_yDispSampled} shows representative values of $y_{disp}$ normalized by the hydrofoil chord $C$ with increasing $U_r$ at $\alpha = 5^{\circ}$. We define the amplitude of oscillations $A_y$ as the peak-peak amplitude of $y_{disp}$.
 \noindent 
 Figure~\ref{fig:ampCompAll} shows the compiled values of $A_y$ across the range of $U_r \in \left[1, 14\right]$ for different $\alpha$ under cavitating and non-cavitating conditions. For the non-cavitating conditions, no significant oscillations are observed.  This is not unexpected for streamlined bodies at low $\alpha$ where flow unsteadiness is low. In contrast, for the cavitating conditions as $U_r$ increases a consistent increase in $A_y$ is observed culminating in a distinct peak. On further increasing $U_r$, the oscillation amplitude drops to a nearly constant value. 
\noindent 

\begin{figure}[htbp]
\centering
\begin{subfigure}[t]{0.49\textwidth}
    \centering
    \includegraphics[width=\linewidth]{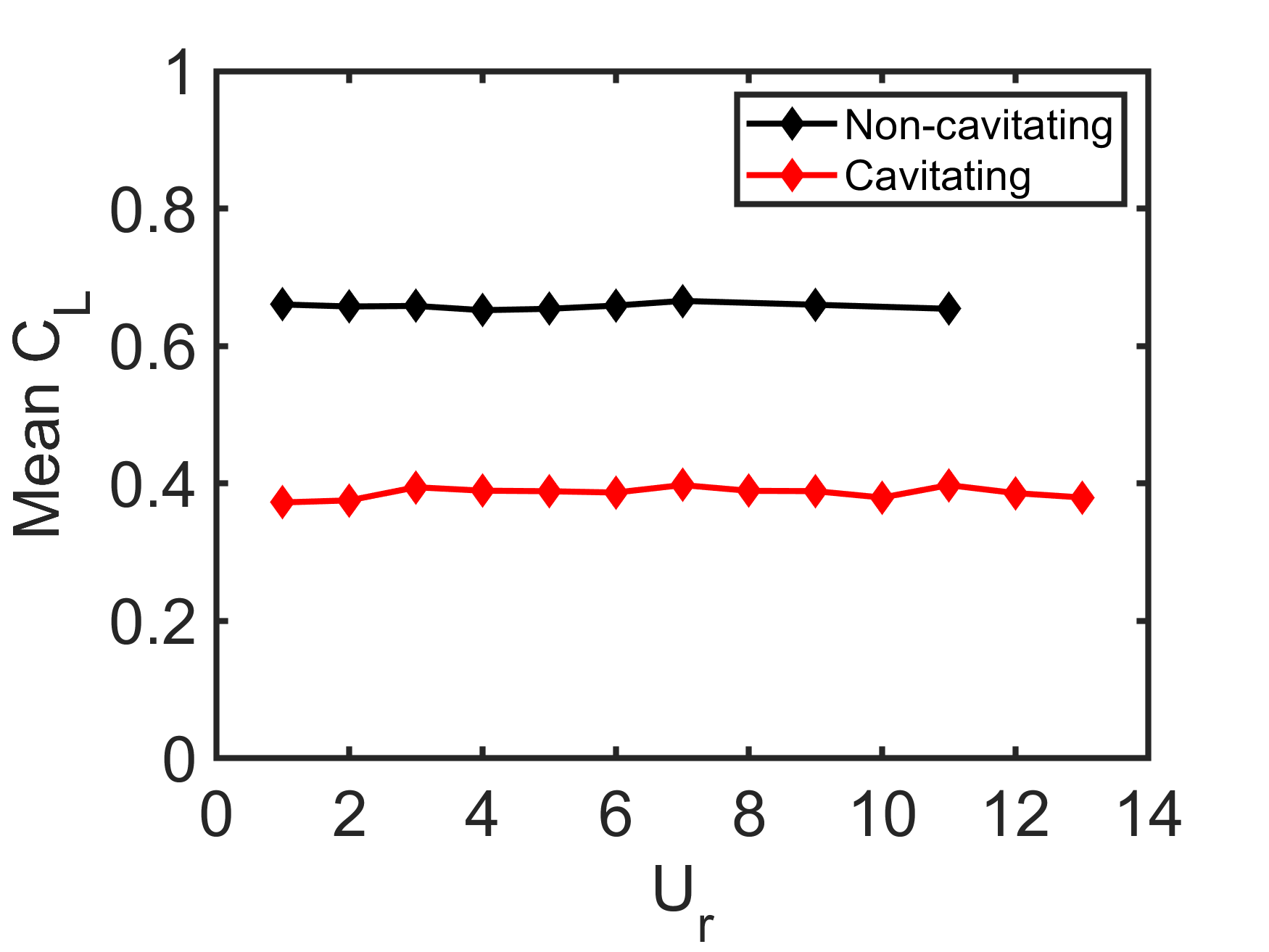}
    \caption{Time averaged lift coefficient $C_L$ at $\alpha = 5^{\circ}$}
    \label{fig:ClMean}
\end{subfigure}\hfill
\begin{subfigure}[t]{0.49\textwidth}
    \centering
    \includegraphics[width=\linewidth]{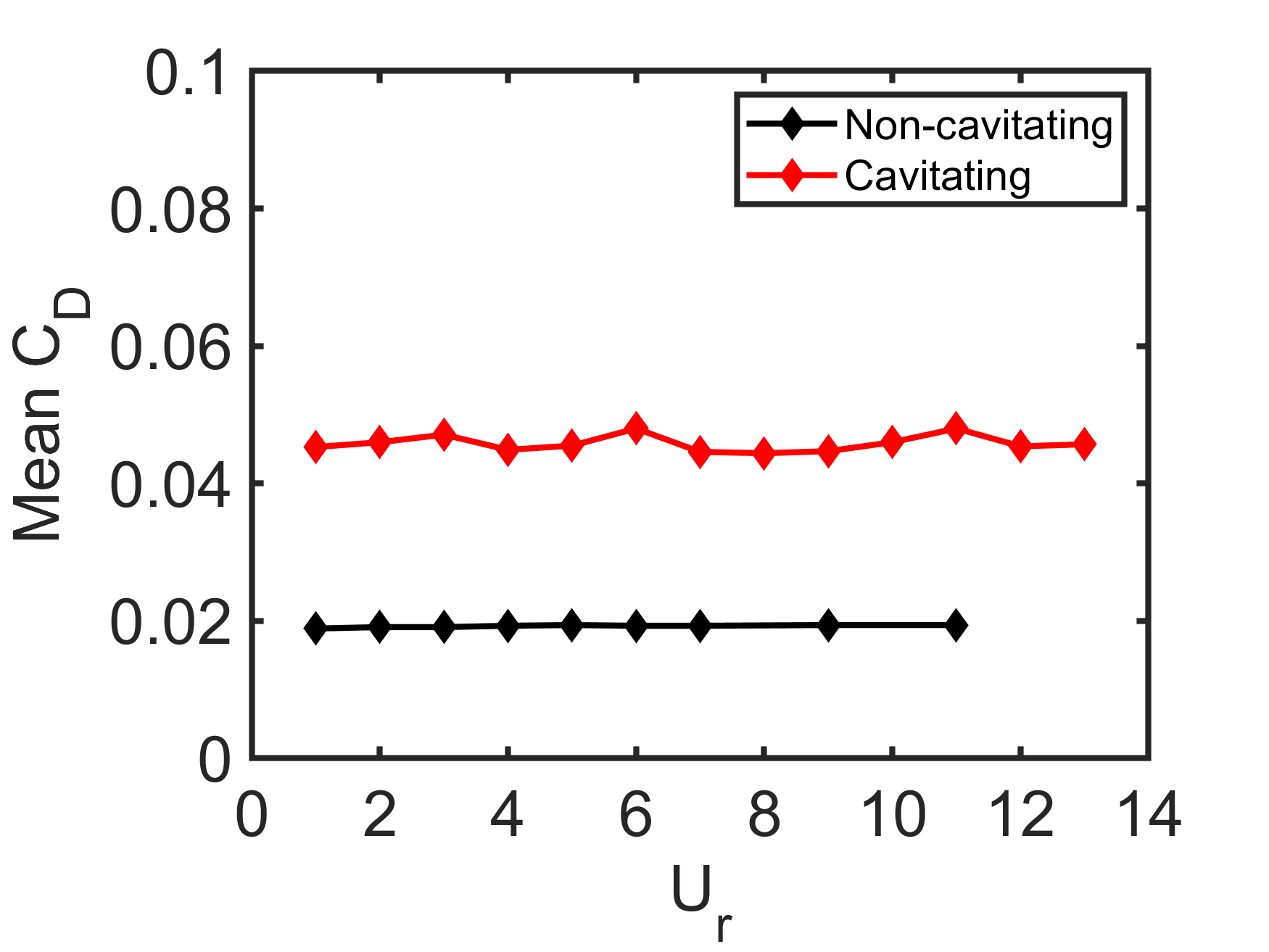}
    \caption{Time averaged drag coefficient $C_D$ at $\alpha = 5^{\circ}$}
    \label{fig:CdMean}
\end{subfigure}
\caption{Comparison of the force coefficients $C_L$ and $C_D$ for the cavitating ($\alpha = 5^{\circ}, \sigma = 0.52$) and non-cavitating ($\alpha = 5^{\circ}, \sigma = 8$) conditions: (a) mean lift $C_L$ showing a drop of $\approx 37\%$ during cavitating conditions, and (b) mean $C_D$ showing a doubling of mean drag forces during cavitating conditions compared to the non-cavitating. Similar observations are made at other angles of attack.}
\label{fig:ClCd_mean}
\end{figure}

\begin{figure}[htbp]
\centering
\begin{subfigure}[t]{0.49\textwidth}
    \centering
    \includegraphics[width=\linewidth]{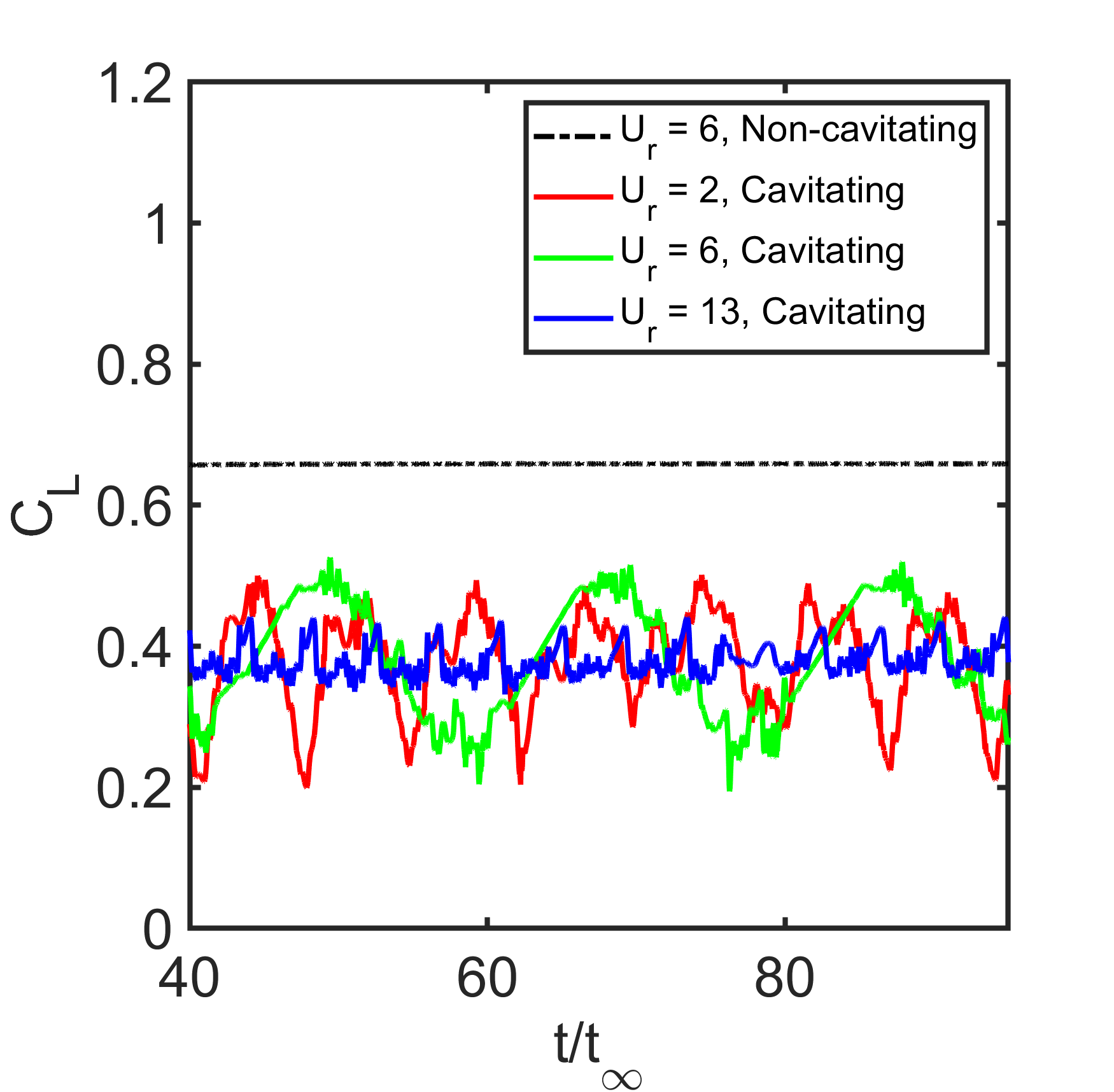}
    \caption{Time evolution of $C_L$ for select $U_r$, $\alpha = 5^{\circ}$}
    \label{fig:figure14_4}
\end{subfigure}\hfill
\begin{subfigure}[t]{0.49\textwidth}
    \centering
    \includegraphics[width=\linewidth]{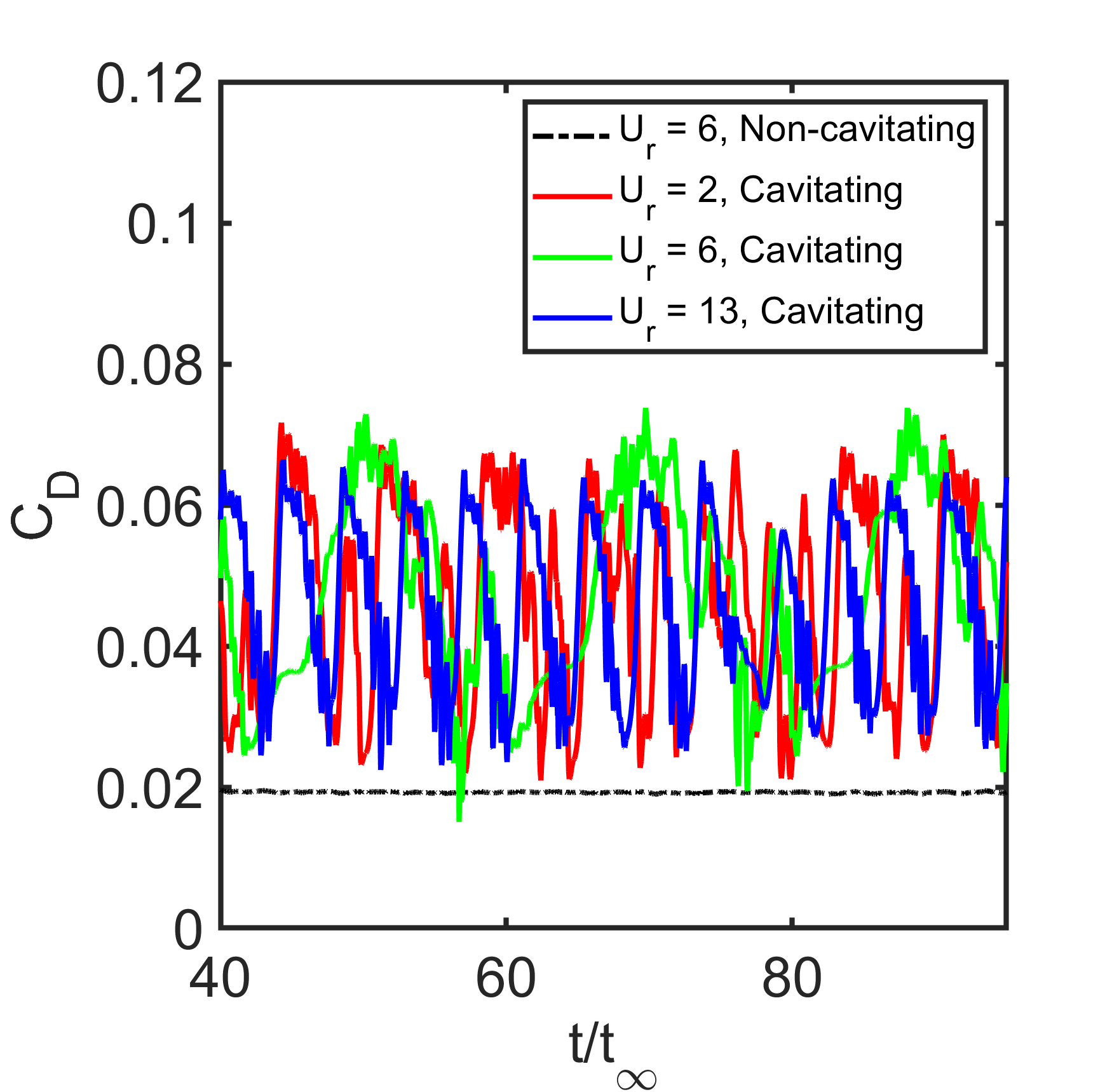}
    \caption{Time evolution of $C_D$ for select $U_r$, $\alpha = 5^{\circ}$}
    \label{fig:figure14_5}
\end{subfigure}
\caption{Time evolution of the force coefficients over a freely vibrating hydrofoil: (a) $C_L$ and (b) $C_D$ for the cavitating ($\alpha = 5^{\circ}, \sigma = 0.52$) and non-cavitating ($\alpha = 5^{\circ}, \sigma = 8$) conditions. Periodic instabilities in the forces are observed for the cavitating case across the range of $U_r$ which are absent for the non-cavitating.}
\label{fig:ClCd_timeSeries}
\end{figure}

The hydrofoil displacement is driven by the unsteady fluid forces due to vorticity and cavity interactions. To understand the coupled behavior, we first compare the force coefficients $C_L$ and $C_D$ for the cavitating and non-cavitating cases. Figures \ref{fig:ClMean} and \ref{fig:CdMean} compare the mean lift $C_L$ and drag $C_D$ of the hydrofoil for cavitating and non-cavitating conditions across the range of $U_r$ at $\alpha = 5^{\circ}$. Around $40\%$ drop in the mean lift is observed, accompanied by a near doubling of the mean drag. Notably, highly unsteady periodic oscillations in $C_D$ and $C_L$ are observed for the cavitating conditions in Fig~\ref{fig:ClCd_timeSeries}, which is absent in the non-cavitating flow. The presence of these unsteady oscillations creates an avenue for the possible lock-in of lift force to the natural frequency $f_N$ of the structure.

\subsection{Frequency lock-in}
\begin{figure}
\centering
\begin{subfigure}{.5\textwidth}
  \centering
  \includegraphics[width=\linewidth]{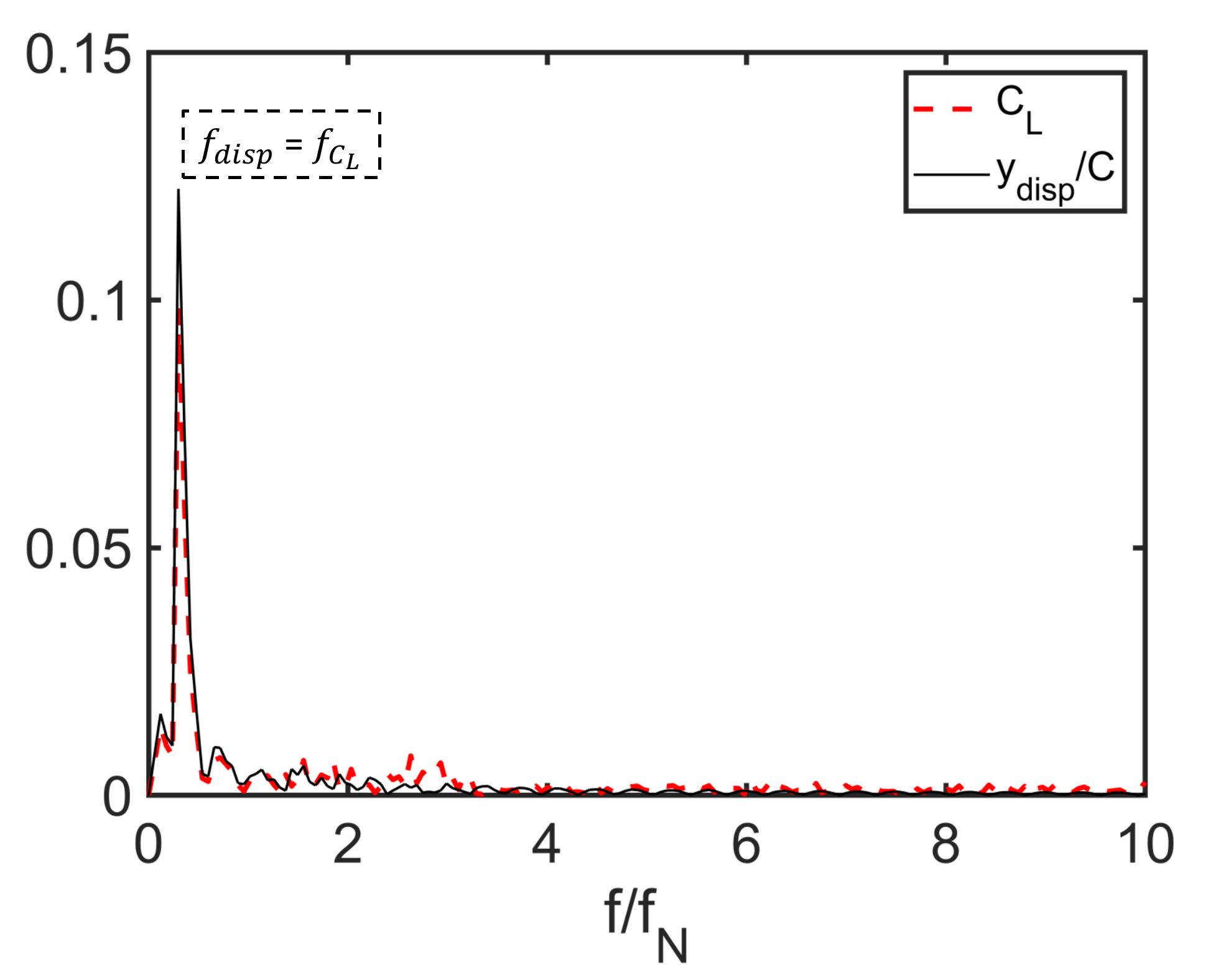}
  \caption{$\alpha = 5^{\circ}, \sigma = 0.52, U_r = 6$}
  \label{fig:A5_ur6fftInc}
\end{subfigure}%
\begin{subfigure}{.5\textwidth}
  \centering
  \includegraphics[width=\linewidth]{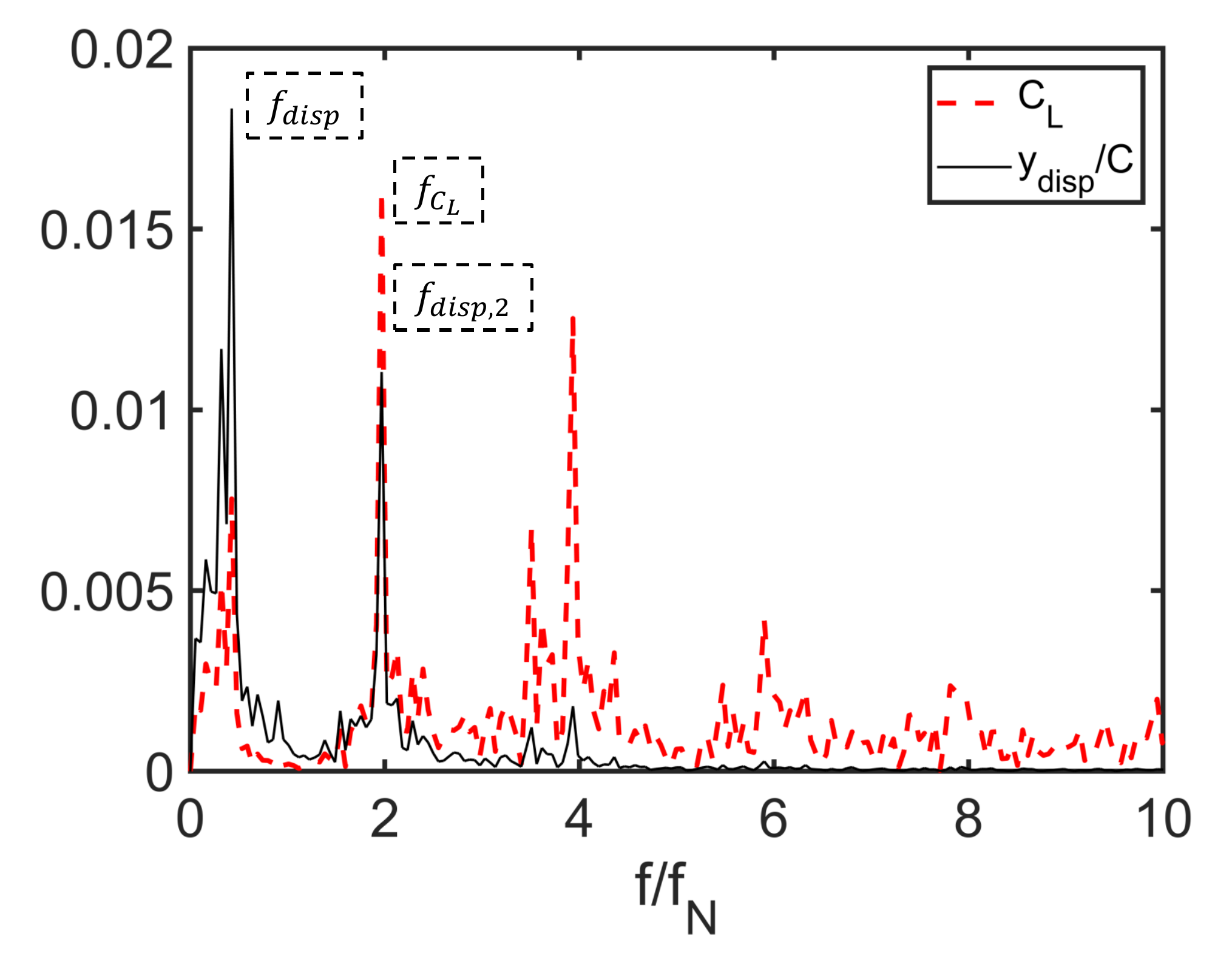}
  \caption{$\alpha = 5^{\circ}, \sigma = 0.52, U_r = 8$}
  \label{fig:A5_ur8fftInc}
\end{subfigure}
\begin{subfigure}{\textwidth}
  \centering
  \includegraphics[width=0.5\linewidth]{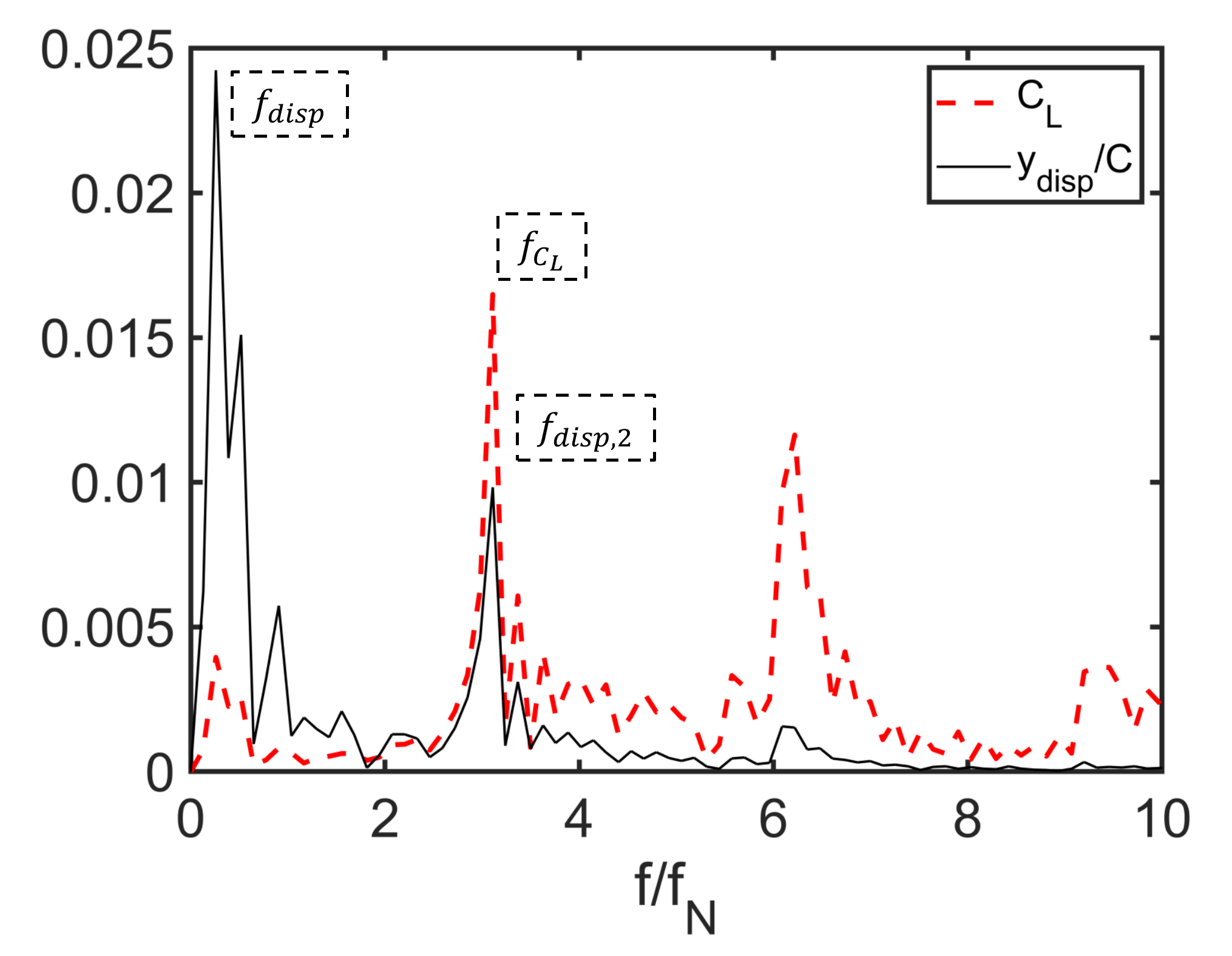}
  \caption{$\alpha = 5^{\circ}, \sigma = 0.52, U_r = 13$}
\end{subfigure}
\caption{Frequency spectra of the lift coefficient $C_L$ and the hydrofoil transverse displacement $y_{disp}$ for the cavitating condition at $\alpha = 5^{\circ}, \sigma = 0.52$ at $U_r =$: (a) 6,  (b) 8  and  (c) 13. 
While $U_r =6$ corresponds to large amplitude oscillations and  the dominant frequencies for $C_L$ and $y_{disp}$ coincide $\approx 0.3f_N$, 
$U_r =8$ and 13 represent the low amplitude oscillation regime in the post-lock-in.}
\label{fig:fftCompWOcav}
\end{figure}

The phenomenon of frequency lock-in occurs when the unsteady flow frequency coincides with the one of the harmonics of the combined fluid-structure system. The presence of high-amplitude oscillations at a specific range of reduced velocities is generally indicative of lock-in of the unsteady flow forces with the natural frequency of the structure. In Fig.~\ref{fig:fftCompWOcav}, we take the representative case $\alpha = 5^{\circ}$ and compare the frequency spectra of the lift coefficient $C_L$ and the structural displacement $y_{disp}$ at $U_r = 6$ where peak oscillations are observed, and at $U_r = 8$ and $U_r = 13$ where the oscillation amplitude is significantly reduced. For $U_r = 6$, the dominant frequencies are observed to synchronize at a sub-harmonic of the wetted natural frequency $f_N$. For $U_r = 8$ and $U_r = 13$, the dominant frequencies of $C_L$ and $y_{disp}$ are increasingly distinct. A secondary frequency $Y_{disp,2}$ of the displacement is seen to synchronize with $C_L$. This is in the form of a secondary high frequency albeit low amplitude vibration observed at higher $U_r$ and can be found in Fig.~\ref{fig:a5_yDispSampled}.

\begin{figure}
\begin{subfigure}{.5\textwidth}
  \centering
  \includegraphics[width=\linewidth]{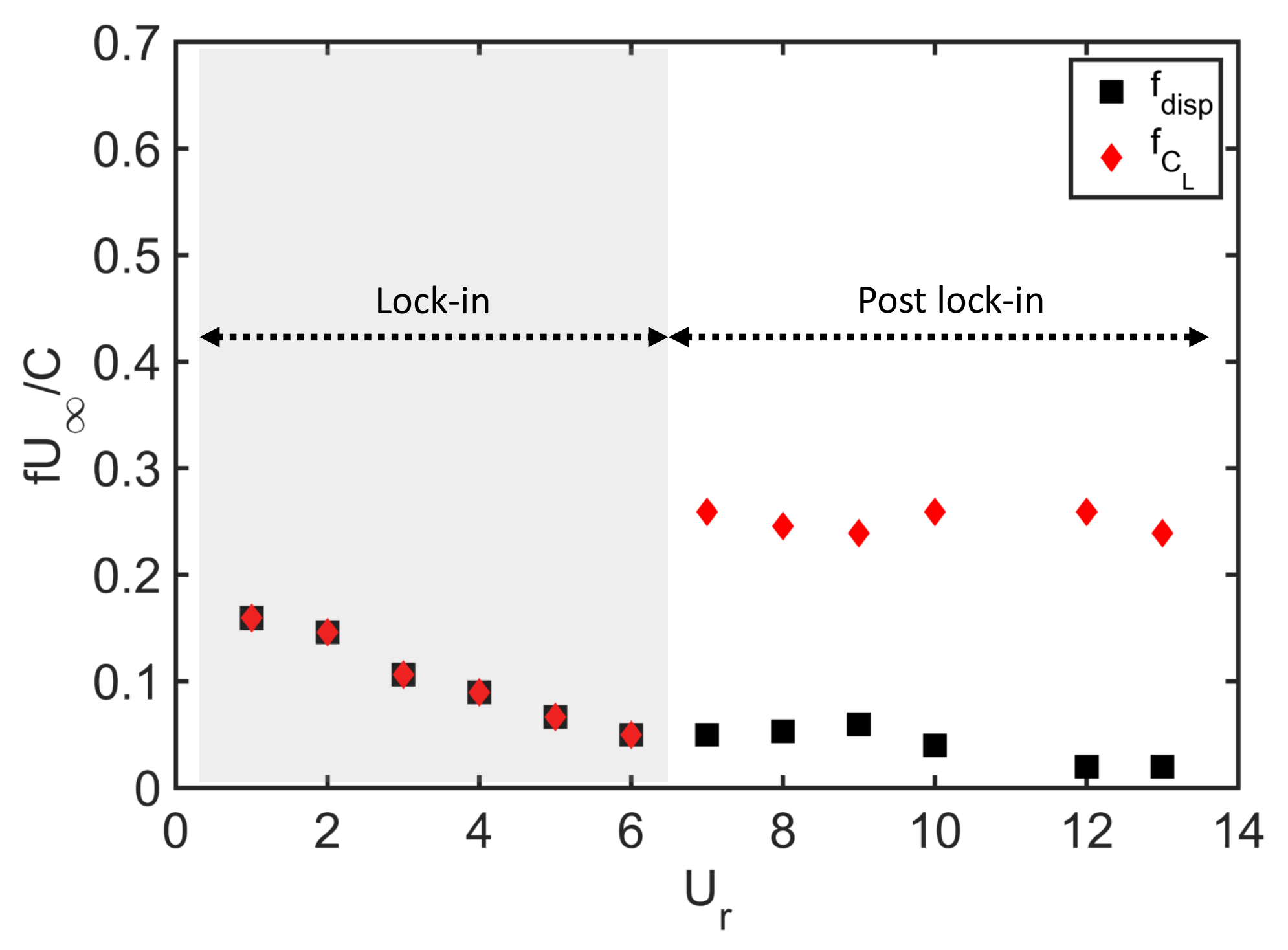}
  \label{fig:A5fubyc}
  \caption{}
\end{subfigure}
\begin{subfigure}{.5\textwidth}
  \centering
  \includegraphics[width=\linewidth]{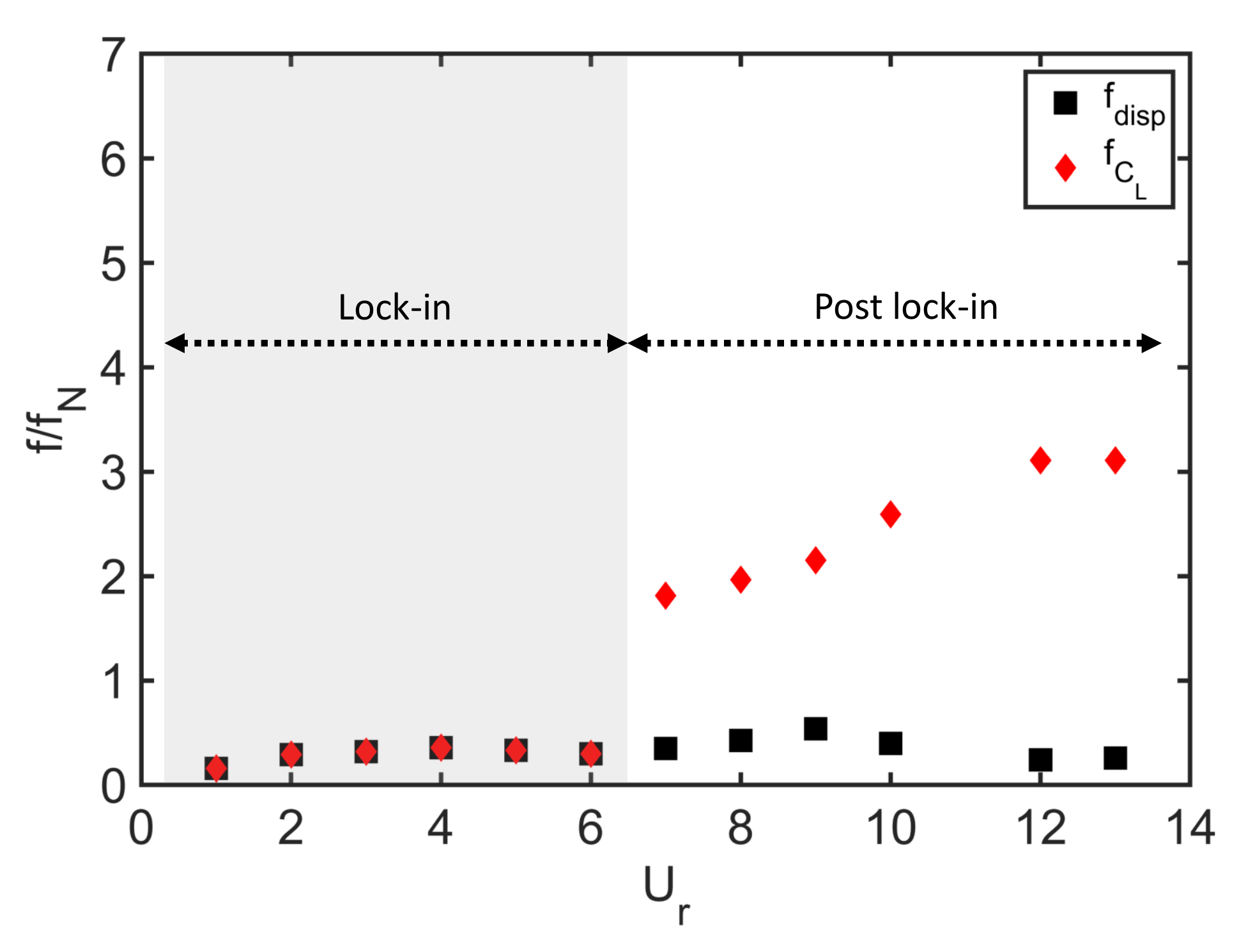}
  \label{fig:A5fbyfn}
  \caption{}
\end{subfigure}
\caption{Frequencies of oscillation of the hydrofoil transverse displacement ($f_{disp}$) and the coefficient of lift ($f_{C_L}$) for $\alpha = 5^{\circ}, \sigma = 0.52$: (a) Normalized by the flow time scales and (b) Normalized by the wetted natural frequency $f_N$. In the lock-in regime $f_{disp} = f_{C_L} \approx 0.3f_N$. Post lock-in there is a distinct difference between the two frequencies with $\dfrac{f_{C_L}U_\infty}{C}$ observed to be largely constant $\approx 0.25$.} 
\label{fig:A5fComp}
\end{figure}

In Fig.~\ref{fig:A5fComp}, we observe the dominant frequencies across the range of $U_r$ for $\alpha = 5^{\circ}, \sigma = 0.52$. In the regime $U_r \in \left[ 1, 6 \right]$ we find a single prominent frequency $f=f_{disp}=f_{C_L}$, defined as the \textit{lock-in} regime. For $U_r > 6$, a nearly constant $f_{C_L}  \neq f_{disp}$ is observed. We define this as the \textit{post lock-in} regime.
Notably, during lock-in we observe that the frequencies $f_{disp} = f_{C_L}$ are a nearly linear function of the natural frequency of the structure $\approx 0.3f_N$. This indicates that during cavitating conditions the dominant frequencies lock-in to a sub-harmonic of the natural frequency. 

\begin{figure}
\begin{subfigure}{.5\textwidth}
  \centering
  \includegraphics[width=\linewidth]{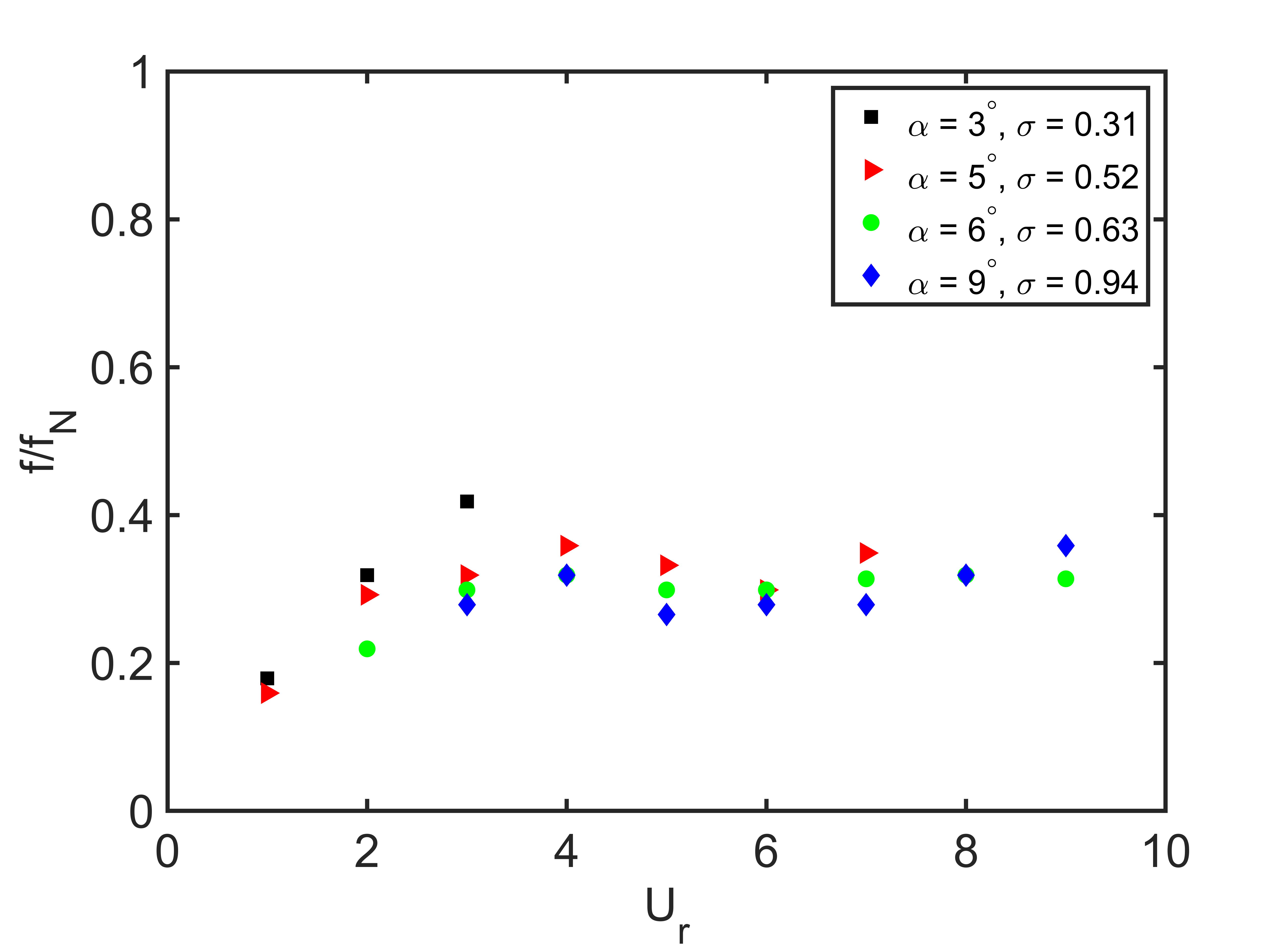}
  \caption{\label{fig:fbyfnAll} }
\end{subfigure}
\begin{subfigure}{.5\textwidth}
  \centering
  \includegraphics[width=\linewidth]{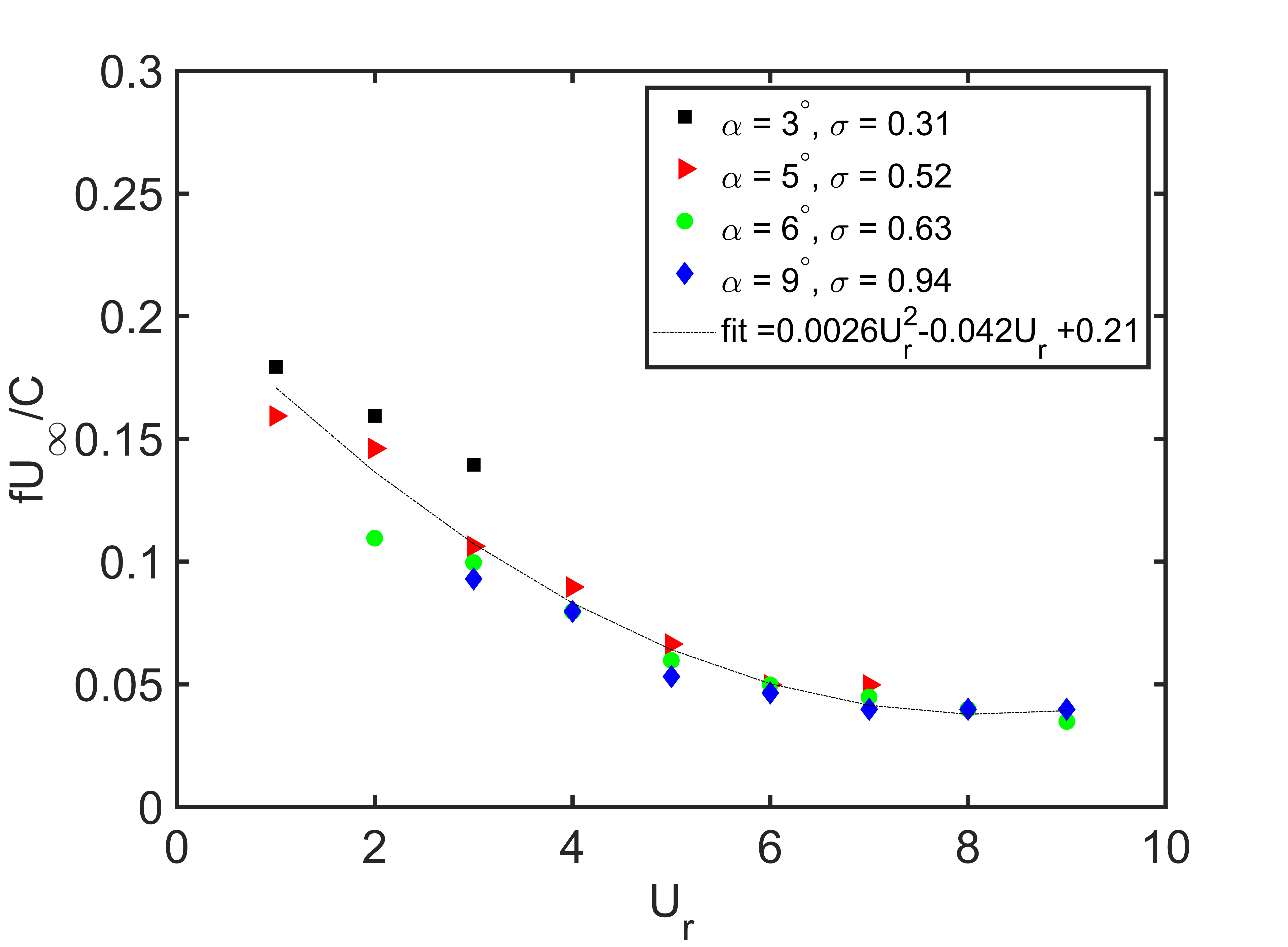}
  \caption{\label{fig:fUbyCAll}}
\end{subfigure}
\caption{Dominant vibration frequencies $f = f_{disp} = f_{C_L}$ in the lock-in regime shown for different angles of attack $\alpha$ of the hydrofoil, (a) normalized by the natural frequency $f_N$ of the elastically mounted hydrofoil. These frequencies are largely observed in a narrow band around $\approx 0.3f_N$. (b) normalized by the flow time-scales where $U_\infty$ is the velocity far upstream and $C$ is the hydrofoil chord length. Across range of angles of attack $\alpha$ and the reduced velocity $U_r$, $f$ is observed to collapse well onto a second-order fit.}
\end{figure}

Figure~\ref{fig:fbyfnAll} shows that this observation is consistent for other values of $\alpha$ in the respective lock-in regimes. Particularly, for the range of flow configurations studied the frequencies are observed to lock into a narrow band $\approx 0.3-0.4f_N$. In Fig.~\ref{fig:fUbyCAll} we see that a second-order polynomial fit can capture the variation of the frequency well for a range of $\alpha$. This demonstrates that distinct patterns in the lock-in regime exist and is susceptible to generalization within regular operational design ranges of the lifting surfaces. This is an encouraging result and can potentially be used for targeted control strategies for cavitation noise mitigation in hydrofoil operation. The exact nature of this fit in practical engineering applications can depend on several additional factors including the blade design and variation in flow regimes. The response surface, in this case, needs to be mapped out using further experimental/computational studies and is scope for future work. 
Thus we observe that lock-in of the fluctuating lift forces is a contributing mechanism to the amplified hydrofoil vibrations during partial cavitating conditions at low-moderate $\alpha$. However, this brings us to another question: What induces this dramatic instability in the lift forces in the presence of cavitation? In the next section, we make an attempt to investigate and explain this instability.

\subsection{Cavitation influence on vorticity and lift generation}\label{sec:cavVort}
We next investigate the source of high unsteadiness in the lift forces observed in the cavitating flow. The assessment of lift force and its connection with the vorticity generation is summarized in Appendix A. To understand how cavitation influences vorticity and in turn the unsteady lift forces, we monitor the temporal-spatial evolution of vorticity dynamics in the domain. 
In the absence of external body forces, the vorticity transport equation can be written as
\begin{equation}
    \frac{D \boldsymbol{\omega}}{D t} = \underbrace{(\boldsymbol{\omega} \cdot \boldsymbol{\nabla}) \boldsymbol{u}}_\text{\clap{$\mathcal{A}$}} - \underbrace{\boldsymbol{\omega}(\boldsymbol{\nabla} \cdot \boldsymbol{u})}_\text{\clap{$\mathcal{B}$}} + \underbrace{\frac{1}{\rho^{2}} \boldsymbol{\nabla} \rho \times \boldsymbol{\nabla} p}_\text{\clap{$\mathcal{C}$}} + \underbrace{\boldsymbol{\nabla} \times\left(\frac{\boldsymbol{\nabla} \cdot \tau}{\rho}\right)}_\text{\clap{$\mathcal{D}$}},
    \label{eq:vortTransEq}
\end{equation}
where $\dfrac{D \boldsymbol{\omega}}{D t}$ denotes the material derivative of the vorticity field. The first term on the right $(\boldsymbol{\omega} \cdot \boldsymbol{\nabla})\boldsymbol{u}$ indicates the stretching of vorticity because of the flow velocity field, playing an important role in turbulence. In the current study, we focus on the largely two-dimensional flow around the mid-span of the blade away from root and tip effects, and exclude the spanwise variation in the flow. Thus this term identically goes to zero. The second term $\boldsymbol{\omega}(\nabla \cdot \boldsymbol{u})$ indicates the stretching of vorticity due to flow compressibility effects. This term can play an important role in our study due to non-zero divergence of velocity in the compressible two phase mixture. The third term $\dfrac{1}{\rho^{2}} \boldsymbol{\nabla} \rho \times \boldsymbol{\nabla} p$ is the baroclinic torque, and is significant in regions where the local density gradients are orthogonal to the local pressure gradients. The fourth term $\boldsymbol{\nabla} \times\left(\dfrac{\boldsymbol{\nabla} \cdot \tau}{\rho}\right)$ represents the viscous diffusion of vorticity. For convenience, we shall denote these terms as $\mathcal{A}$, $\mathcal{B}$, $\mathcal{C}$ and $\mathcal{D}$ respectively as marked in Eq.~(\ref{eq:vortTransEq}).

\begin{figure}[!h]
\centering
\includegraphics[width=\columnwidth]{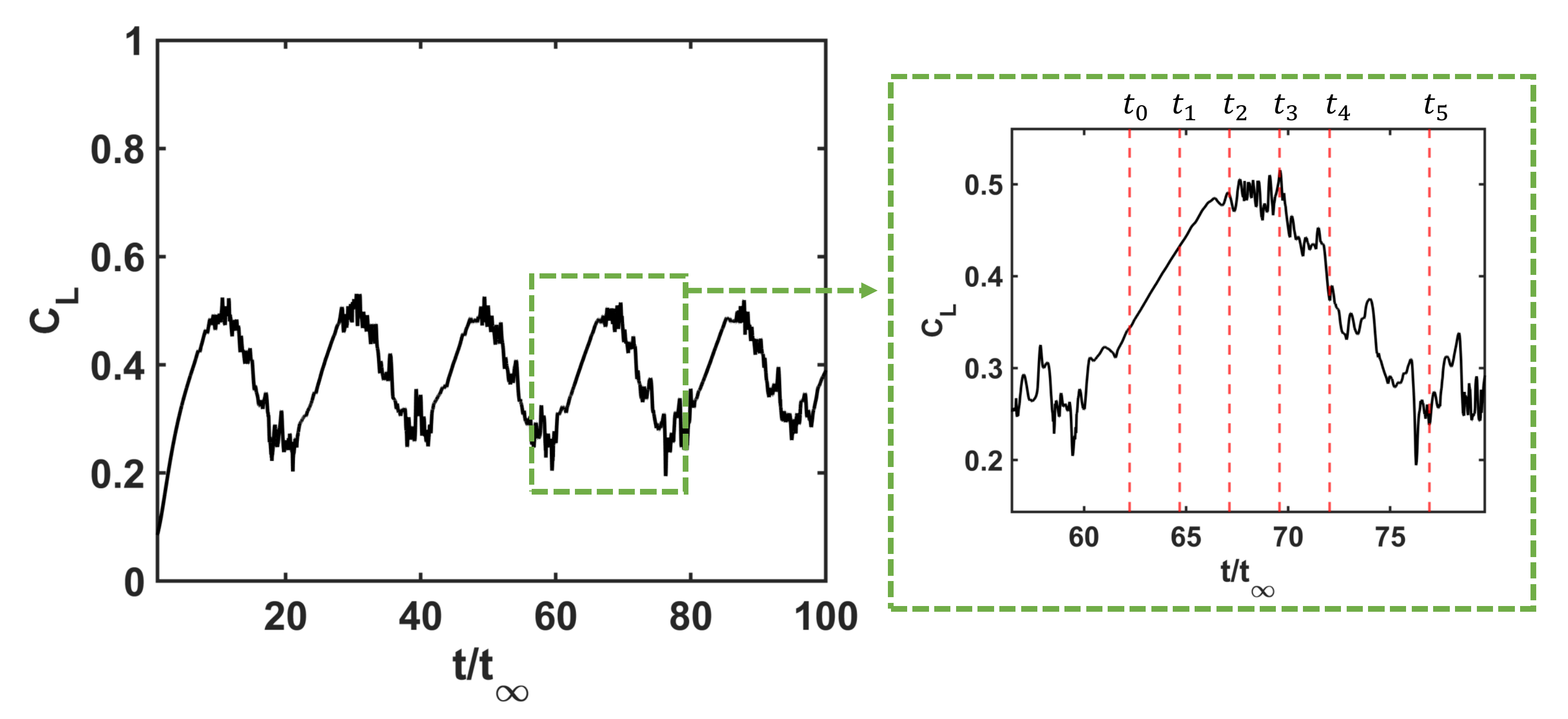}
\caption{Evolution of lift coefficient $C_L$ with time for the cavitating condition $\alpha = 5^{\circ}, \sigma = 0.52$, $U_r = 6$. Inlay showing one typical cycle of lift fluctuations. Some key stages marked by non-dimensional times $t_n$ for discussion in text.}
\label{fig:A5_6_CL_TS}        
\end{figure}

\begin{figure}
    \centering
    
\vspace*{-0.25in}
\hspace*{-0.45in}
\begin{tabular}{c@{}c@{}c}
 \rotatebox[origin=l]{90}{\makebox[0.8in]{$t_0$}}%
 \quad
 \fbox{\includegraphics[width=0.33\textwidth]{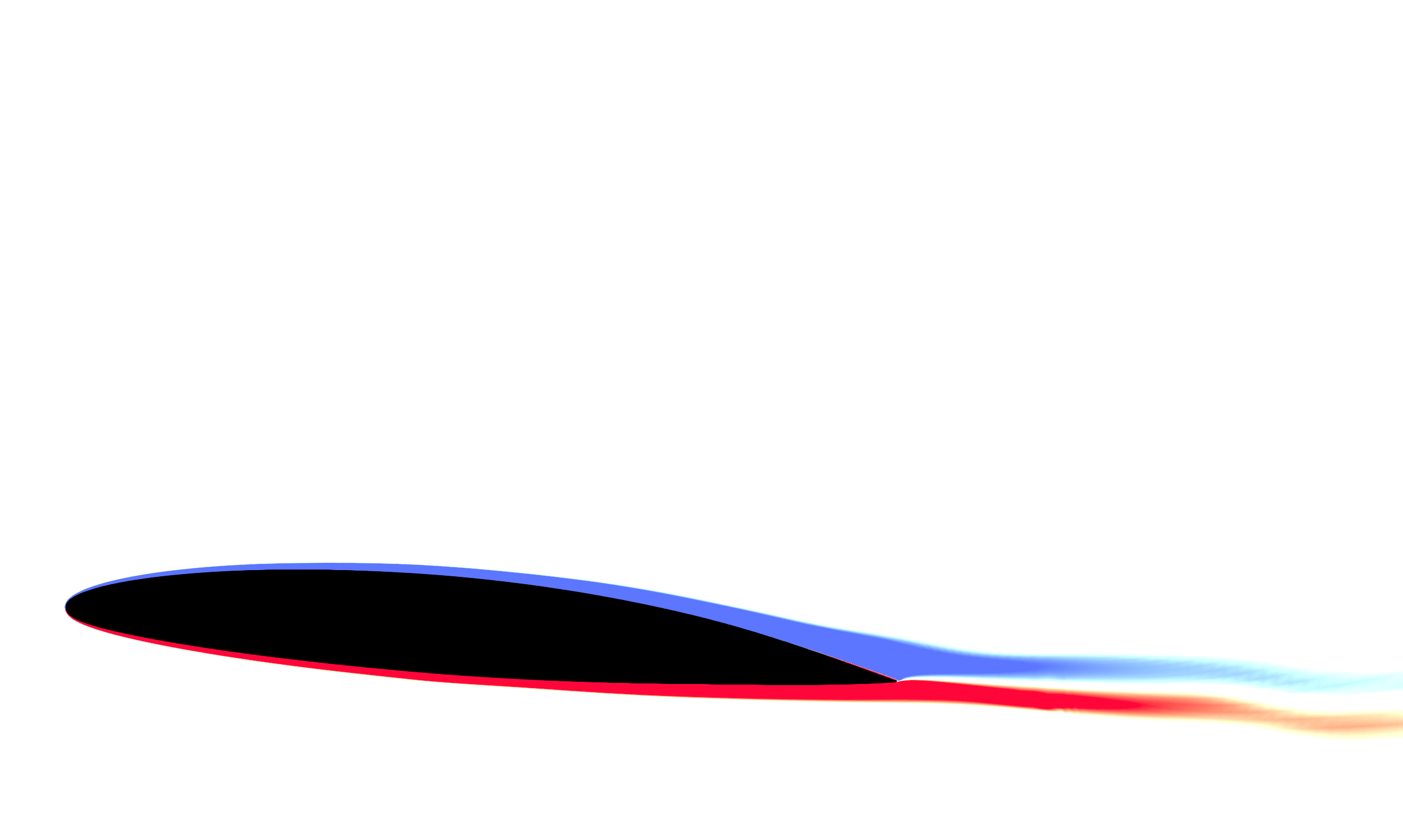}}&%
 \fbox{\includegraphics[width=0.33\textwidth]{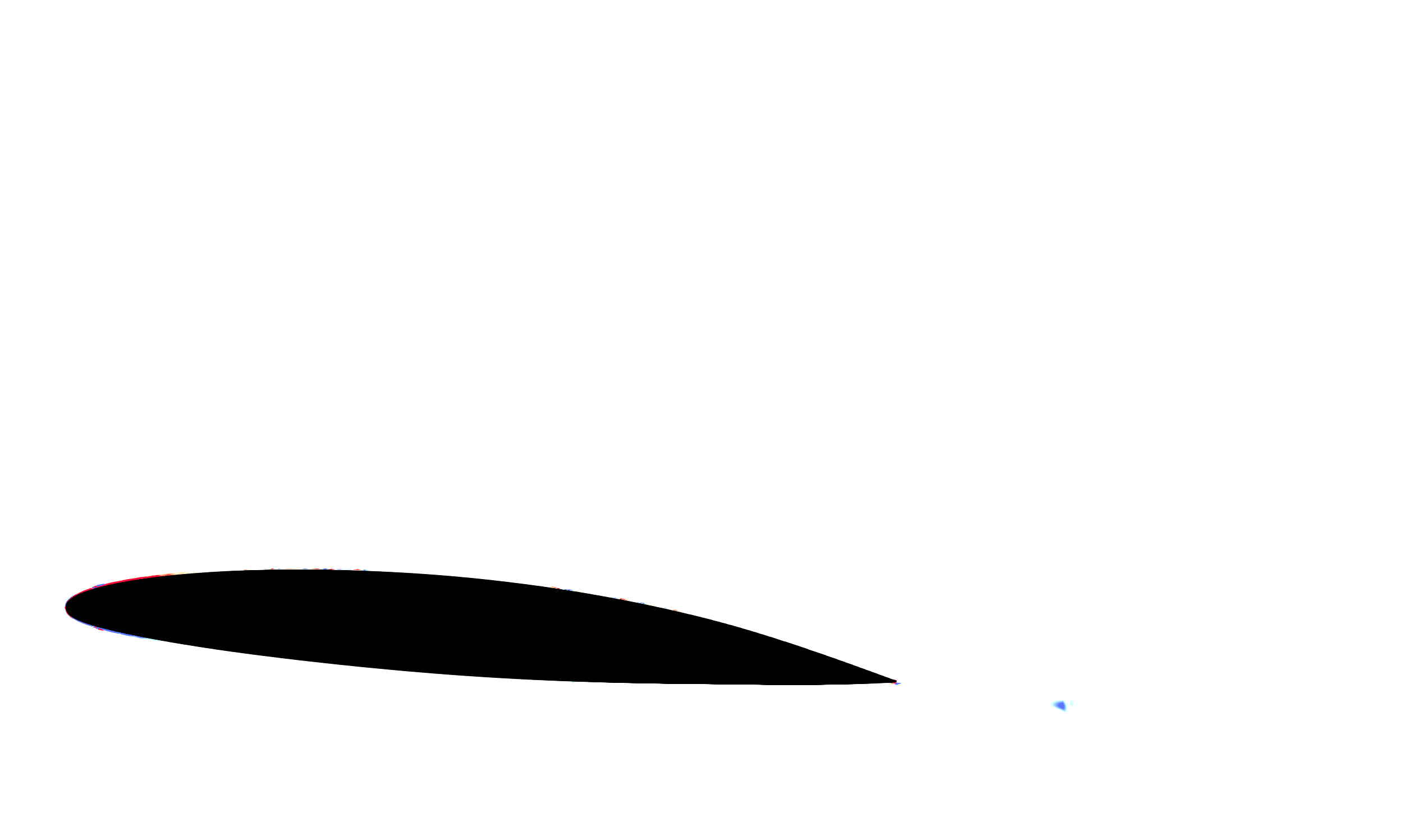}}&
 \fbox{\includegraphics[width=0.33\textwidth]{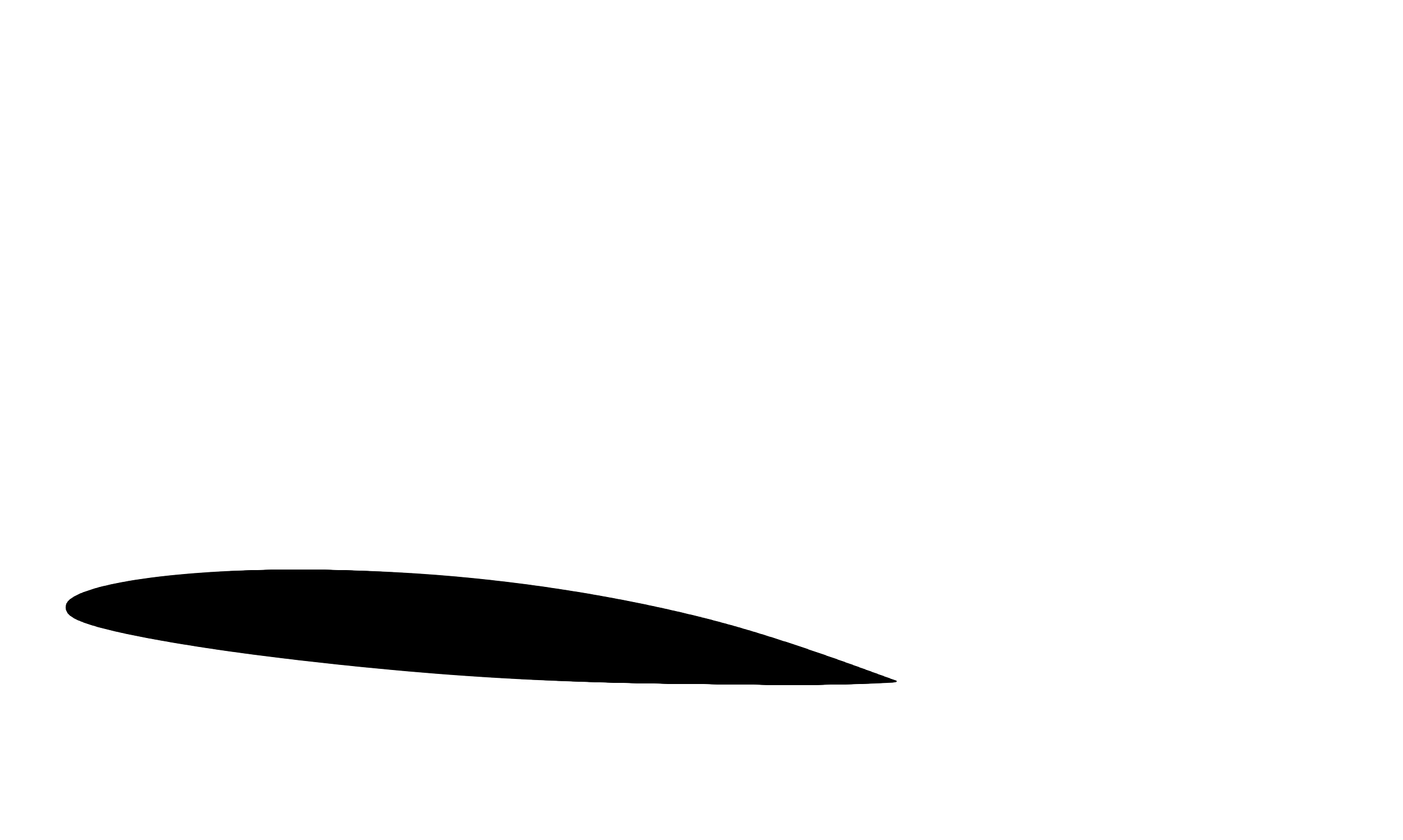}}\\
 \rotatebox[origin=l]{90}{\makebox[0.8in]{$t_1 = t_0 + 0.5/t_\infty$}}%
 \quad
 \fbox{\includegraphics[width=0.33\textwidth]{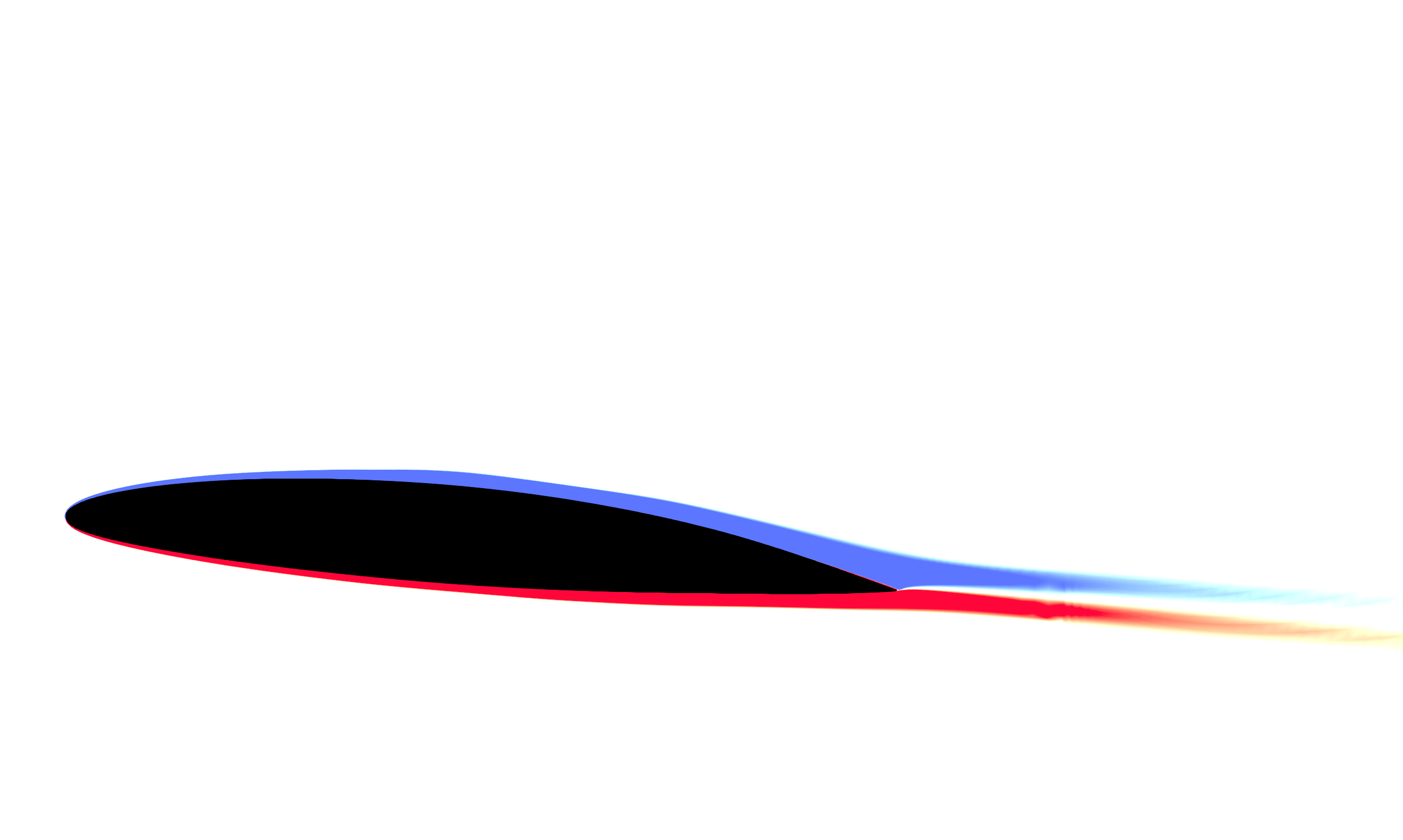}}&%
 \fbox{\includegraphics[width=0.33\textwidth]{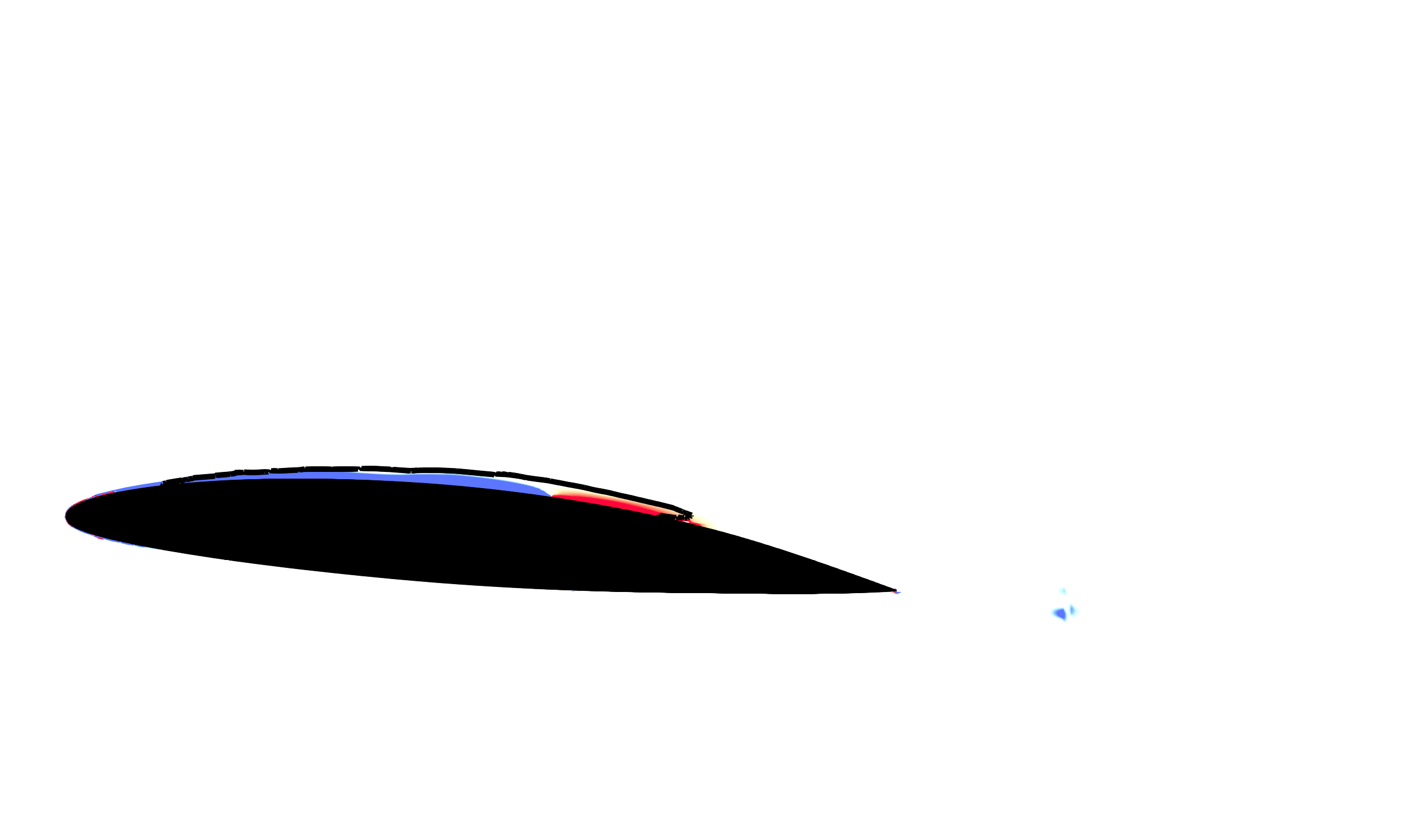}}&
 \fbox{\includegraphics[width=0.33\textwidth]{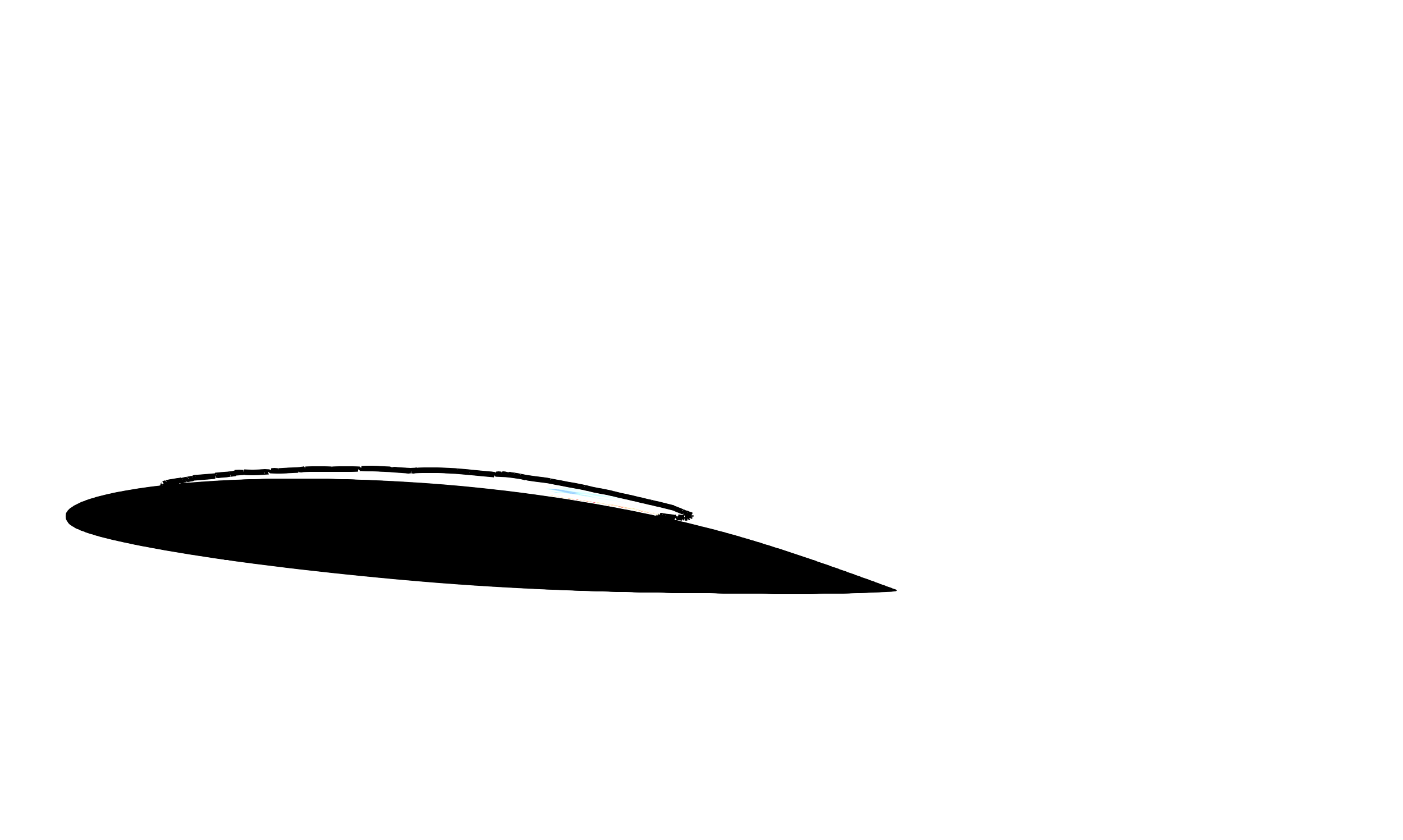}}\\
 \rotatebox[origin=l]{90}{\makebox[0.8in]{$t_2 = t_0 + 1/t_\infty$}}%
 \quad
 \fbox{\includegraphics[width=0.33\textwidth]{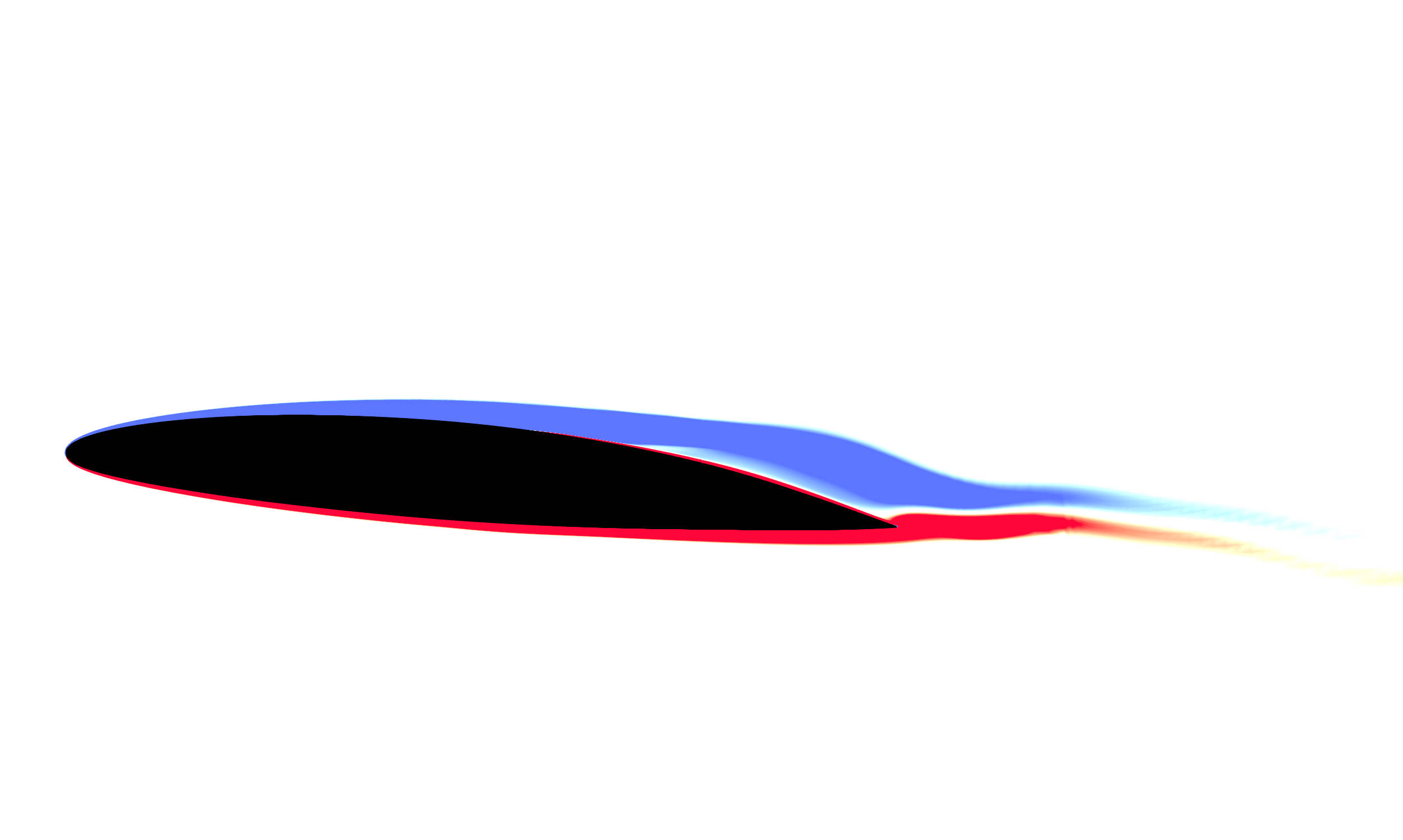}}&%
 \fbox{\includegraphics[width=0.33\textwidth]{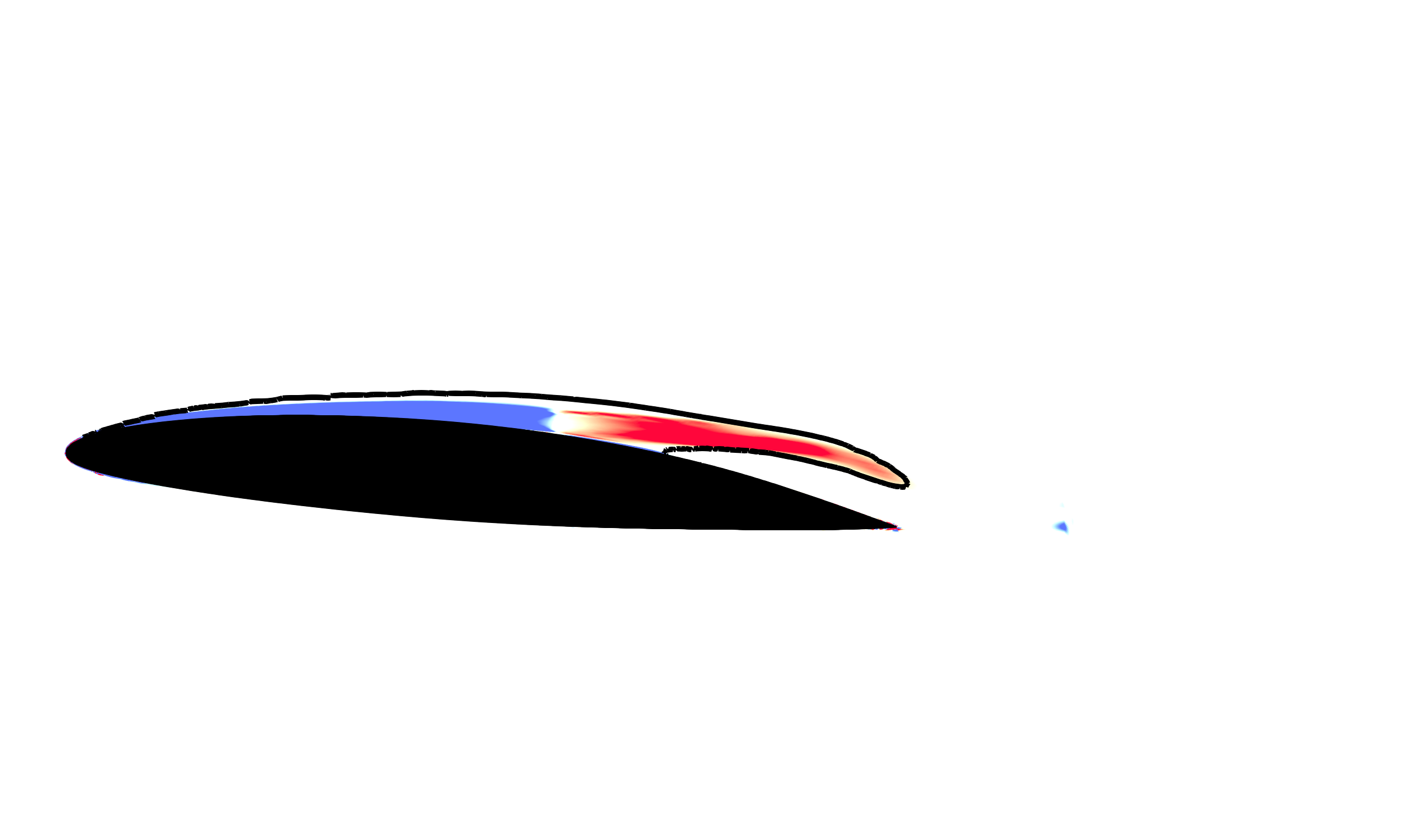}}&
 \fbox{\includegraphics[width=0.33\textwidth]{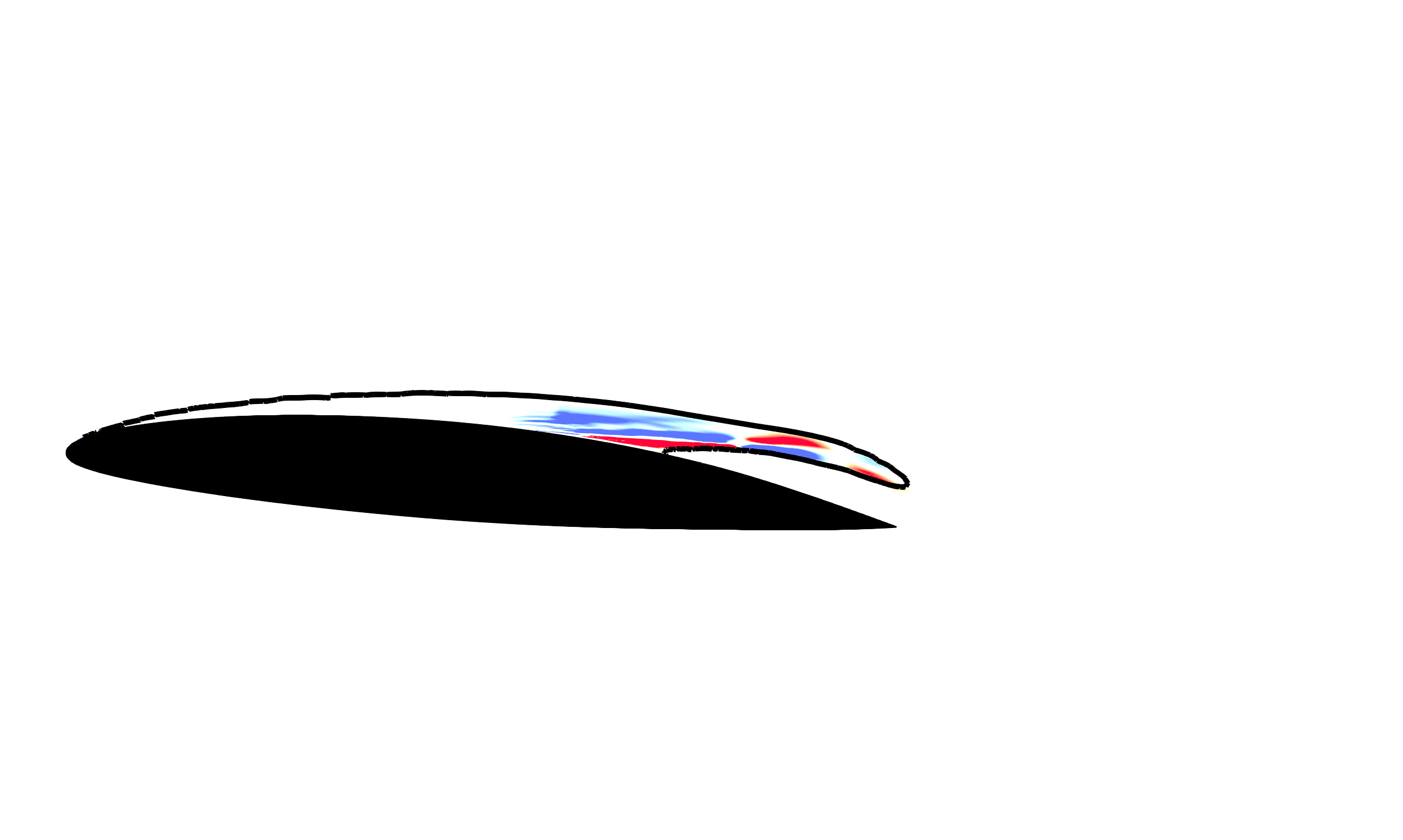}}\\
 \rotatebox[origin=l]{90}{\makebox[0.8in]{$t_3 = t_0 + 1.5/t_\infty$}}%
 \quad
 \fbox{\includegraphics[width=0.33\textwidth]{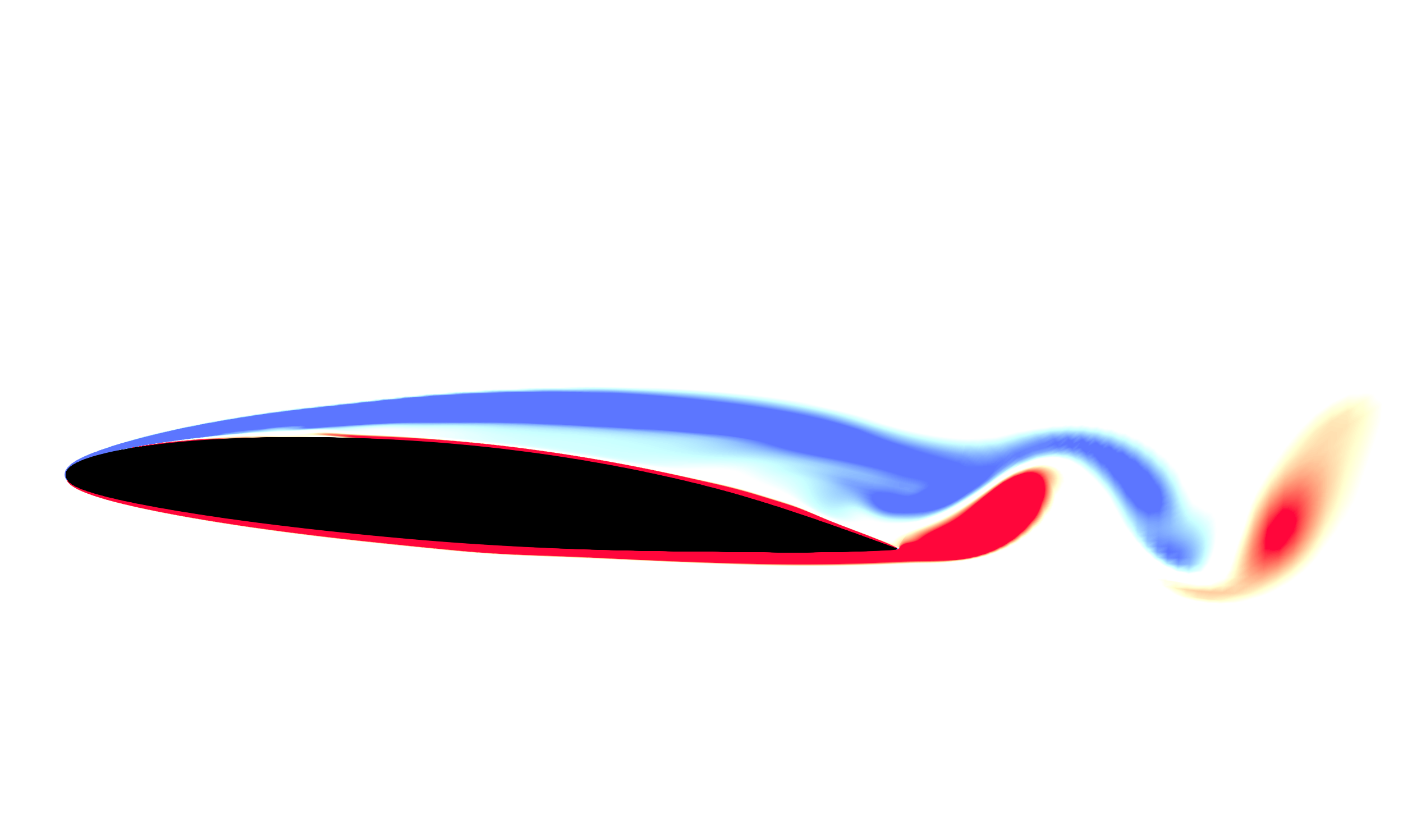}}&%
 \fbox{\includegraphics[width=0.33\textwidth]{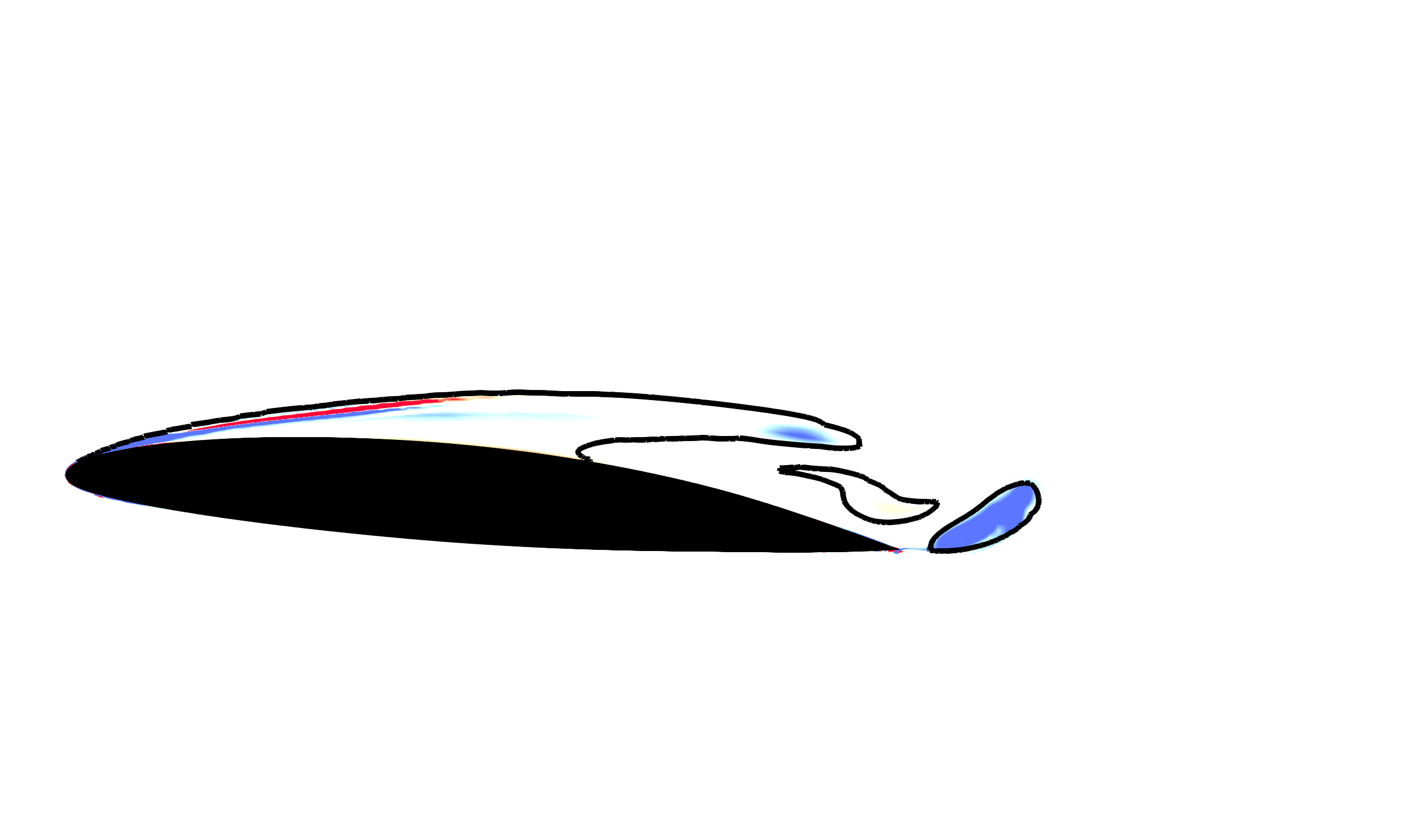}}&
 \fbox{\includegraphics[width=0.33\textwidth]{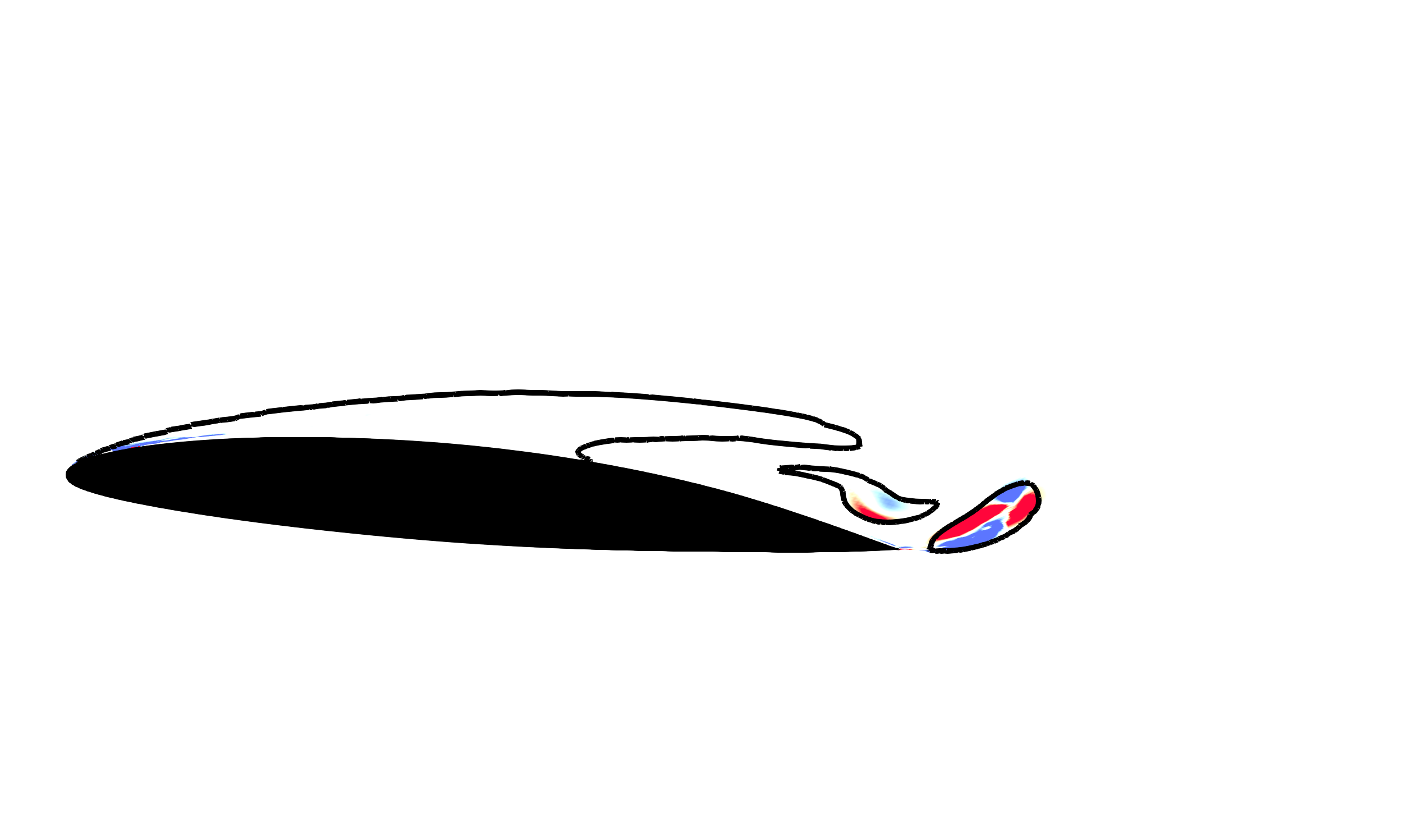}}\\
 \rotatebox[origin=l]{90}{\makebox[0.8in]{$t_4 = t_0 + 2/t_\infty$}}%
 \quad
 \fbox{\includegraphics[width=0.33\textwidth]{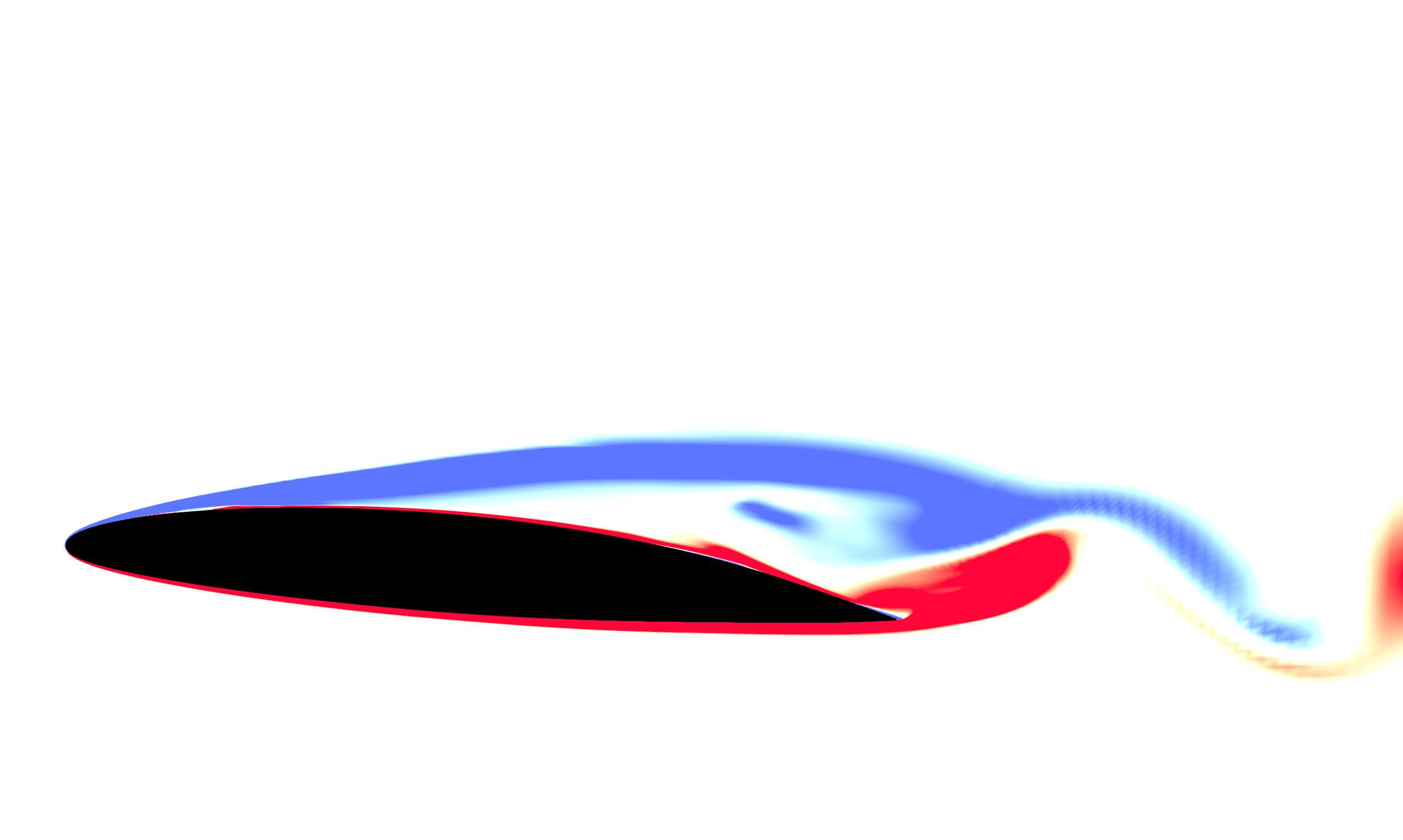}}&%
 \fbox{\includegraphics[width=0.33\textwidth]{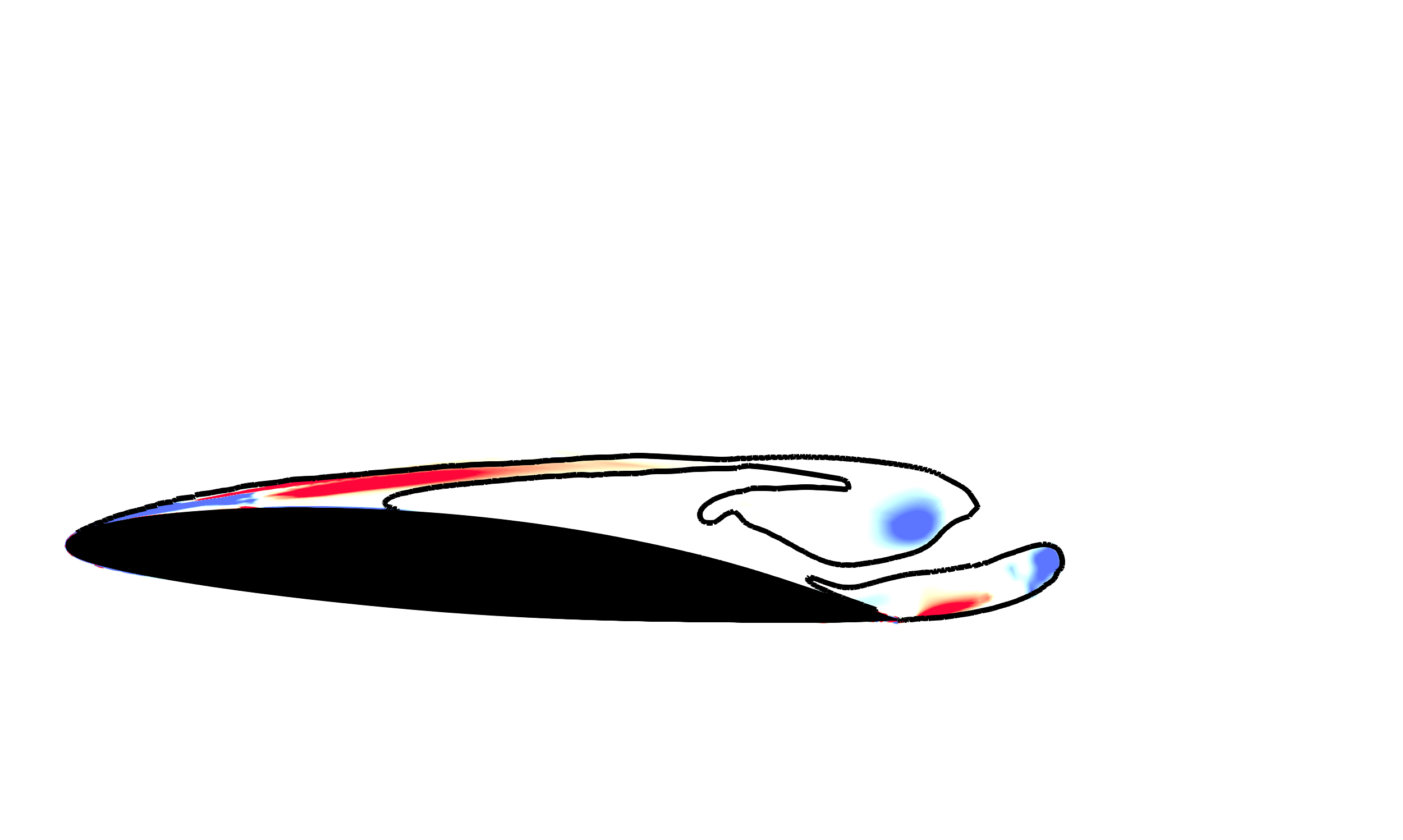}}&
 \fbox{\includegraphics[width=0.33\textwidth]{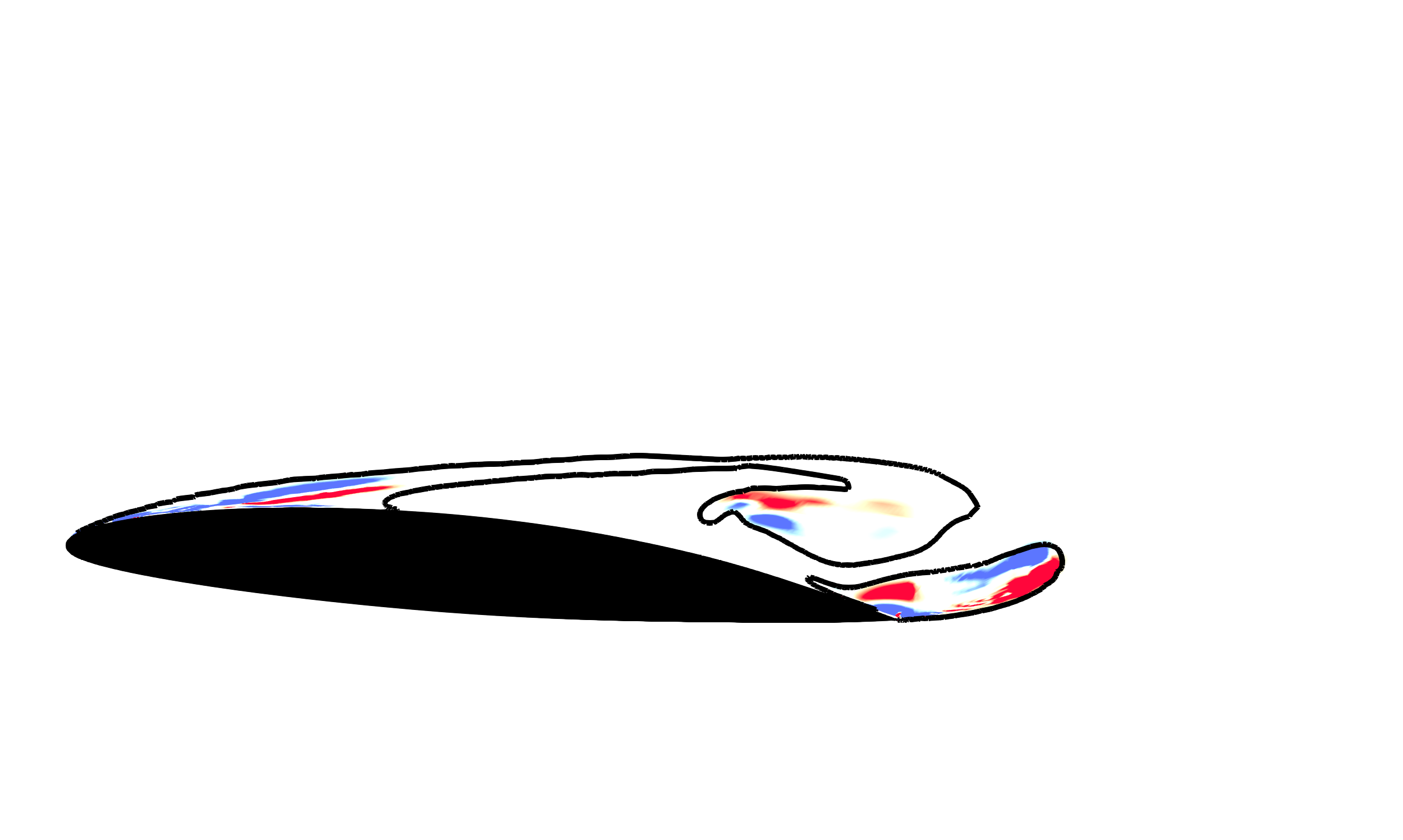}}\\
 \rotatebox[origin=l]{90}{\makebox[0.8in]{$t_5 = t_0 + 3/t_\infty$}}%
 \quad
 \fbox{\includegraphics[width=0.33\textwidth]{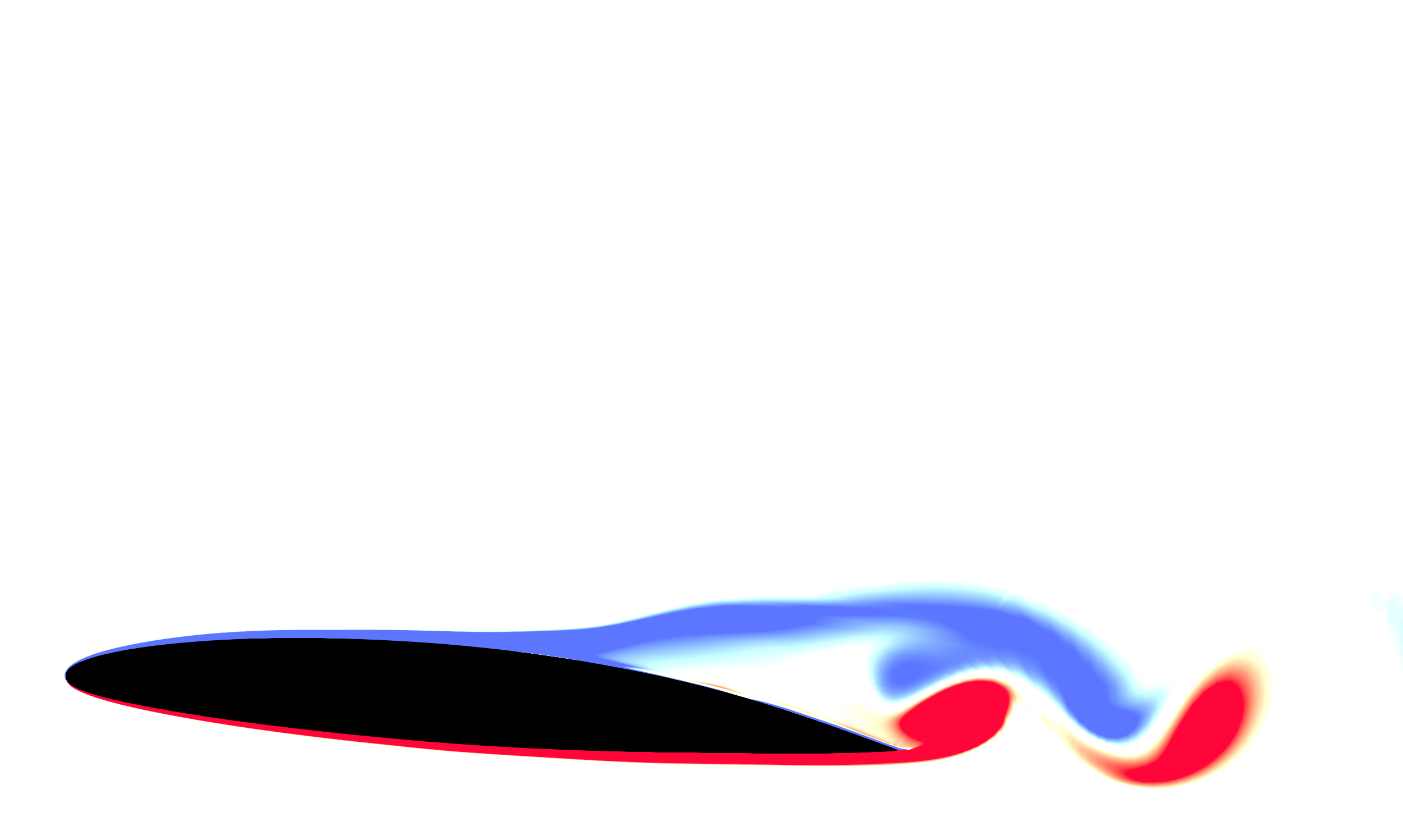}}&%
 \fbox{\includegraphics[width=0.33\textwidth]{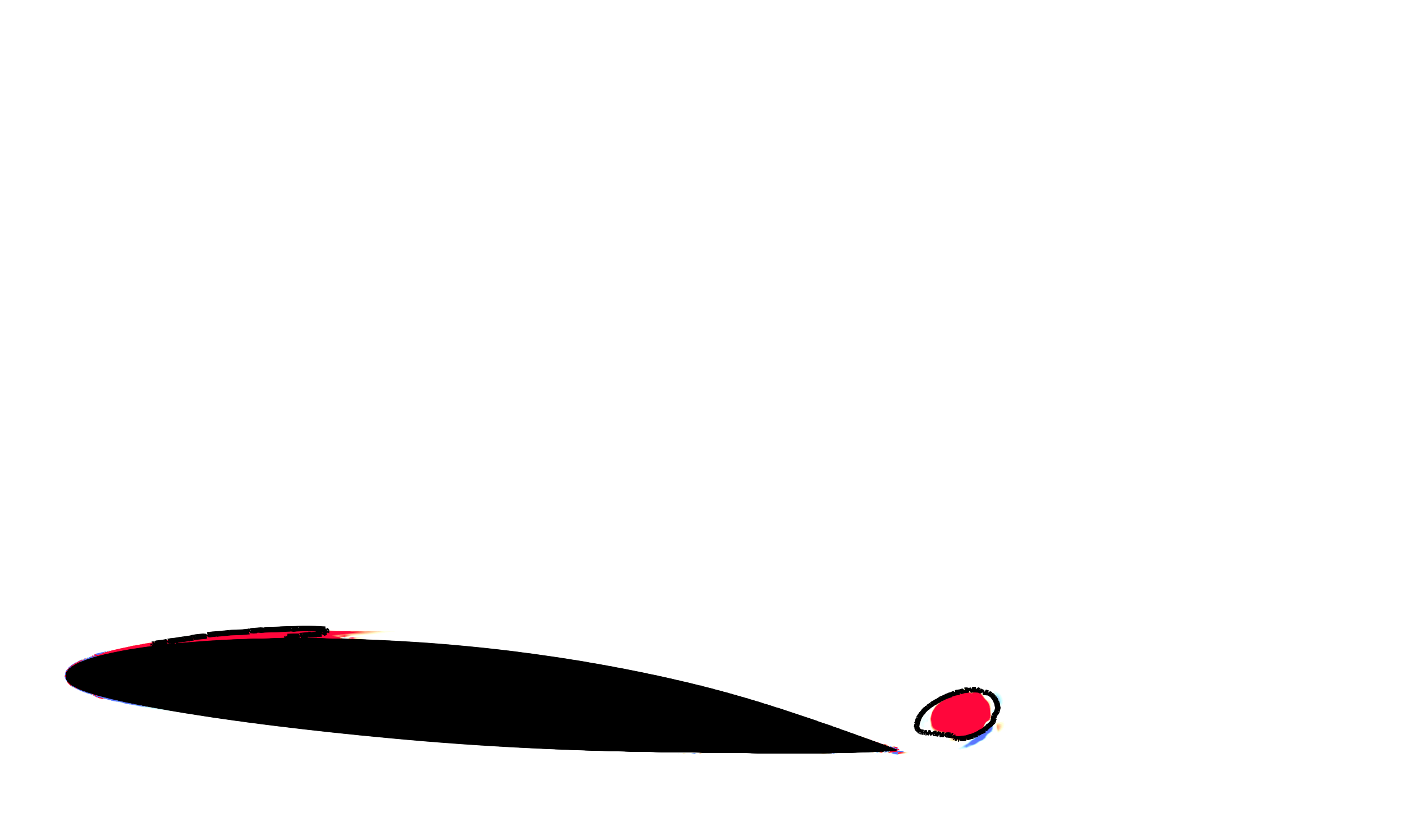}}&
 \fbox{\includegraphics[width=0.33\textwidth]{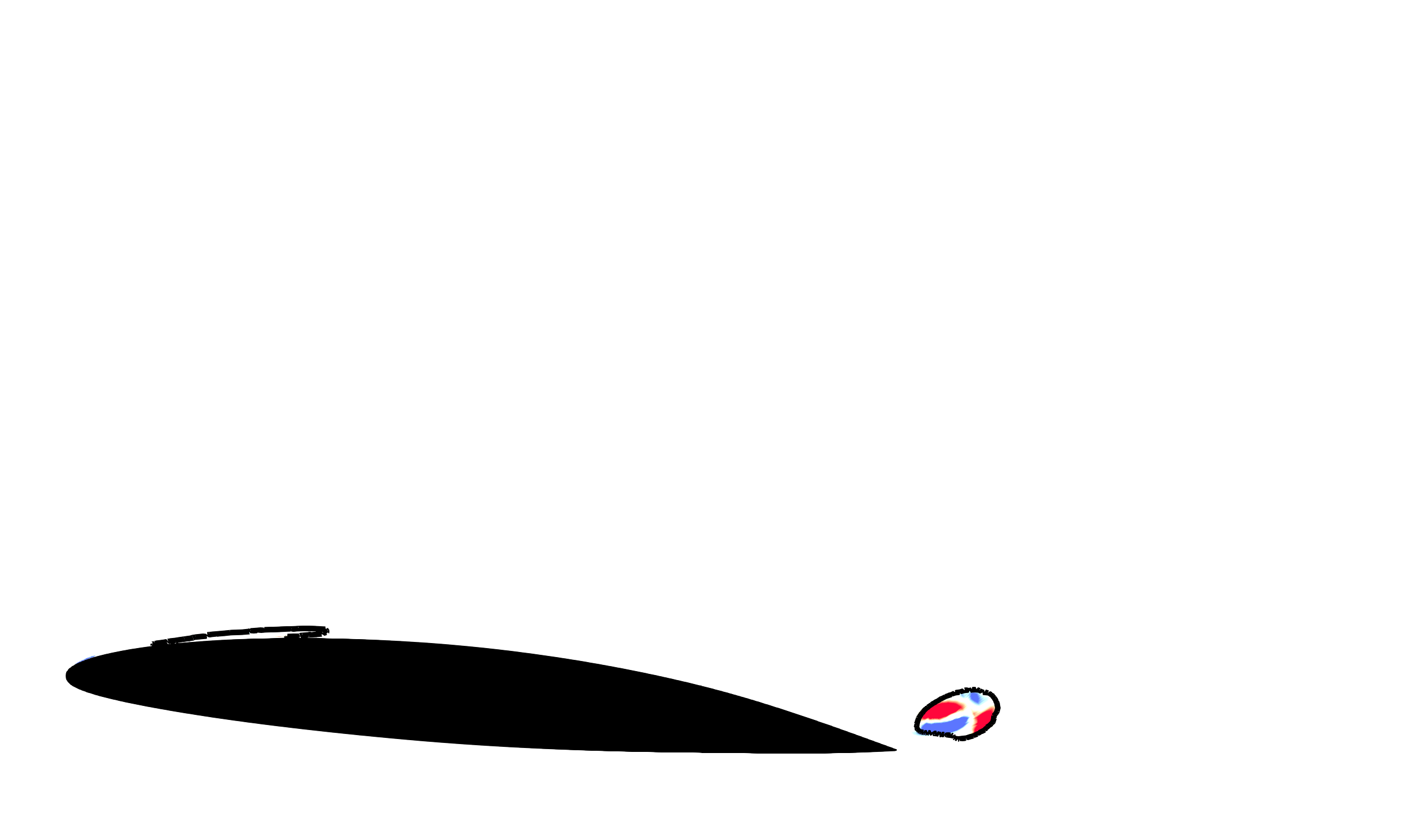}}\\
 \rotatebox[origin=l]{90}{\makebox[0.4in]{}}%
 \quad
 \quad
 \fbox{\includegraphics[width=0.33\textwidth]{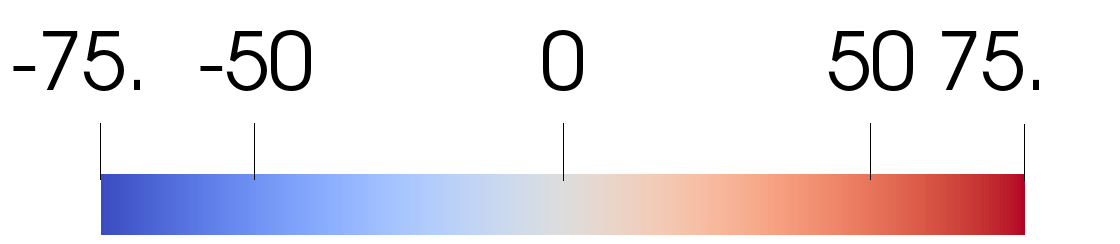}}&%
 \fbox{\includegraphics[width=0.33\textwidth]{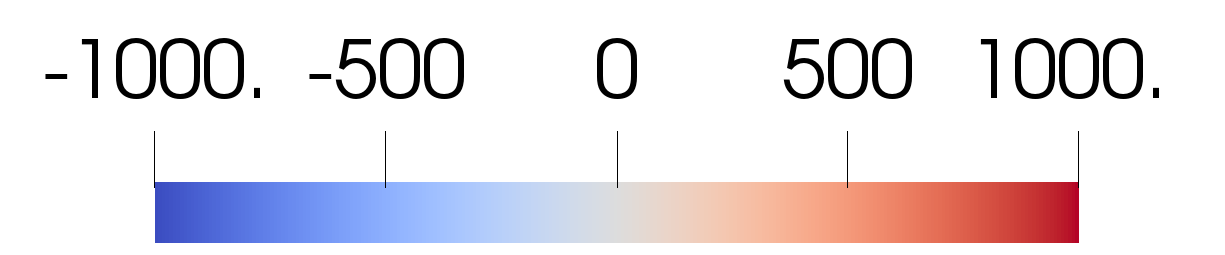}}&
 \fbox{\includegraphics[width=0.33\textwidth]{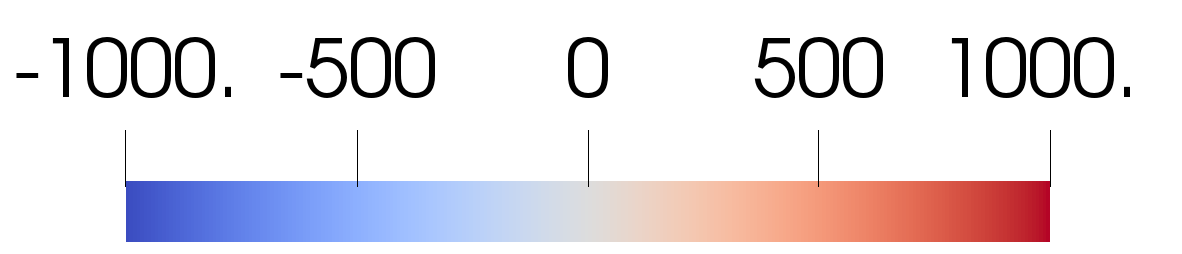}}\\
 $\boldsymbol{\omega}_z$&
 $\left( \boldsymbol{\omega}(\nabla \cdot \mathbf{u}) \right)_z$&
 $\left( \dfrac{\boldsymbol{\nabla} \rho \times \boldsymbol{\nabla} p}{\rho^2} \right)_z$\\
\end{tabular}
\caption{\label{fig:vortTE}Flow features at key stages of lift cycle depicted in Fig.~\ref{fig:A5_6_CL_TS}. (a) $Z$-component of the vorticity $\boldsymbol{\omega}$, (b) and (c) $Z$-components of the terms $\mathcal{B}$ and $\mathcal{C}$ in Eq.~(\ref{eq:vortTransEq}). The cavity outline is marked in solid black line. Positive (counter-clockwise) $\omega_z$ is out of the plane and negative (clockwise) $\omega_z$ is into the plane of the paper.}
\end{figure}

\begin{figure}[htbp]
\vspace*{-0.5in}
\centering
\quad \quad \quad \quad
\begin{subfigure}[t]{0.3\textwidth}
    \includegraphics[width=\linewidth]{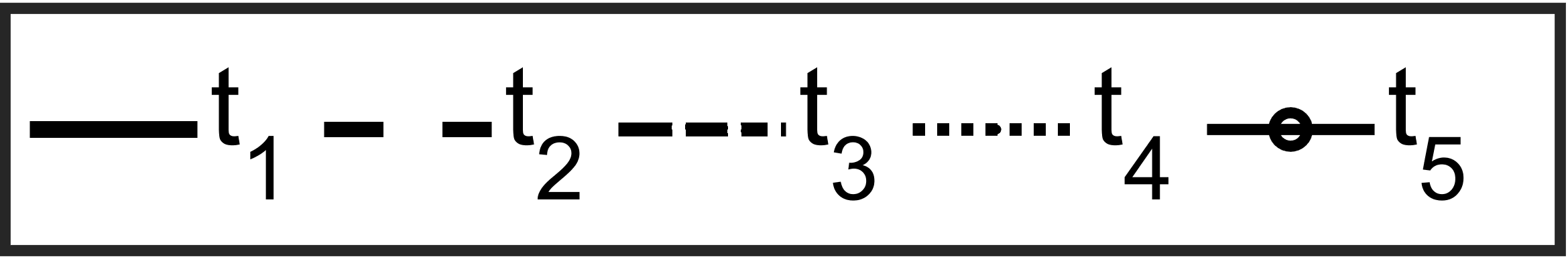}
\end{subfigure}

\begin{subfigure}[t]{0.75\textwidth}
    \includegraphics[width=\linewidth]{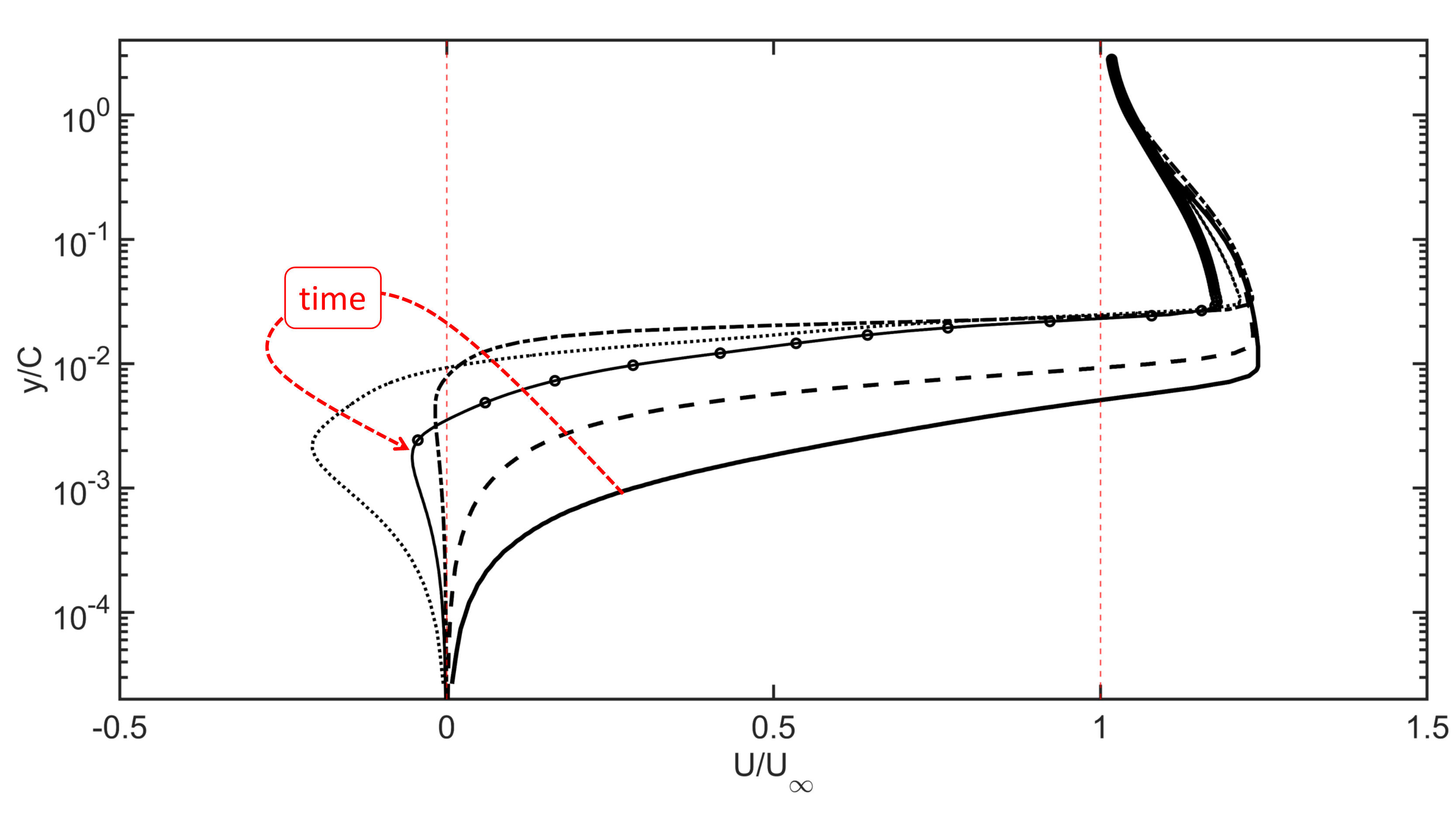}
\caption{$x/C = 0.25$}
\end{subfigure}

\begin{subfigure}[t]{0.75\textwidth}
    \includegraphics[width=\linewidth]{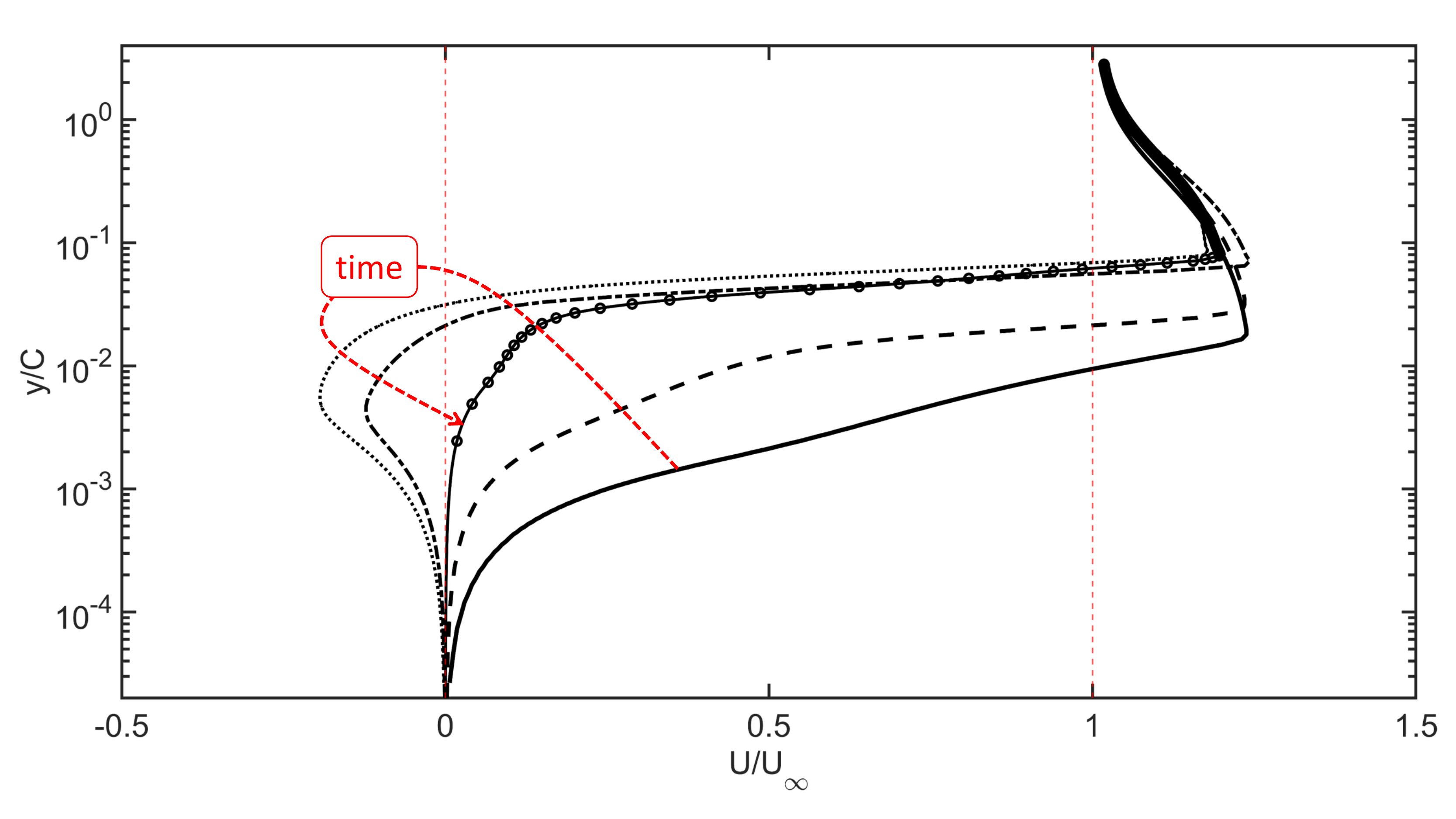}
\caption{$x/C = 0.5$}
\end{subfigure}

\begin{subfigure}[t]{0.75\textwidth}
    \includegraphics[width=\linewidth]{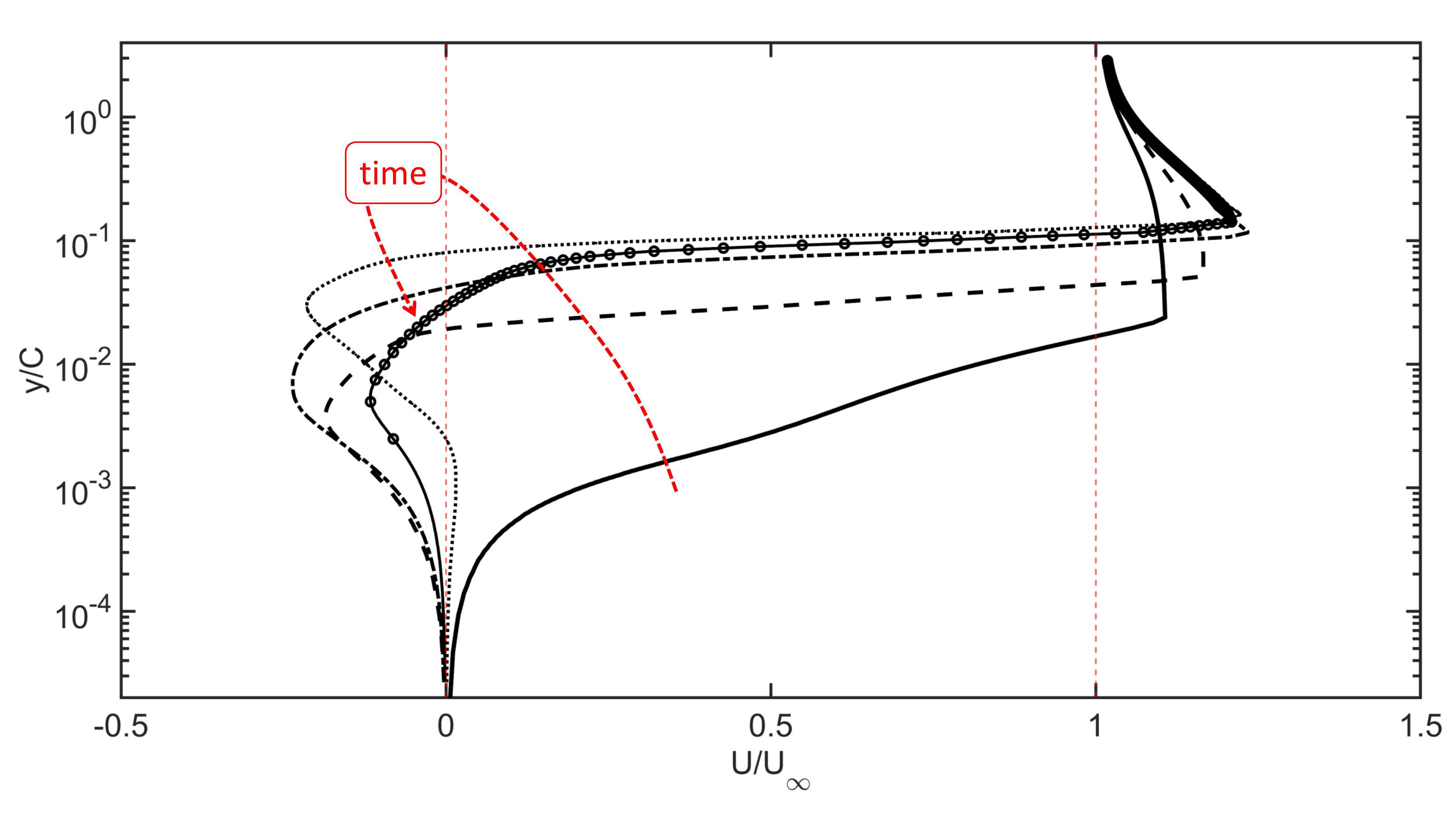}
    \caption{$x/C = 0.75$}
\end{subfigure}

\caption{Streamwise velocity on the hydrofoil suction surface at the times $t_1-t_5$ marked in Fig.~\ref{fig:A5_6_CL_TS} at (a) quarter chord $x/C = 0.25$, (b) mid chord $x/C = 0.5$ and (c) three-quarter chord $x/C = 0.75$. Significant increase in the shear layer thickness is seen in the presence of cavitation. The reverse flow originating near the trailing edge propagates upstream over the hydrofoil surface.}
\label{fig:BLevolution}
\end{figure}

Figure~\ref{fig:A5_6_CL_TS} shows a representative time-series evolution of the lift coefficient $C_L$ for the cavitating lock-in condition at $\alpha = 5^{\circ}, \sigma = 0.52, U_r = 6$. We take one typical cycle of the fluctuating lift and mark key time-stamps for discussion.
Figure~\ref{fig:vortTE} shows the evolution of the vorticity component $\boldsymbol{\omega}_z$, and the terms $\mathcal{B}$ and $\mathcal{C}$ of the transport equation (\ref{eq:vortTransEq}) at these time-stamps. Also shown is the cavity outline marked by the iso-contour of ${\phi}^{\mathrm{f}} = 0.95$. Figure~\ref{fig:BLevolution} shows corresponding changes in the boundary layer over the hydrofoil suction surface. Based on the observations, we propose a mechanism for the unsteady lift forces observed in cavitating flows.

\begin{enumerate}
    \item \textbf{Pre-inception ($t_0$)}: At the time $t_0$, cavity inception has not yet started. Vorticity $\boldsymbol{\omega}_z$ is concentrated primarily in a thin shear layer along the hydrofoil surface and wake. There is no significant presence of the terms $\mathcal{B}$ and $\mathcal{C}$ in the domain. During this stage, there is a steady increase in the lift forces.
    
    \item \textbf{Attached cavity growth ($t_1$)}: At the time $t_1$, a thin attached cavity layer has developed over the hydrofoil suction surface and growing in the streamline direction towards the trailing edge. A concentration of the vortex stretching term $\mathcal{B}$ is observed inside the two-phase compressible liquid-vapor mixture and is zero in the liquid away from the cavity. The magnitude of the baroclinic term $\mathcal{C}$ is relatively low and is limited primarily to the cavity trailing end. The boundary layer thickness increases with the growth of the cavity over the hydrofoil surface. This influences the effective hydrofoil curvature encountered by the incoming flow. The lift forces continue to increase at this stage.
    
    \item \textbf{Trailing edge interaction and cavity instability ($t_2 - t_3$)}: Between times $t_2$ and $t_3$, rapid dynamical changes occur. Just before time $t_2$, the attached cavity has grown to a length $\approx0.8C$ towards the trailing edge of the hydrofoil. Note that the maximum cavity length is longer than this and overhangs over the blade near the trailing end. This is close to the separation point on the hydrofoil suction surface and an adverse pressure gradient exists along the surface in the mean flow direction. The cavity trailing end starts growing away from the hydrofoil surface, followed by the overall growth of the thickness of the rest of the cavity. There is a significant increase in the vortex stretching term $\mathcal{B}$ inside the expanding cavity. The vortex stretching term redistributes the existing vorticity. Due to the presence of the negative vorticity shear layer on the surface, this primarily results in a positive addition to the vorticity transport. This is accompanied by a thickening of the shear layer as seen in Fig.~\ref{fig:BLevolution}. On the other hand, the magnitude of the baroclinic term $\mathcal{C}$ increases towards the trailing end of the cavity. As the cavity grows away from the hydrofoil surface, density gradients exist in the transverse flow direction. The density gradient interacts with the reverse pressure gradient in the orthogonal streamline direction near the separation point resulting in baroclinic torque. The specific alignment of the density gradients and pressure gradients at this stage contributes primarily to the generation of clockwise (negative) $\boldsymbol{\omega}_z$ near the trailing edge. By conservation of angular momentum, there should be a counter-clockwise circulation of flow around the hydrofoil to balance the generated baroclinic vorticity. This is observed in the next time frame.
    
    \item \textbf{Reverse flow and cavity detachment ($t_3 - t_4$)}: Between times $t_3$ and $t_4$, the positive vorticity shear layer on the hydrofoil pressure surface starts curving around the trailing edge and moves upstream along the suction surface. The interaction of the shear layers destabilizes the flow and alternating vortices are shed. There is a reverse flow along the surface and the separation point moves upstream, detaching the cavity. The detached cavities are shed along with the shedding vortices and convected by the mean flow. The attached cavity length reaches a minimum and experiences an unstable rebound growth. This period also observes a rapid loss of the lift force experienced by the hydrofoil, caused by the counter-clockwise circulation of fluid near the hydrofoil surface.
    
    \item \textbf{Cavity rebound and collapse ($t_4 - t_5$)}: Between times $t_4$ and $t_5$, the cavity undergoes rapid cycles of rebound growth and collapse. These are unsustainable and accompany unstable streamline pressure gradients as the upstream-moving adverse pressure gradients interact with an opposite pressure gradient at the leading edge of the hydrofoil. After a few cycles of alternating growth and collapse, there is a complete collapse of the cavity.
    
\end{enumerate}

\begin{figure}[htbp]
\begin{subfigure}{\textwidth}
  \centering
  \includegraphics[width=0.6\linewidth]{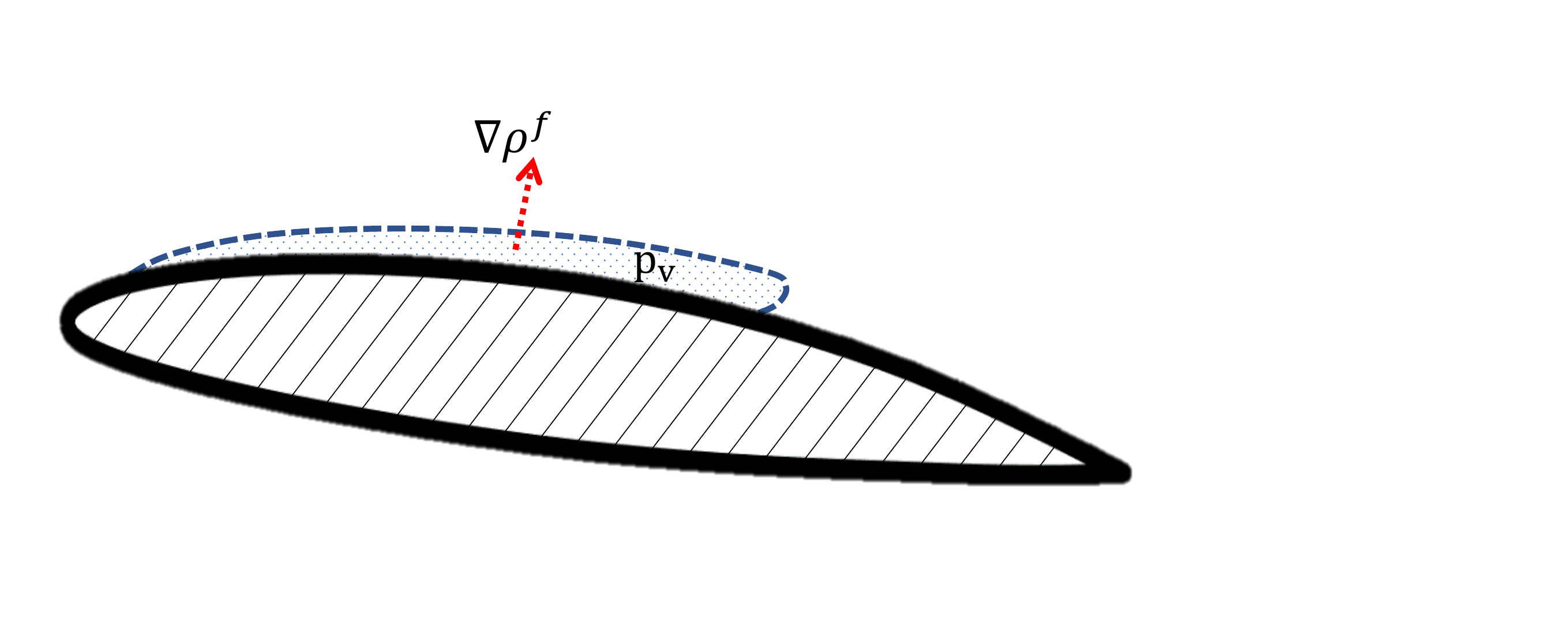}
  \caption{Growth stage of the attached cavity: The expanding cavity contributes to a positive divergence of velocity in the compressible mixture. With the presence of a negative magnitude vorticity shear layer on the suction surface, the overall contribution of the term $\left( \boldsymbol{\omega}(\nabla \cdot \mathbf{u}) \right)$ to the vorticity transport is positive. Within the bulk of the cavity the pressure $p = p_v$. Density gradients are out primarily normal to the cavity interface into the incompressible liquid. No significant orthogonality between the density and pressure gradients is observed apart from the trailing end of the cavity. The lift is observed to increase steadily at this stage.}
  \label{fig:mechanism1}
\end{subfigure}
\begin{subfigure}{\textwidth}
  \centering
  \includegraphics[width=0.6\linewidth]{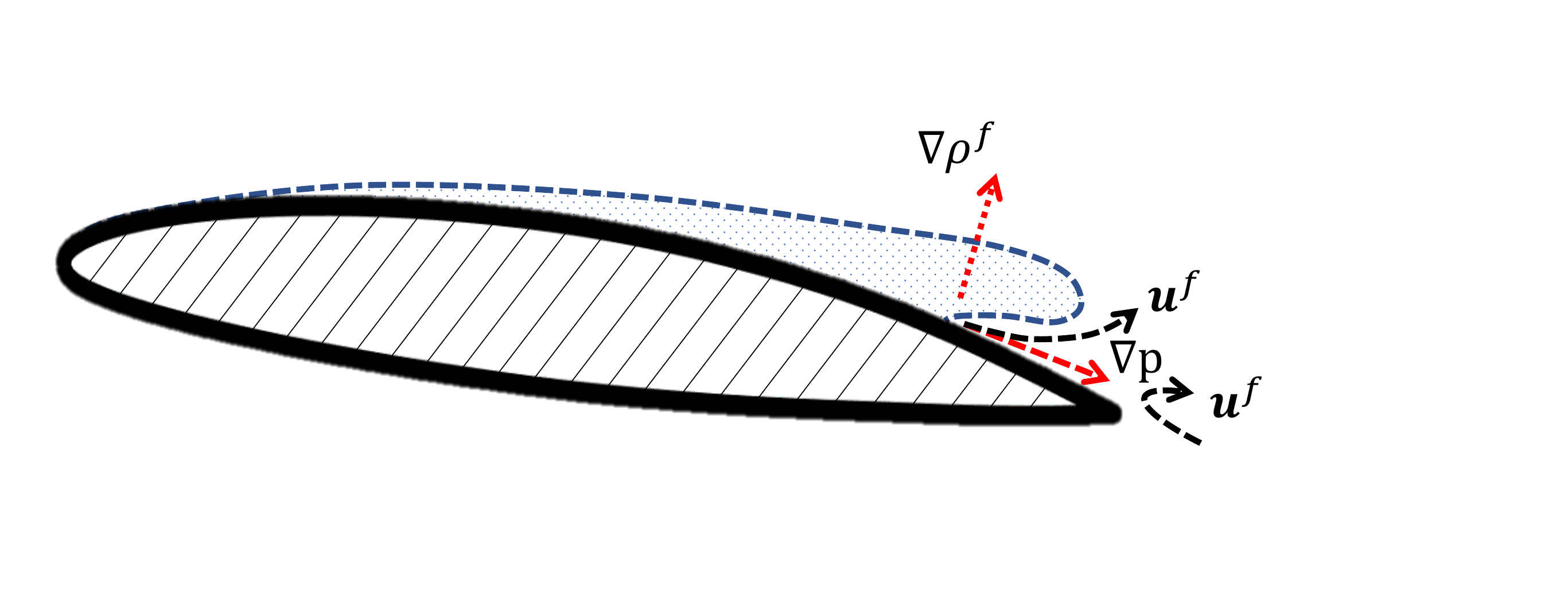}
  \caption{As the cavity reaches the trailing edge of the hydrofoil it encounters an adverse pressure gradient. Depending on the strength of the adverse pressure gradient a flow separation point may exist near the trailing edge lifting the trailing end of the cavity. In both cases orthogonal components of the density and pressure gradients now exist near the trailing end of the cavity, contributing to clockwise baroclinic torque.}
  \label{fig:mechanism2}
\end{subfigure}
\begin{subfigure}{\textwidth}
  \centering
  \includegraphics[width=0.6\linewidth]{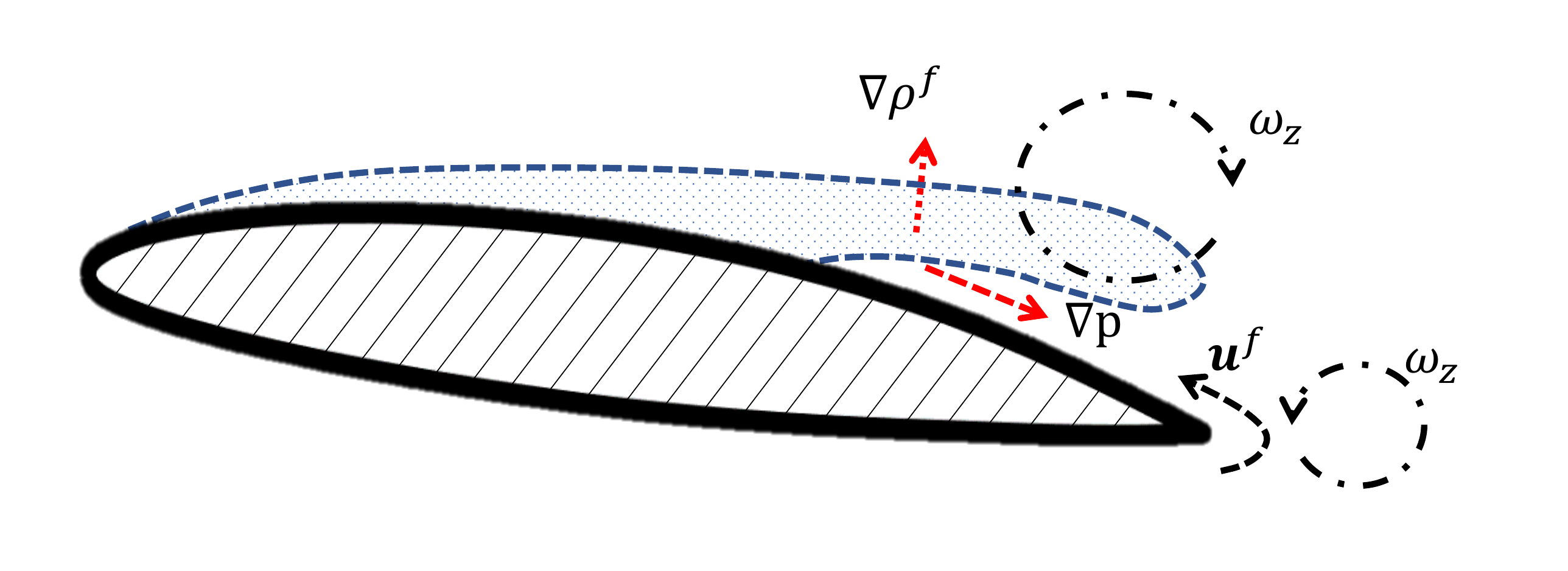} 
  \caption{Generation of strong clockwise vorticity near the cavity trailing end can lead to a counter-clockwise flow circulation around the hydrofoil for the conservation of angular momentum. This results in a reverse flow on the hydrofoil suction surface along with the adverse pressure moving further upstream, further detaching the cavity and generation of negative vorticity. The counter-clockwise flow circulation near the hydrofoil can lead to loss of lift which is observed at this stage. Along with the flow, a positive vorticity shear layer moves upstream along the hydrofoil suction surface.}
  \label{fig:mechanism3}
\end{subfigure}
\begin{subfigure}{\textwidth}
  \centering
  \includegraphics[width=0.6\linewidth]{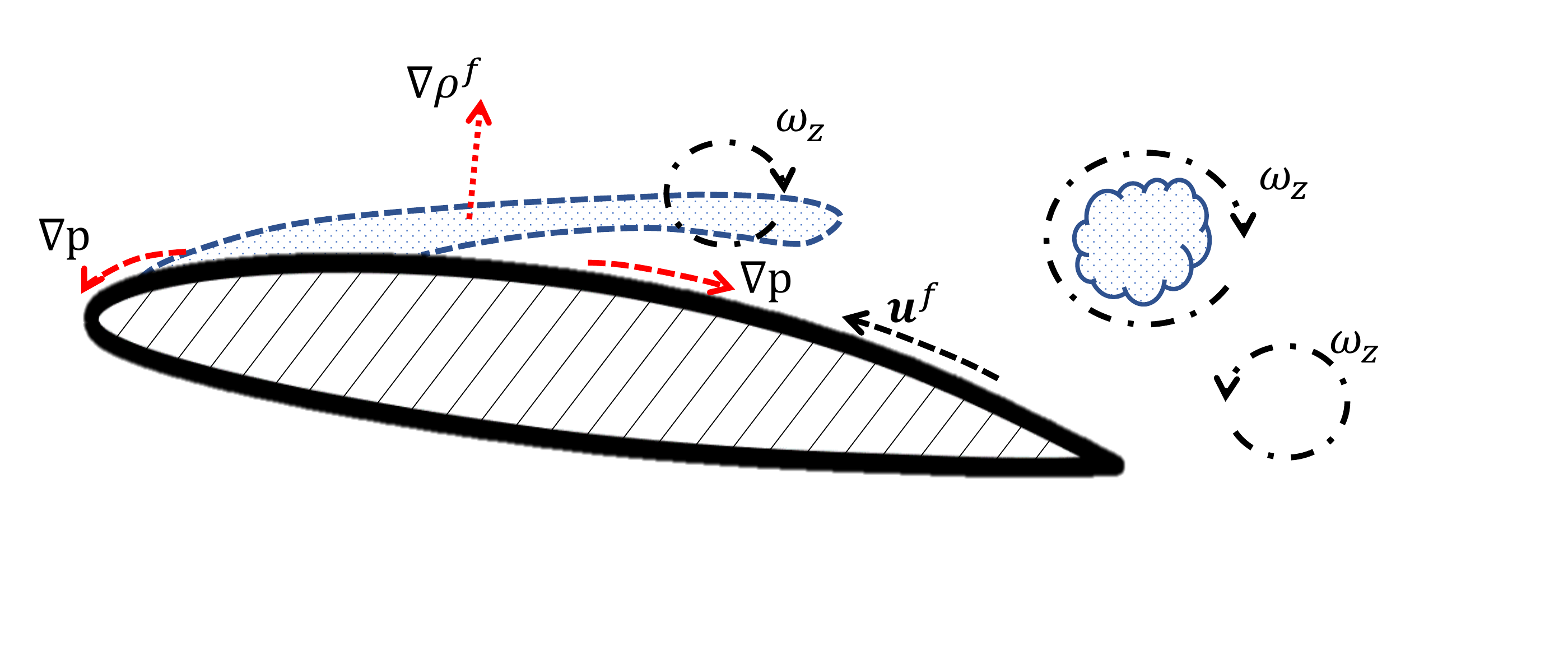} 
  \caption{As the reverse flow cuts through the cavity and reaches the leading edge it encounters the opposite pressure gradient near the leading edge stagnation point. At this stage, the cavity can be completely detached or undergo unsustainable rebound cycles before the complete collapse. Towards the hydrofoil trailing edge the mixing of the shear layers can result in flow destabilization and shedding of vortices, often collocated with shedding cavities.}
  \label{fig:mechanism4}
\end{subfigure}
\caption{Sketches of flow patterns (a), (b), (c) and (d) highlighting salient stages in the unstable partial cavity cycle and corresponding effects on vorticity and lift. Descriptions are provided in the captions in support of the accompanying text and numerical results presented in Fig.~\ref{fig:vortTE}.}
\label{fig:mechanism}
\end{figure}

We now make an attempt to describe the mechanism using the flow cartoons in Fig.~\ref{fig:mechanism}. Detailed descriptions are provided in the captions of Fig.~\ref{fig:mechanism} whereby flow features salient to the descriptions are shown. Note that we propose a plausible mechanism for this complex interplay between the cavitating flow and the fluid-structure interaction. There is a possibility of multiple mechanisms during this coupled cavitation and fluid-structure system. For example, \citet{smith2020influence} discussed the influence of added-mass effects because of the presence of cavity on the hydrofoil response. 

\begin{figure}
\centering
\begin{subfigure}{.85\textwidth}
  \centering
  \includegraphics[width=\linewidth]{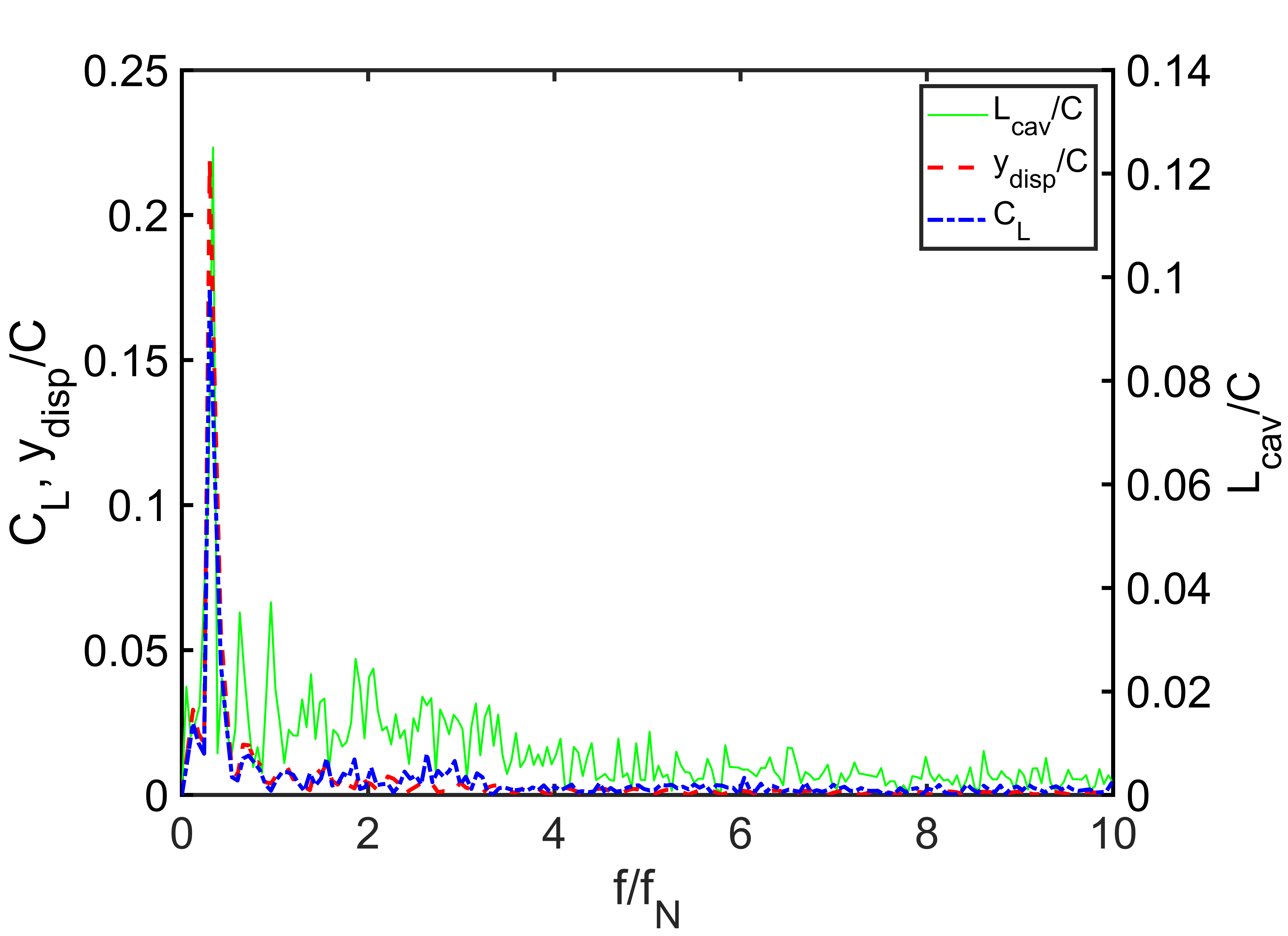}
  \caption{$\alpha = 5^{\circ}, \sigma = 0.52, U_r = 6$}
\end{subfigure}\vfill
\begin{subfigure}{.85\textwidth}
  \centering
  \includegraphics[width=\linewidth]{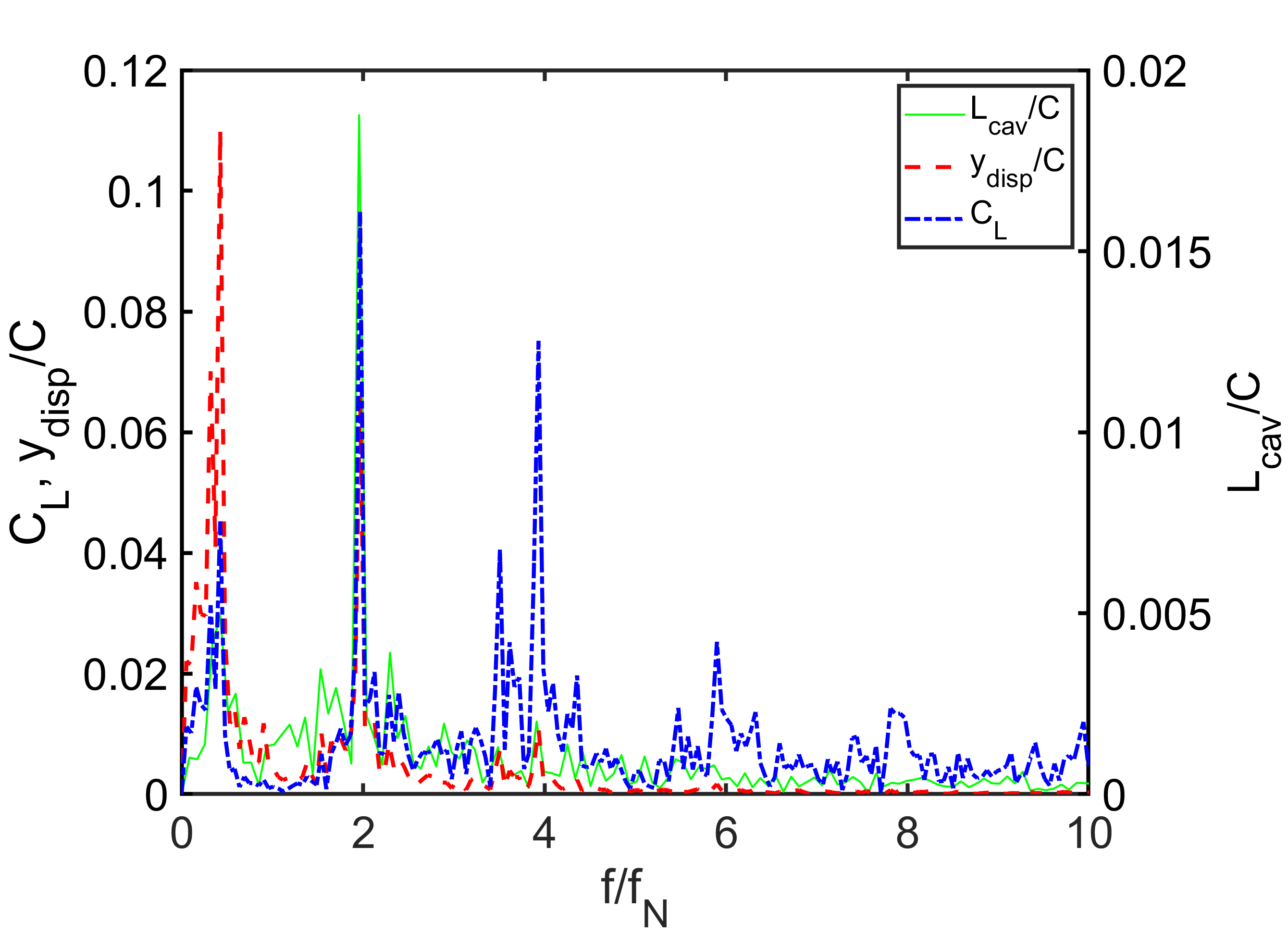}
  \caption{$\alpha = 5^{\circ}, \sigma = 0.52, U_r = 8$}
\end{subfigure}
\caption{Representative frequency spectra for a freely oscillating hydrofoil: (a) lock-in, and (b) post-lock-in. The dominant frequencies for the cavity shedding and the lift coefficient are seen to match in both cases, confirming a synchronization between the two.}
\label{fig:fftComparisonWcav}
\end{figure}

The proposed cyclic process results in rapidly fluctuating lift forces, even at low angles of attack $\alpha$ where lift fluctuations are otherwise observed negligible for non-cavitating flow. This is confirmed in Fig.~\ref{fig:fftComparisonWcav} where the frequency spectra of oscillations in the cavity length $L_{cav}$, the lift coefficient $C_L$ and the transverse displacement $y_{disp}$ are compared in the lock-in and post-lock-in regimes. The cavity shedding frequency $f_{cav}$ is seen to be consistent with $f_{C_L}$ both in the lock-in and post-lock-in regimes. When these periodic fluctuations lock onto the natural frequency of the hydrofoil or its harmonics, it can result in large amplitude transverse oscillations as observed in the previous section. 

\begin{figure}[htbp]
\begin{subfigure}{\textwidth}
  \centering
  \includegraphics[width=0.4\linewidth]{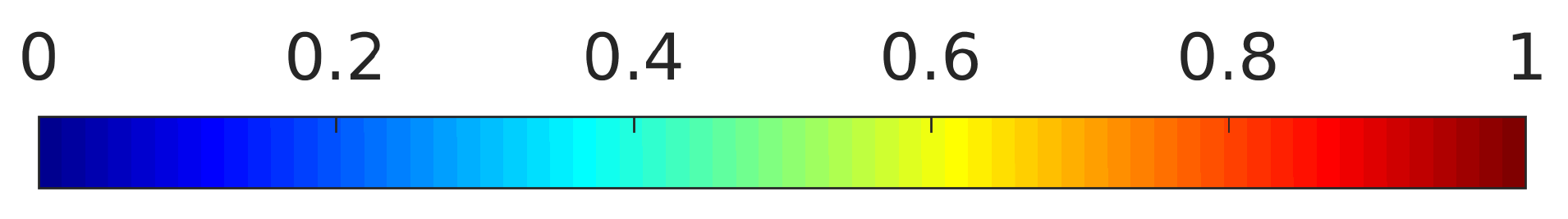}  
\end{subfigure}
\begin{subfigure}{.05\textwidth}
  \centering
  \includegraphics[height=1.775in]{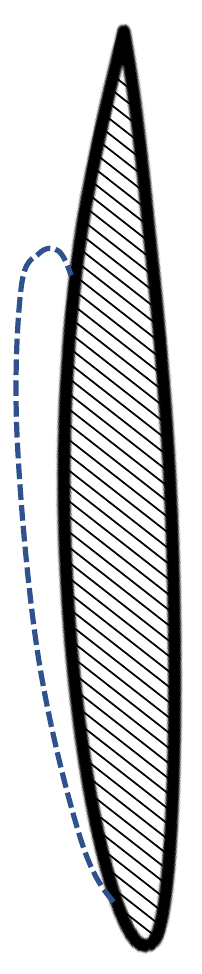}
\end{subfigure}
\begin{subfigure}{.475\textwidth}\vskip 22pt
  \centering
  \includegraphics[width=\linewidth]{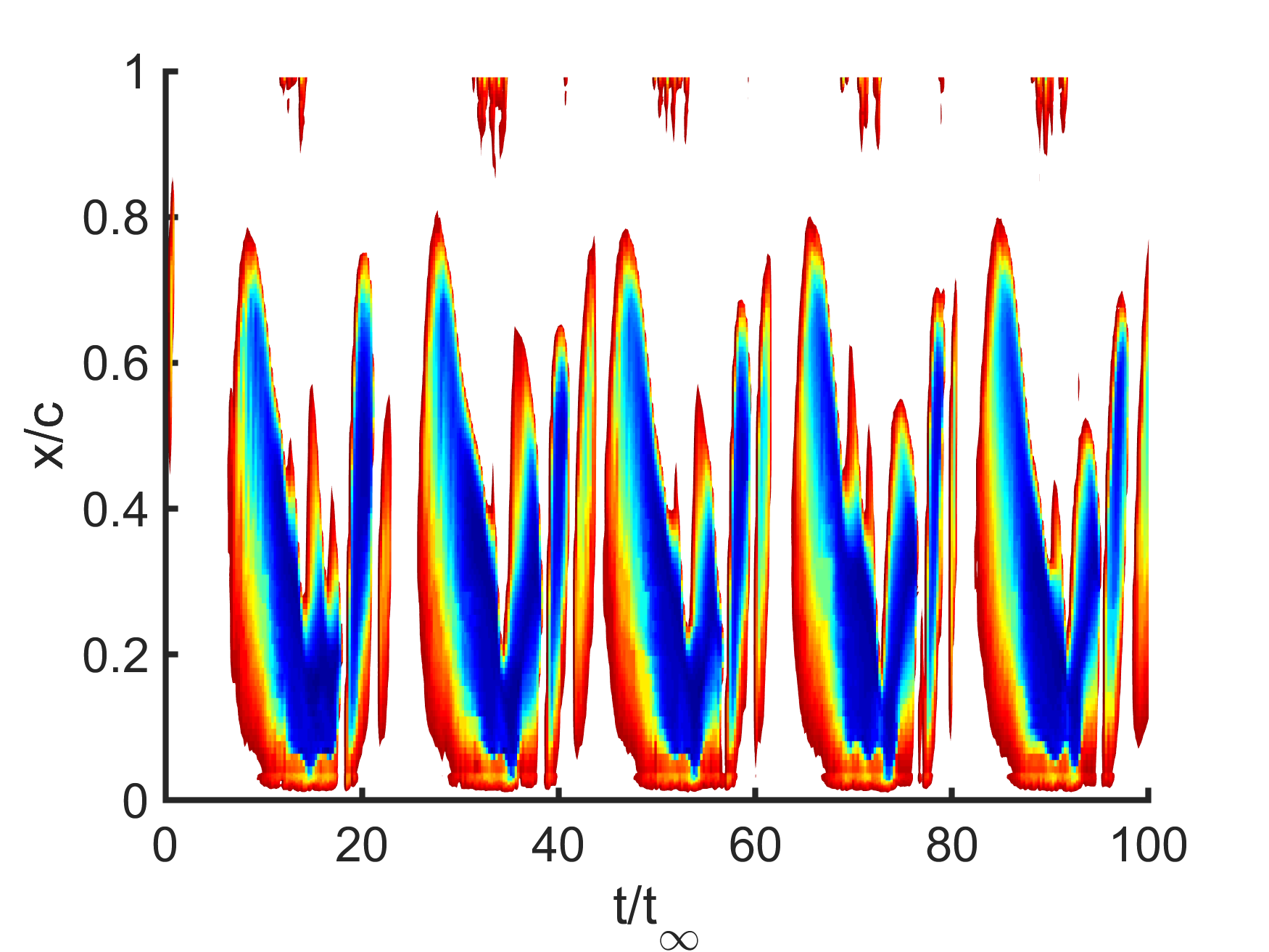}
  \caption{\label{fig:cavContUr6}$\phi^{\mathrm{f}}$ on hydrofoil suction surface $\alpha = 5^{\circ}, \sigma = 0.52, U_r = 6$}
\end{subfigure}
\begin{subfigure}{.475\textwidth}\vskip 22pt
  \centering
  \includegraphics[width=\linewidth]{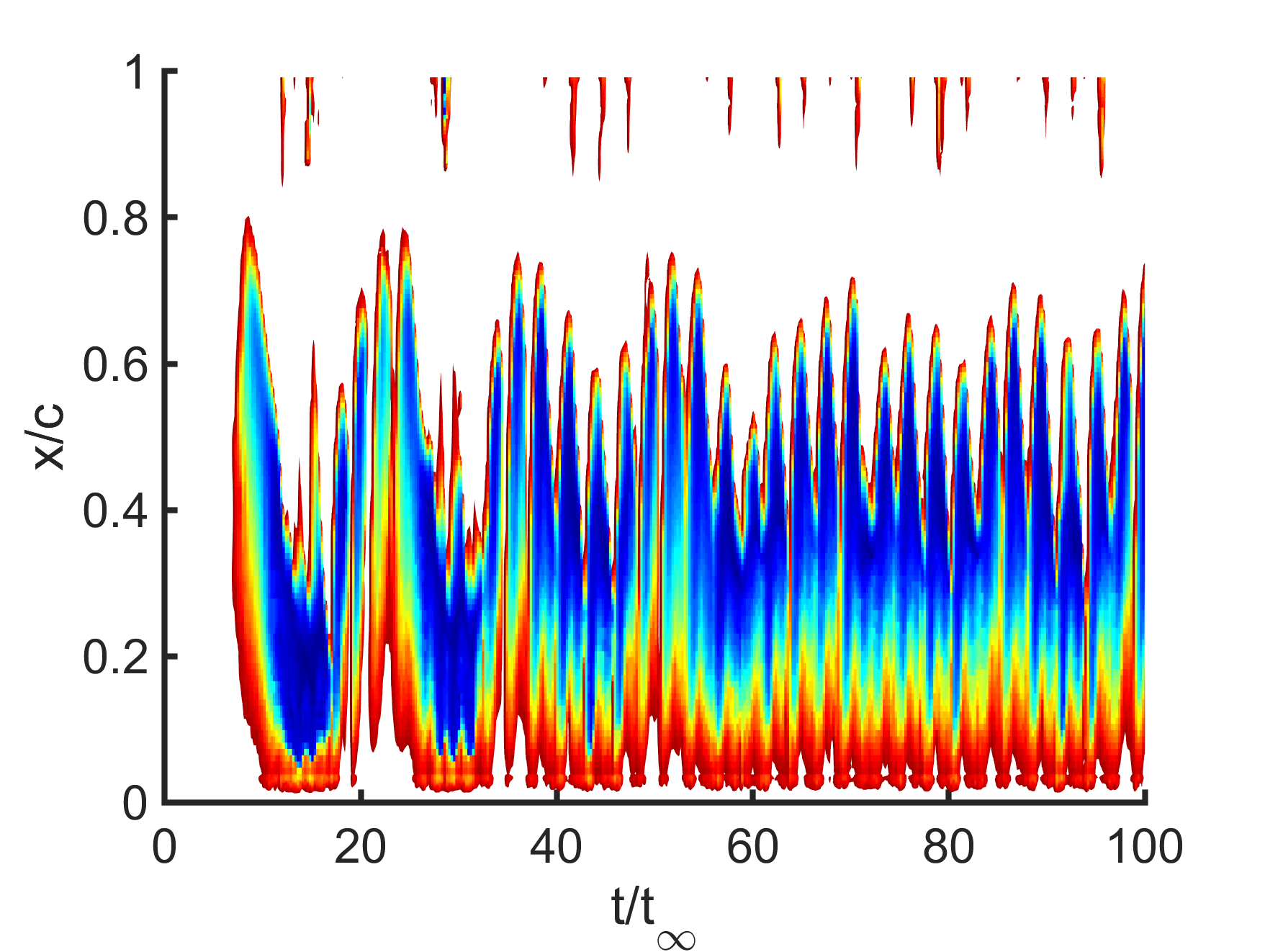}
  \caption{\label{fig:cavContUr8}$\phi^{\mathrm{f}}$ on hydrofoil suction surface $\alpha = 5^{\circ}, \sigma = 0.52, U_r = 8$}
\end{subfigure}
\begin{subfigure}{\textwidth}
  \centering
  \includegraphics[width=0.5\linewidth]{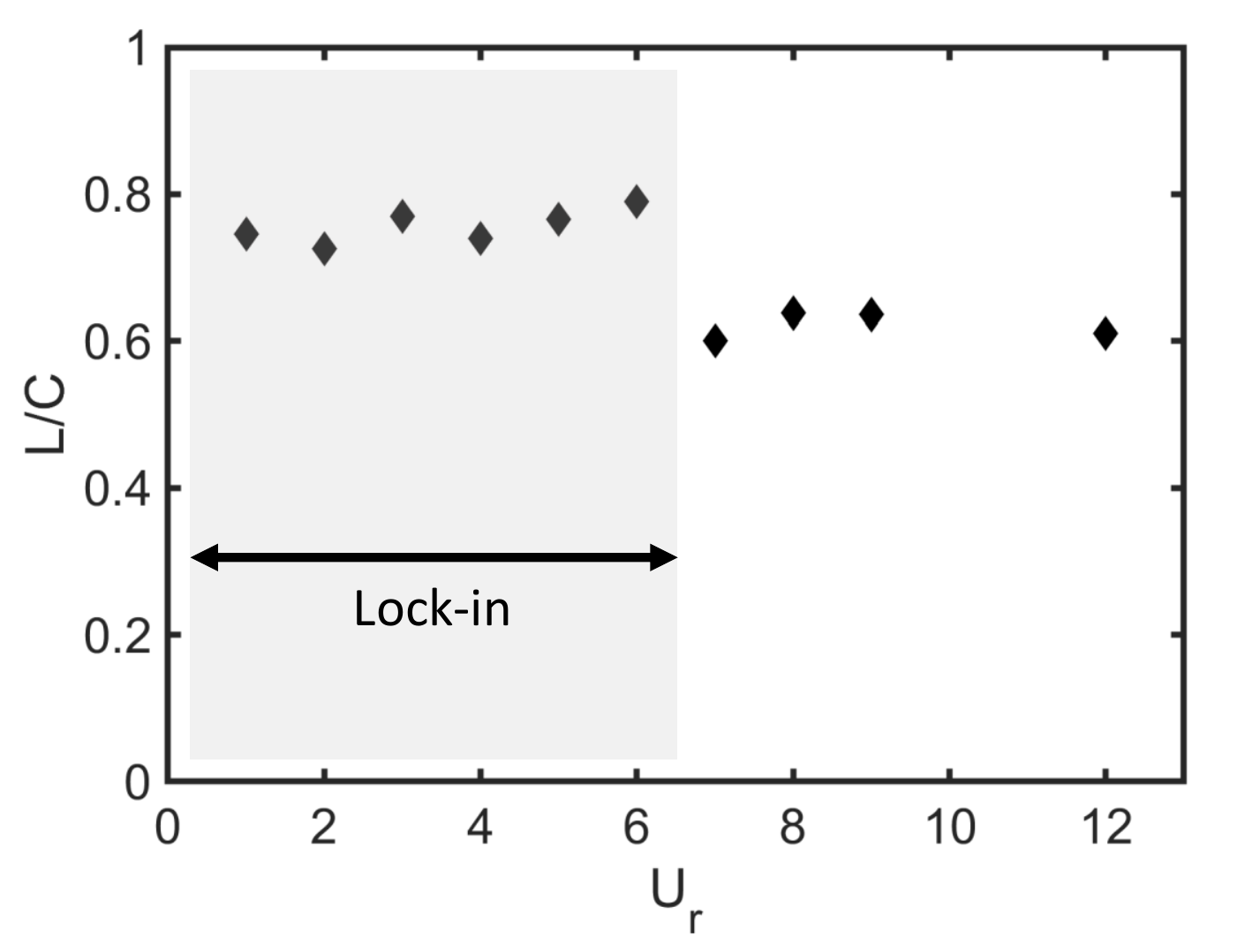}  
  \caption{\label{fig:A5_LbyC}Maximum attached cavity length $\alpha = 5^{\circ}, \sigma = 0.52$}
\end{subfigure}
\caption{Contours of the phase fraction $\phi^{\mathrm{f}}$ along the hydrofoil suction surface demonstrating the cavity evolution during: (a) lock-in and (b) post-lock-in. Here $\phi^{\mathrm{f}}$ is limited to the range $\phi^{\mathrm{f}} \in \left[0, 0.95 \right]$ for easy identification of the cavity. The location $x/C = 0$ corresponds to the hydrofoil leading edge with trailing edge at $x/C = 1$.  (c) Time-averaged values of the maximum attached cavity length $L$ normalized by the chord length $C$ over the range of $U_r$.}
\label{fig:cavBehaviour}
\end{figure}

Before closing, we note that the structural vibration also influences the cavity dynamics, with distinct cavity patterns found in the lock-in and post-lock-in regimes. Figures~\ref{fig:cavContUr6} and \ref{fig:cavContUr8} show the phase fraction ${\phi^{\mathrm{f}}}$ on the suction surface of the hydrofoil. During lock-in, the cavity goes through a full growth-collapse cycle, with periodic unstable rebounds. During the post-lock-in regime, the cavity is seen to undergo rapid oscillations. The frequency of these oscillations matches the secondary low-amplitude vibration frequency. The maximum attached cavity length is observed to be shorter than in the lock-in regime. Rapid cavity shedding is observed but is primarily limited to the tail end of the cavity. Complete cavity collapse is not as frequent. 
There is a distinct difference in the average maximum lengths of the cavities in the two regimes given in Fig.~\ref{fig:A5_LbyC}, with the formation of larger coherent cavitating structures over the hydrofoil during lock-in. In Appendix B, the 3D effect on the cavity dynamics and the hydroelastic is briefly presented which also confirm the adequacy of the 2D investigations.

\section{Conclusion}
A numerical study has been performed to investigate the flow-induced vibration of a freely vibrating hydrofoil in unsteady cavitating conditions. We employed the recently developed unified variational framework for fluid–structure interaction with cavitating flows at high Reynolds number. This study provided several novel insights into the mechanism of flow-induced vibrations in the presence of unsteady cavitation. Unsteady partial cavitating conditions were seen to produce structural vibrations several orders of magnitude higher than non-cavitating counterparts at low angles of attack. We summarize some of the salient findings as follows:
 \begin{itemize}
     \item A lock-in mechanism of unsteady lift forces to the structural frequency is identified to sustain high amplitude oscillations of the hydrofoil. 
     \item During lock-in, the dominant fluid-structure frequencies synchronize at a sub-harmonic of the hydrofoil natural frequency. In the post-lock-in, the fluid frequencies (cavity, vortex shedding and lift) synchronize distinct from the frequency of structural vibration.
    \item The unsteady features of cavity dynamics drive fluctuations in the lift forces via a process of vorticity generation which seems to be a possible mechanism of the lift unsteadiness and the frequency lock-in with large-amplitude oscillations.
    \item Sufficiently distinguishable regimes in the data are present during lock-in and post-lock-in, generalizable over a range of angles of attack. This holds promise for the development of efficient noise mitigation strategies in propellers.
\end{itemize}

During frequency lock-in, large coherent cavitating structures are seen over the hydrofoil suction surface. The cavities undergo a full cavity growth-detachment-collapse cycle in sync with the corresponding changes in the fluctuating lift forces. Post-lock-in shedding is observed to be primarily from the trailing end of the cavity with a reduction in the frequency of complete detachment. The fluctuations in the lift forces are still noted to be influenced by the unsteady cavity dynamics, however, the synchronization with the structural vibration does not exist. 
A key highlight is the identification of distinguishable characteristics in the frequency spectra and the physical behavior of the fluid and structural dynamics during lock-in and post-lock-in. This can be exploited to develop mitigation strategies for propeller noise.

The above insights can prove useful in the development of targeted control mechanisms for the mitigation of flow-induced vibrations of hydrofoils in cavitating conditions. In future work, the authors plan to extend the work to 3D flexible propeller blades with full hydroelastic deformations. Finally, we would like to acknowledge that the numerical studies are conducted using homogeneous mixture with semi-empirical modeling for cavitating flows. Although the mixture-based cavitation models are well validated and successfully employed over the years on a wide variety of flow configurations and geometries, we would like to welcome more experimental or fully-resolved cavitation modeling.

\section*{Acknowledgements}
The authors would like to acknowledge the Natural Sciences and Engineering Research Council of Canada (NSERC) for the funding. This research was enabled in part through computational resources and services  provided by (WestGrid) (https://www.westgrid.ca/), Compute Canada (www.computecanada.ca) and the Advanced Research Computing facility at the University of British Columbia. 

%
%


\bibliography{mybibfile}   

\appendix

\section{Assessment of vorticity and lift generation}
\label{sec:investigation}
The importance of the vorticity field on the hydrodynamic lift force can be demonstrated using the fundamental lifting section theories of Kelvin's circulation theorem and the Kutta-Joukowski theorem. For a simple confirmation, we use the well-known form of the Kutta-Joukowski theorem \cite{anderson2010fundamentals} which relates the lift with the circulation around the hydrofoil as
\begin{equation}
    L^{\prime}=\rho_{\infty} U_{\infty} \Gamma,
    \label{eq:KJ}
\end{equation} 
where $L^{\prime}$ is the lift per unit span of the hydrofoil, $\rho_{\infty}$ and  $U_{\infty}$ are the fluid density and velocity far upstream of the hydrofoil. $\Gamma$ is the circulation given as the line-integral of the fluid velocity $\boldsymbol{u}^{\mathrm{f}}$ along a closed curve $\partial S$ around the hydrofoil
\begin{equation}
    \Gamma=\oint_{\partial S} \boldsymbol{u}^{\mathrm{f}} \cdot \mathrm{d} \mathbf{l}
\end{equation}
Using the Stokes's Theorem, the circulation can be further related to the surface integral of the vorticity $\omega$ as 
\begin{equation}
    \Gamma=\oint_{\partial S} \boldsymbol{u}^{\mathrm{f}} \cdot \mathrm{d} \mathbf{l}=\iint_{S} \nabla \times \boldsymbol{u}^{\mathrm{f}} \cdot \mathrm{d} \boldsymbol{S}=\iint_{S} \boldsymbol{\omega} \cdot \mathrm{d} \boldsymbol{S}
\label{eq:circVorticity}
\end{equation}
This indicates that the lift force is influenced by the Cartesian z-component of vorticity $\boldsymbol{\omega}_z$ in the domain.   
\noindent 
\begin{figure}[!h]
\centering
\includegraphics[width=0.8\columnwidth]{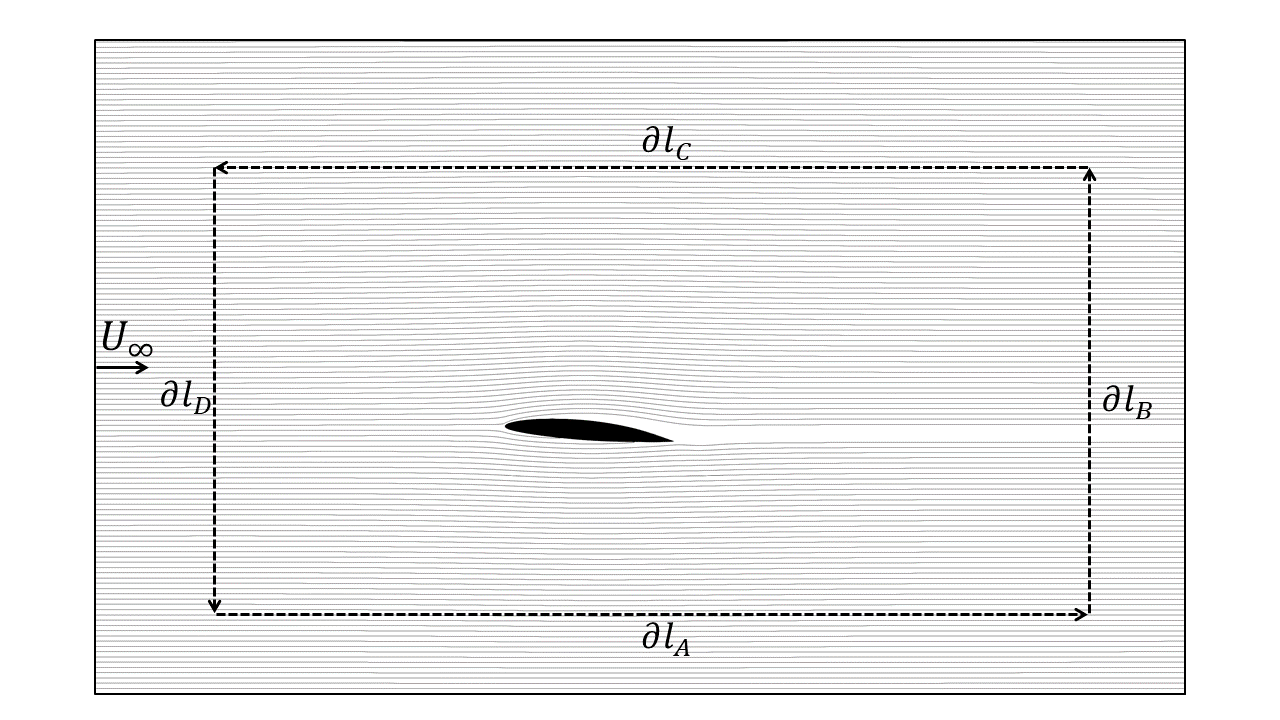}
\caption{Representative domain around the hydrofoil demonstrating stationary curves used for calculating the integral $\oint_{\partial S} \boldsymbol{u}^{\mathrm{f}} \cdot \mathrm{d} \mathbf{l}$. Also shown are the streamlines in the vicinity of the hydrofoil. Note that the curves are adjusted per the flow configuration of $\alpha, \sigma$ and $U_r$, as the mean hydrofoil displacement changes with a variation of these parameters.}
\label{fig:circLineIntgSchematic}        
\end{figure} 
For the numerical confirmation, we evaluate the circulation as the line integral of the velocity along a closed curve sufficiently away from the hydrofoil boundary layer effects. A stationary rectangular curve aligned with the Cartesian axes is chosen for ease of integration as shown in the representative Fig.~\ref{fig:circLineIntgSchematic}. Note that the spatial extents of the curves are chosen such that they encompass the range of oscillatory motion of the hydrofoil for a given flow configuration (combination of $\alpha, \sigma$ and $U_r$). The numerically obtained instantaneous velocity field is integrated along the curve as 
\begin{equation}
    \oint_{\partial S} \boldsymbol{u}^{\mathrm{f}} \cdot \mathrm{d} \mathbf{l} = \int_{A} U \cdot \mathrm{d} \mathbf{l_A} + \int_{B} V \cdot \mathrm{d} \mathbf{l_B} - \int_{C} U \cdot \mathrm{d} \mathbf{l_C} - \int_{D} V \cdot \mathrm{d} \mathbf{l_D}
\end{equation}
where $U$ and $V$ are the Cartesian x- and y-components of the fluid velocity field $\boldsymbol{u}^{\mathrm{f}}$.
\begin{figure}
\begin{subfigure}{\textwidth}
  \centering
  \includegraphics[width=0.3\linewidth]{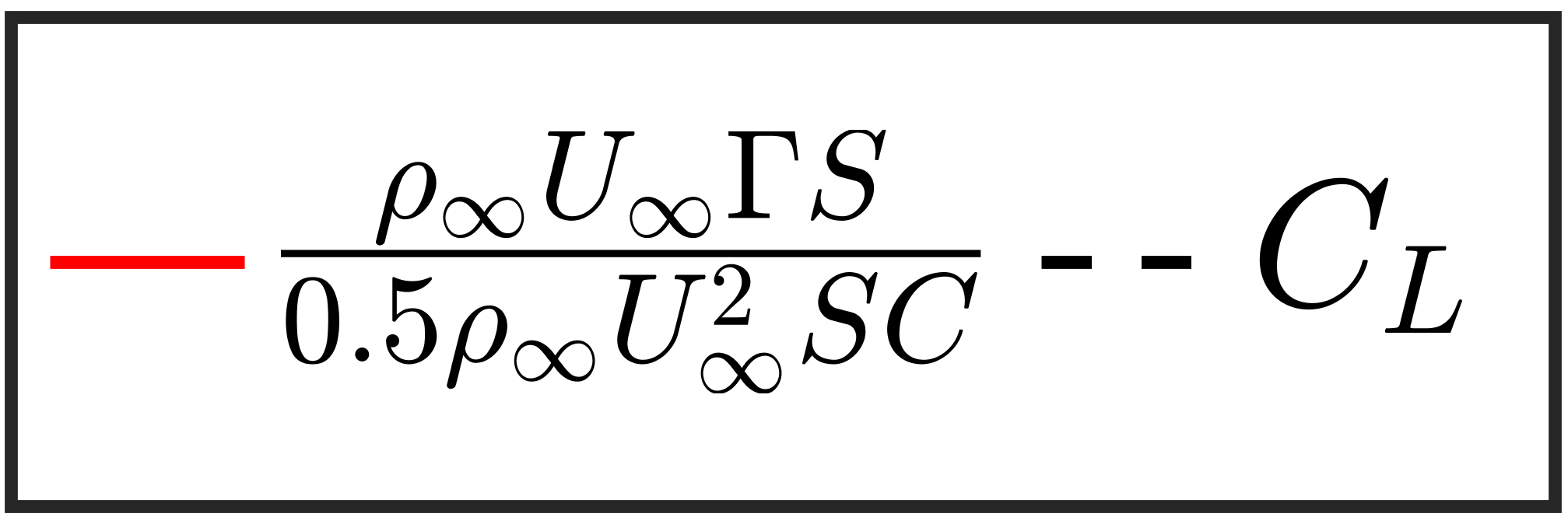}
\end{subfigure}
\begin{subfigure}{\textwidth}
  \centering
  \includegraphics[width=0.8\linewidth]{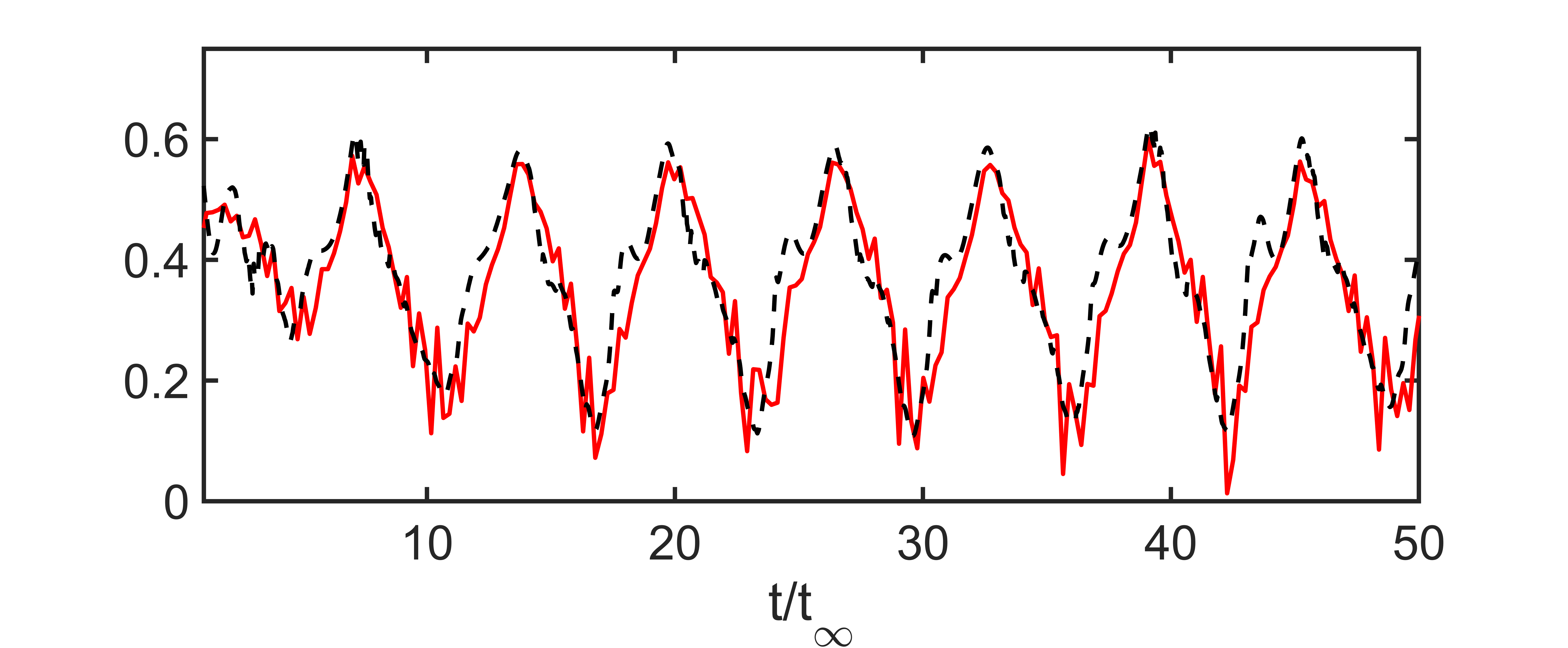}
  \caption{$\alpha = 5^{\circ}, \sigma = 0.52, U_r = 1$}
\end{subfigure}
\begin{subfigure}{\textwidth}
  \centering
  \includegraphics[width=0.8\linewidth]{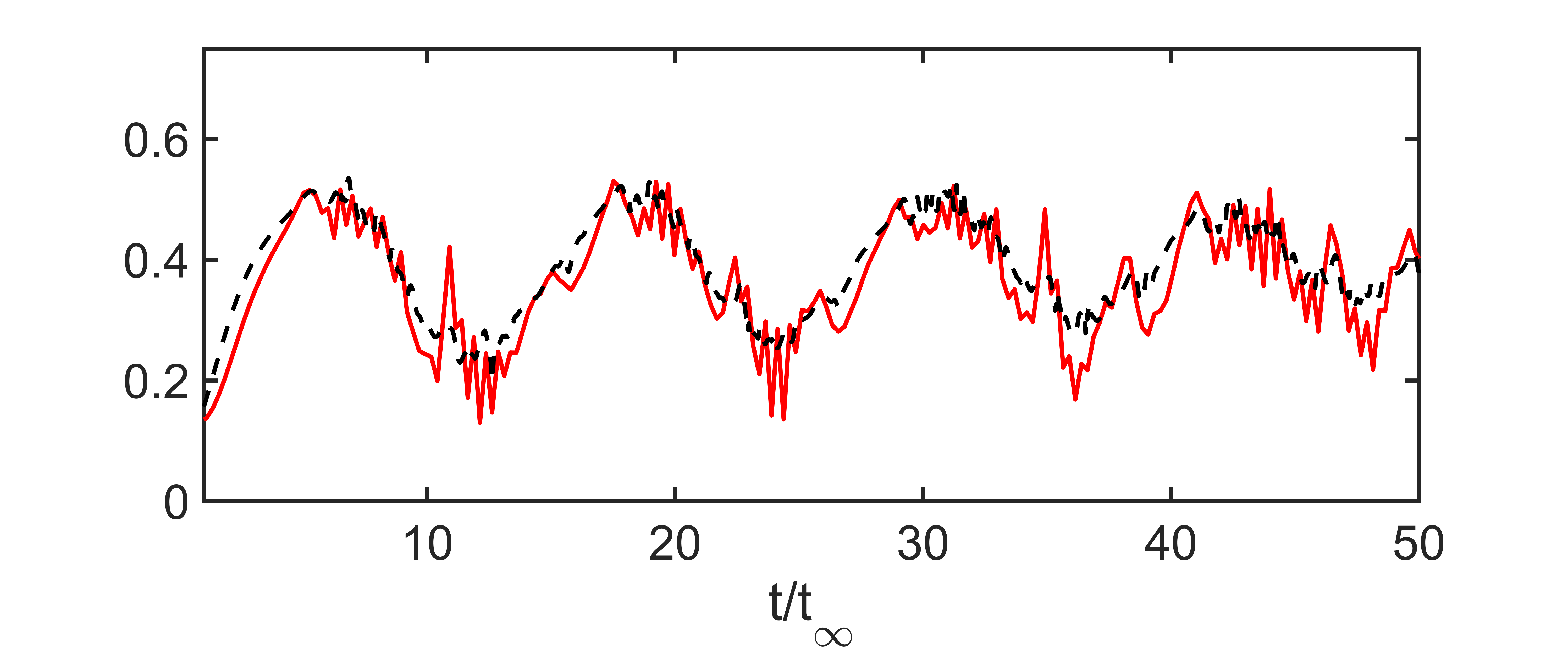}
  \caption{$\alpha = 5^{\circ}, \sigma = 0.52, U_r = 4$}
\end{subfigure}
\begin{subfigure}{\textwidth}
  \centering
  \includegraphics[width=0.8\linewidth]{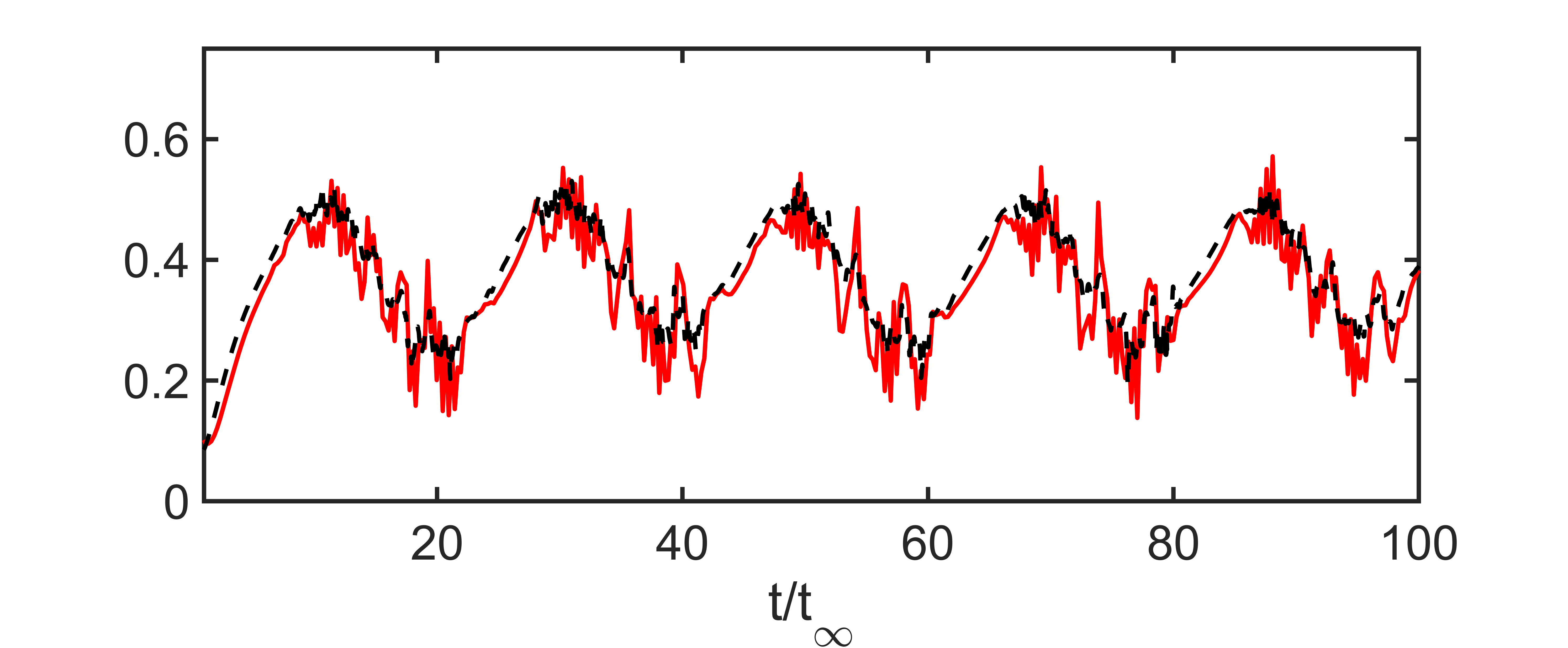} 
  \caption{$\alpha = 5^{\circ}, \sigma = 0.52, U_r = 6$}
\end{subfigure}
\caption{Comparison of the lift coefficient predicted by numerical integration of the surface traction on the hydrofoil surface via Eq.~(\ref{eq:forceCoefficients}) and linear function of the circulation around the hydrofoil using Eq.~(\ref{eq:KJ}).}
\label{fig:circLineIntg}
\end{figure}
Figure~\ref{fig:circLineIntg} compares the lift coefficient predicted by Eq.~(\ref{eq:KJ}) with that obtained by numerical integration of the surface traction as given by Eq.~(\ref{eq:forceCoefficients}). We note that Eq.~(\ref{eq:KJ}) was developed for inviscid, steady unseparated flows. However, it agrees remarkably well with the instantaneous numerically integrated lift coefficient. Thus we shall use it as a first step to make some key insights into the fluctuating lift forces.

We note that a drop in $C_L$ corresponds to a drop in the surface integral of vorticity $\iint_{S} \boldsymbol{\omega} \cdot \mathrm{d} \mathbf{S}$ in the domain. We further note that the dominant phase and frequency of oscillation in the lift obtained by the two approaches matches closely. This confirms that periodic changes in $\boldsymbol{\omega}_z$ are reflected in periodic oscillations of the lift force.

\section{Effect of 3D flow dynamics on hydroelastic response}
\begin{figure}[!h]
\centering
\includegraphics[width=0.75\textwidth]{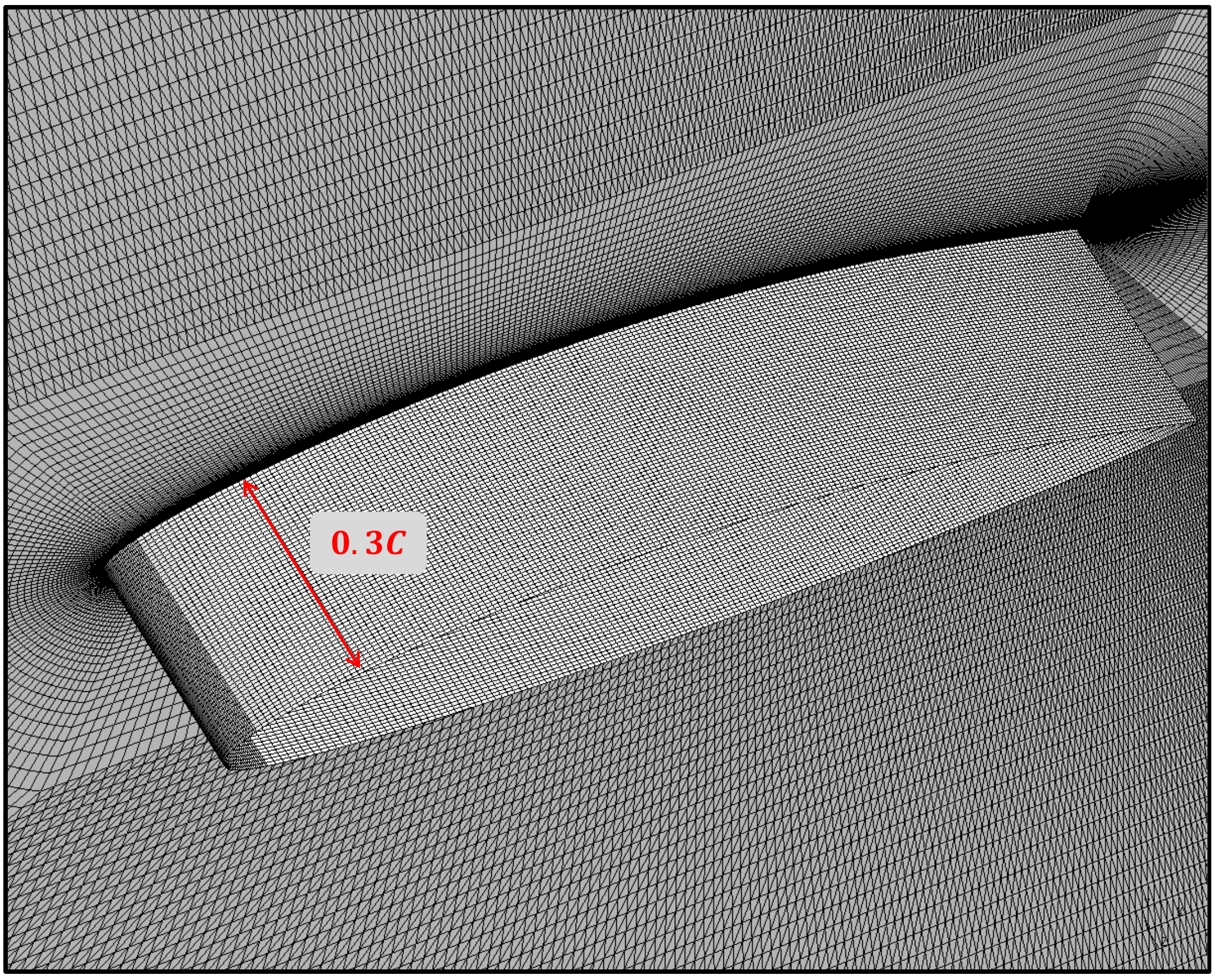}
\caption{Computational grid in the vicinity of the hydrofoil for the 3D studies.}
\label{fig:grid3D}
\end{figure}

In the current work, we use 2D periodic conditions to represent flow configurations for parametric studies. This is done with the assumption that away from root and tip effects spanwise variations in the flow field over the hydrofoil are low compared to the streamwise and transverse components. Furthermore, the objective of the study is to model an infinitesimal section of the hydrofoil to eliminate any spanwise variations in the bending moments, and the local hydroelastic response is instead governed by the mass-spring-damper characteristics. However, 3D dynamics in the flow field cannot be completely ignored. Thus for the sake of completeness, we inspect the influence of 3D effects on our observations. For this purpose, we model the hydrofoil as a section of finite spanwise thickness. To resolve the streamwise vortex, a hydrofoil span $S = 0.3C$ is considered which is more than twice the maximum hydrofoil thickness \cite{ji2015large}. Figure~\ref{fig:grid3D} shows the computational grid in the vicinity of the hydrofoil. We take 60 nodes to resolve the spanwise direction. 

\begin{figure}
\centering
\begin{subfigure}{.65\textwidth}
  \centering
  \includegraphics[width=\linewidth]{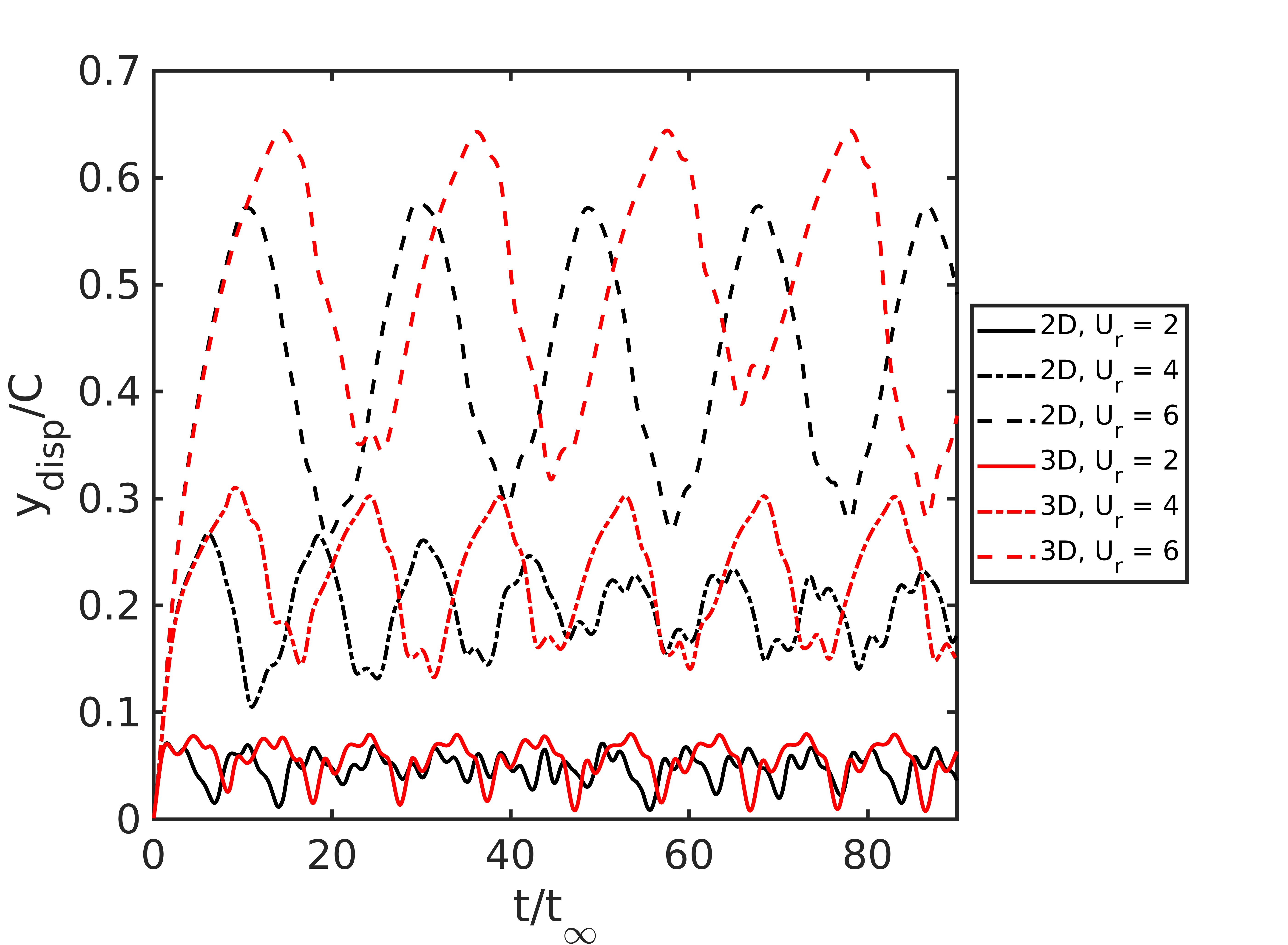}
  \caption{$\alpha = 5^{\circ}, \sigma = 0.52$}
\end{subfigure}\vfill
\begin{subfigure}{.5\textwidth}
  \centering
  \includegraphics[width=\linewidth]{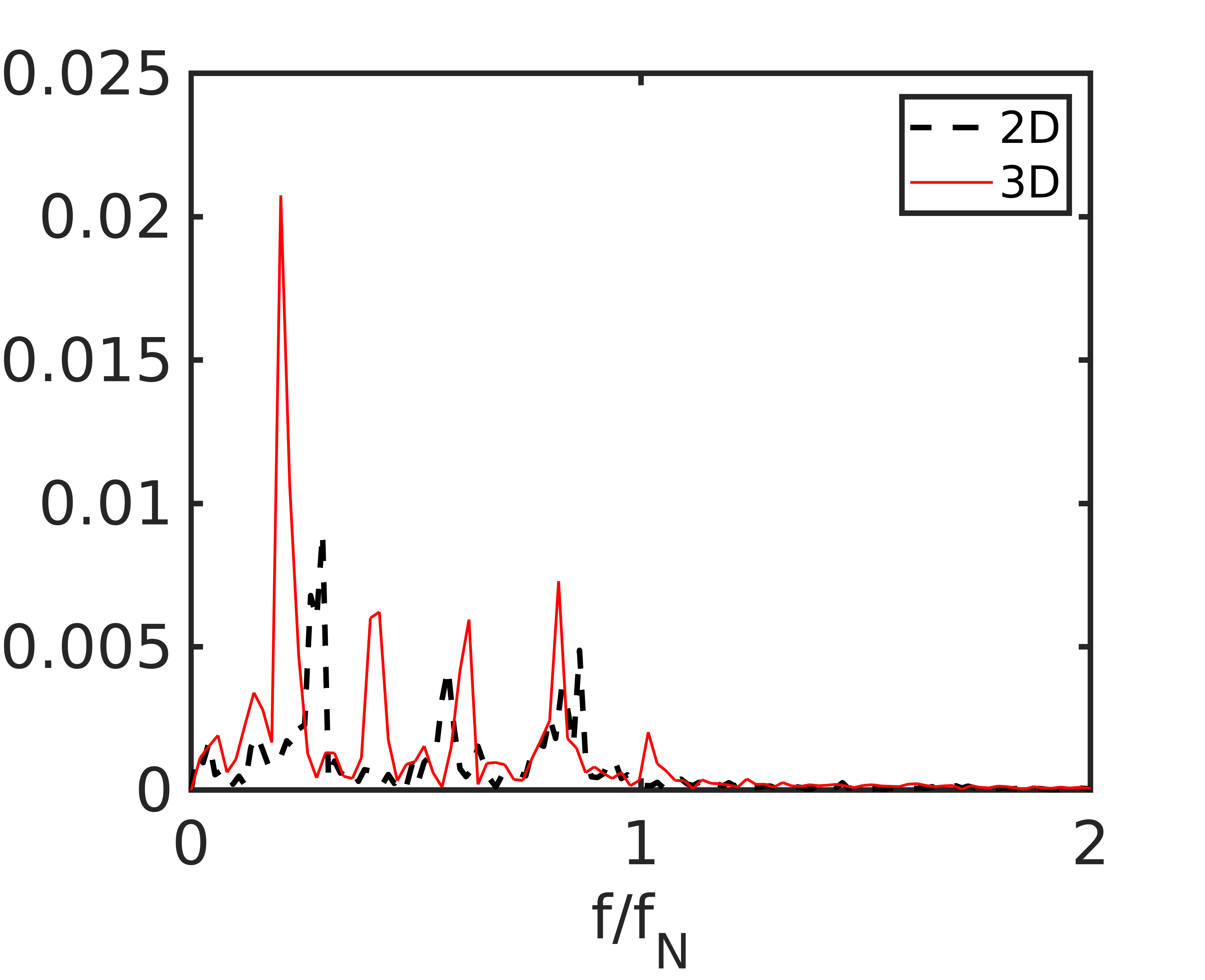}
  \caption{$\alpha = 5^{\circ}, \sigma = 0.52, U_r = 2$}
\end{subfigure}%
\begin{subfigure}{.5\textwidth}
  \centering
  \includegraphics[width=\linewidth]{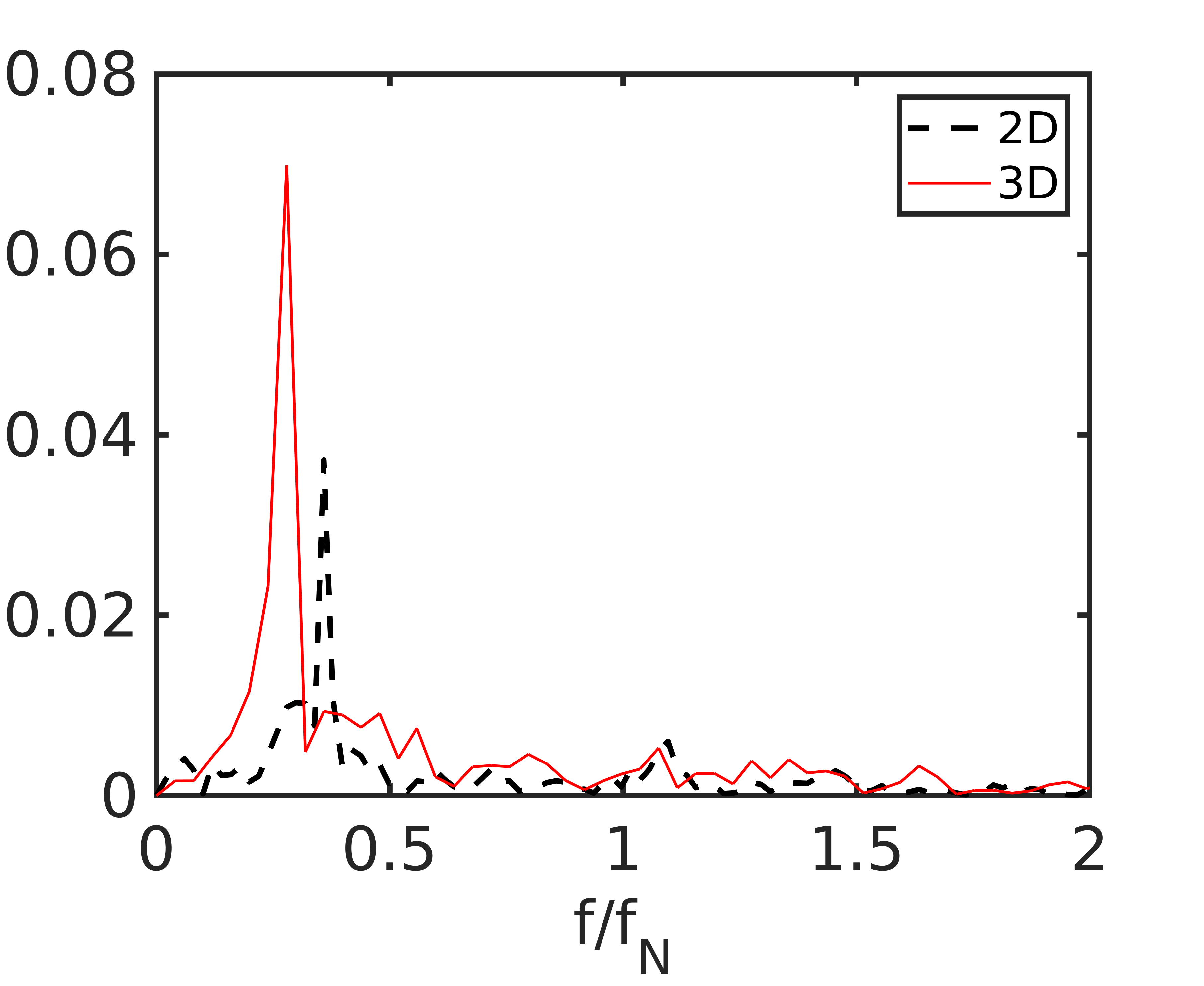}
  \caption{$\alpha = 5^{\circ}, \sigma = 0.52, U_r = 4$}
\end{subfigure}
\begin{subfigure}{\textwidth}
  \centering
  \includegraphics[width=0.5\linewidth]{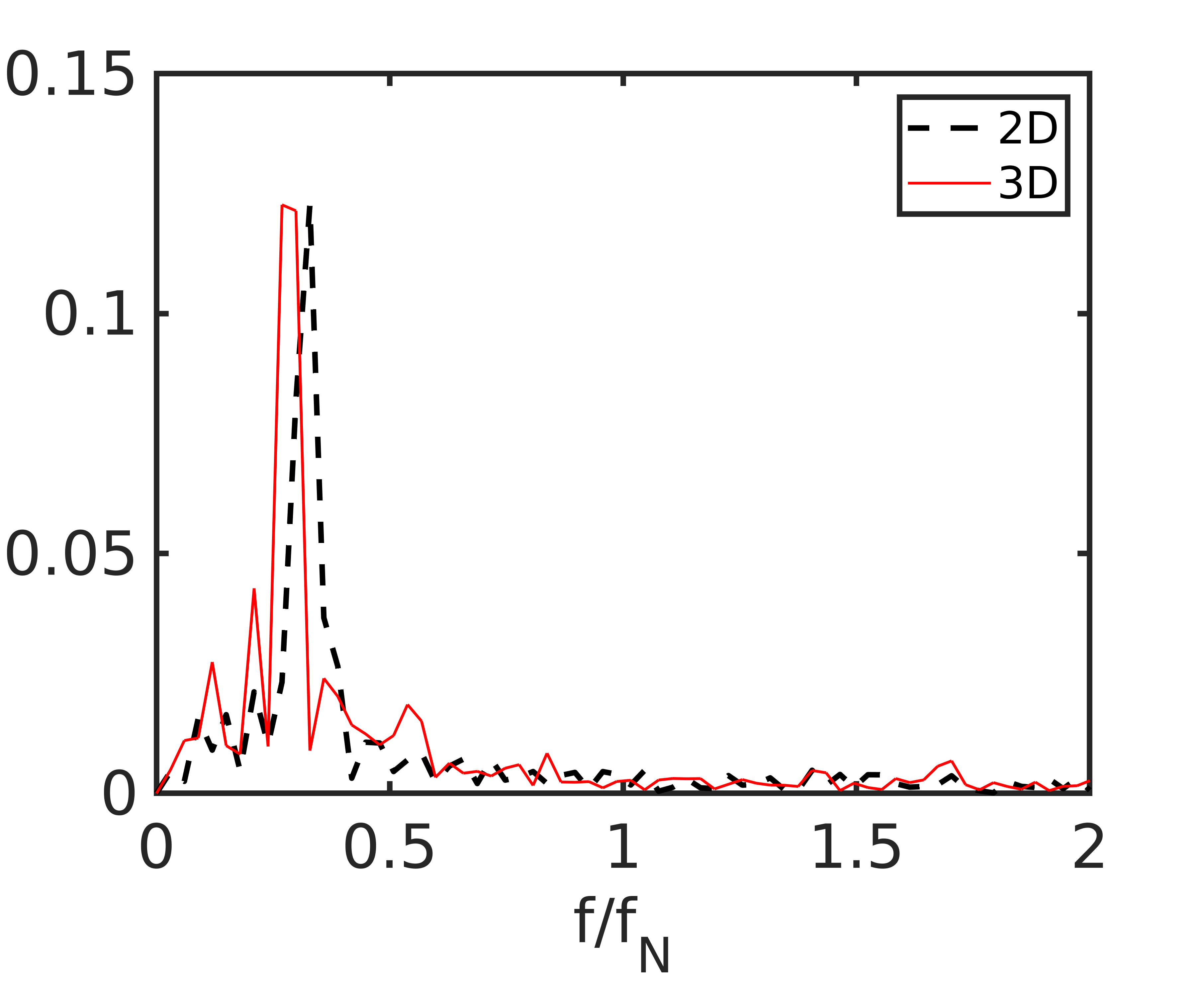}
  \caption{$\alpha = 5^{\circ}, \sigma = 0.52, U_r = 6$}
\end{subfigure}
\caption{Comparison of the hydrofoil transverse displacement $y_{disp}$ for the 2D and 3D computations. (a) Time-series evolution of $y_{disp}$ for the cavitating condition $\alpha = 5^{\circ}, \sigma = 0.52$ and $U_r = 2, 4$ and $6$. (b), (c) and (d) show the corresponding frequency spectra of $y_{disp}$.}
\label{fig:2D3Dcomp}
\end{figure}

We take the representative cavitating configurations $U_r = 2, 4$ and $6$ at $\alpha = 5^{\circ}, \sigma = 0.52$. Figure~\ref{fig:2D3Dcomp} shows the comparison of the time-series evolution of the hydrofoil displacement and the corresponding frequency spectra for the 2D and 3D computations. We see that the 2D computations underestimate the transverse displacement compared to the 3D studies. A possible explanation could be an overestimation of the spanwise $\boldsymbol{\omega_z}$ by 2D periodic boundary conditions. The corresponding oscillation frequencies are a little higher compared to 3D cases. At the same time, we note that the general hydrodynamic flow features and the hydroelastic response of the hydrofoil are captured well by the 2D computations. 

\begin{figure}
\vspace*{-0.45in}
\hspace*{0.1in}
\begin{subfigure}{\textwidth}
  \centering
  \includegraphics[width=0.25\linewidth]{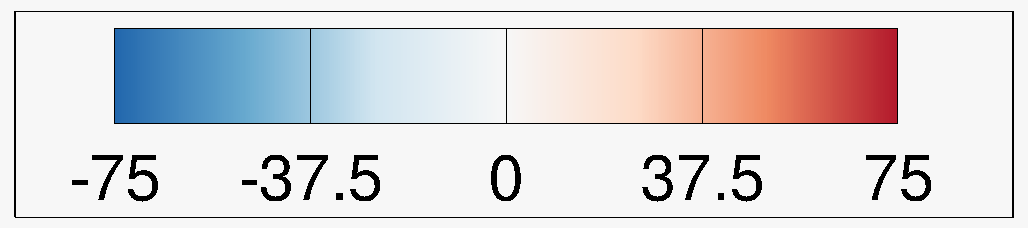}
\end{subfigure}
\hspace*{1.25in}\rotatebox[origin=c]{90}{\makebox[0.2in]{$t_0$}}
\hspace*{-1.25in}
\begin{subfigure}{\textwidth}
  \centering
  \includegraphics[width=0.55\linewidth]{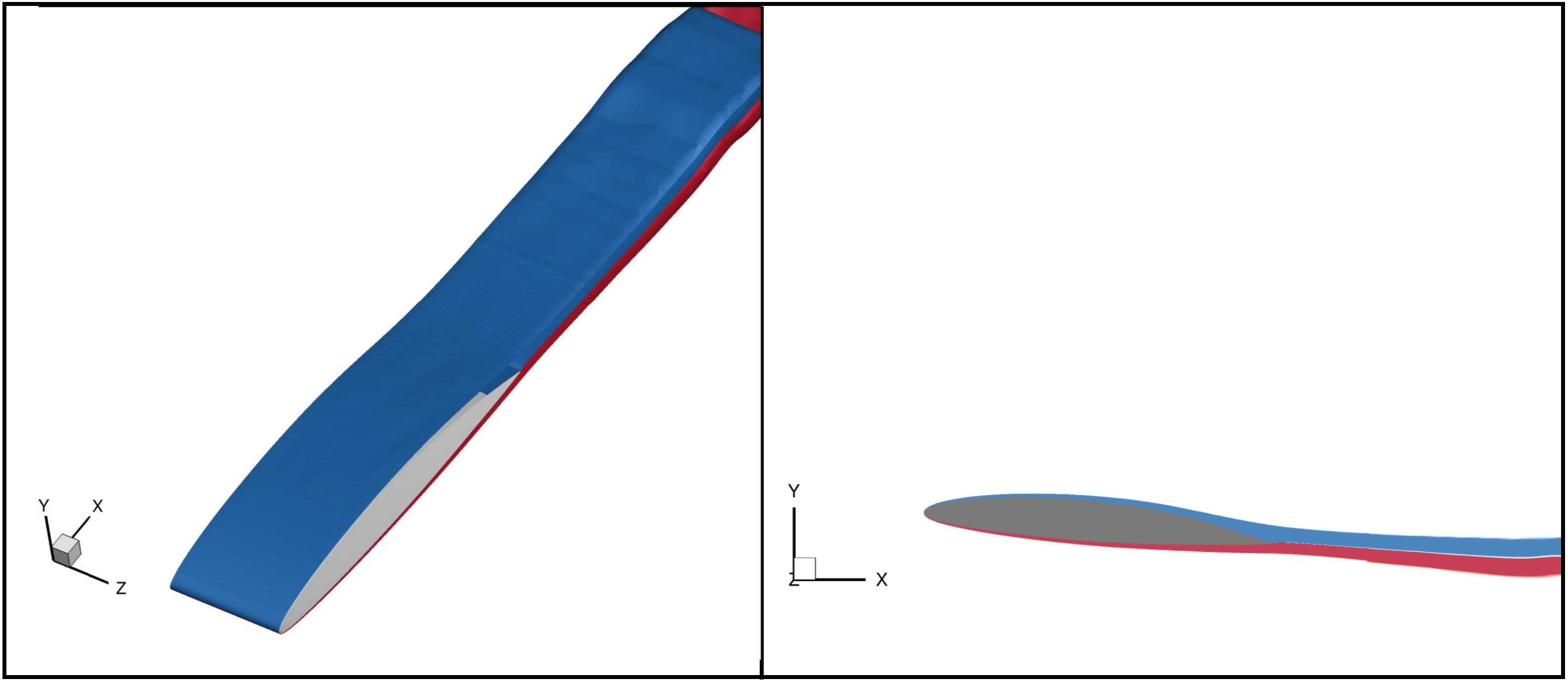}
\end{subfigure}
\hspace*{1.25in}\rotatebox[origin=c]{90}{\makebox[0.35in]{$t_1 = t_0 + 1.0/t_\infty$}}
\hspace*{-1.275in}
\begin{subfigure}{\textwidth}
  \centering
  \includegraphics[width=0.55\linewidth]{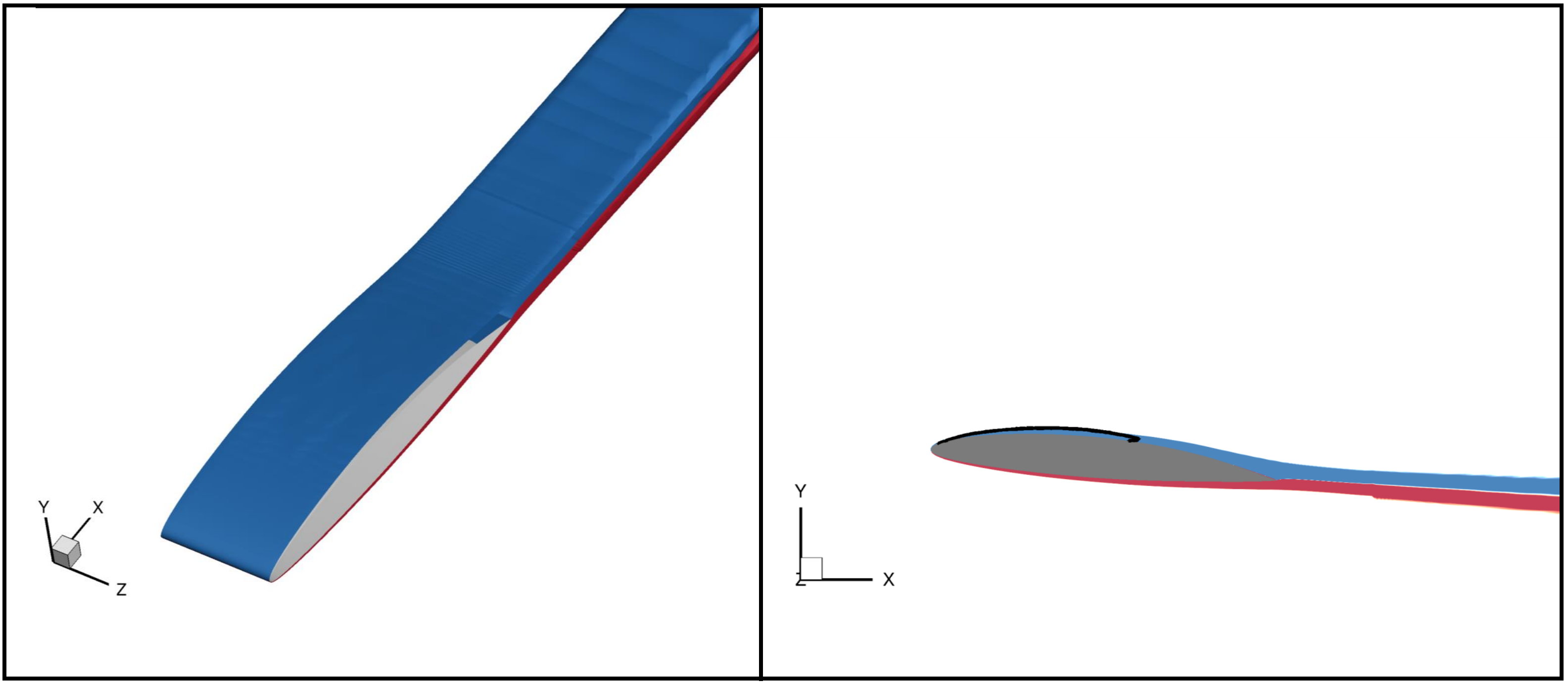}
\end{subfigure}
\hspace*{1.25in}\rotatebox[origin=c]{90}{\makebox[0.35in]{$t_1 = t_0 + 1.5/t_\infty$}}
\hspace*{-1.275in}
\begin{subfigure}{\textwidth}
  \centering
  \includegraphics[width=0.55\linewidth]{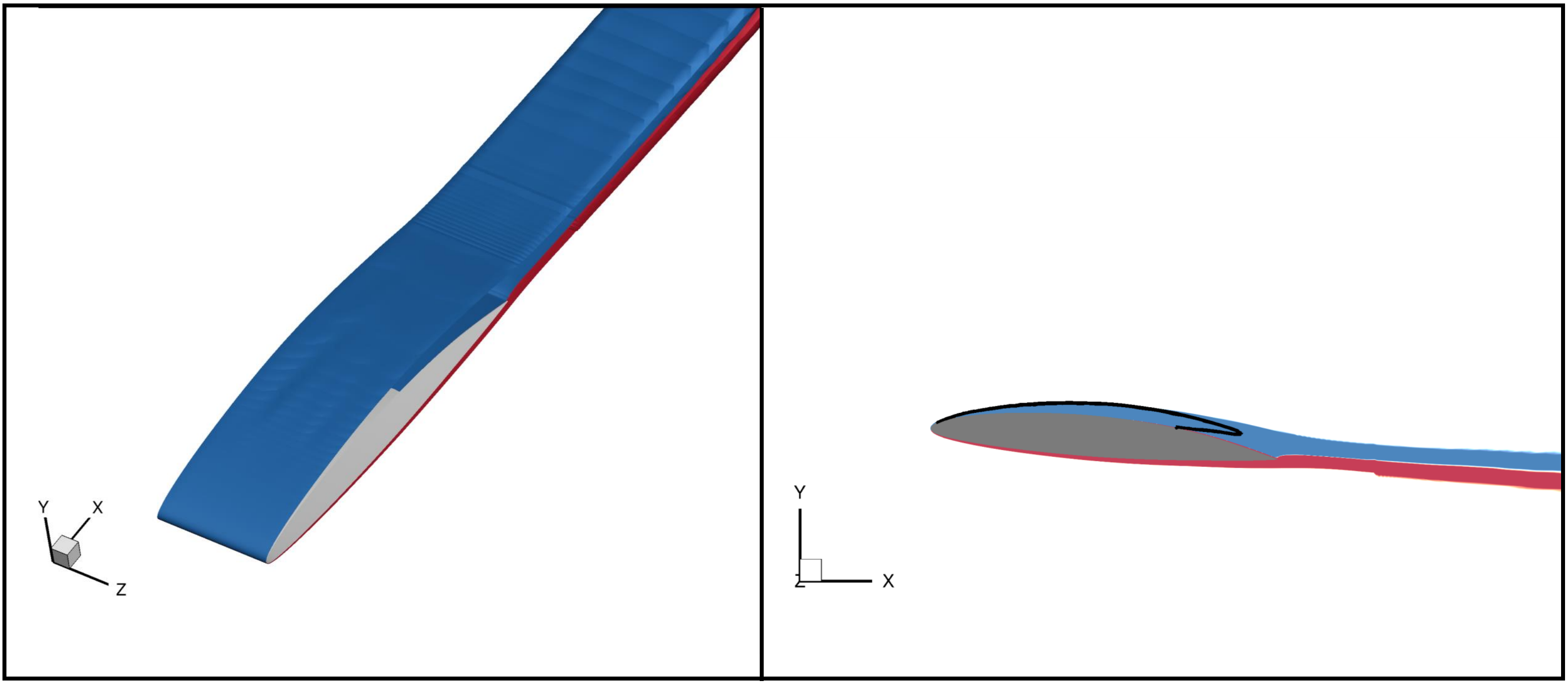}
\end{subfigure}
\hspace*{1.25in}\rotatebox[origin=c]{90}{\makebox[0.35in]{$t_1 = t_0 + 2.0/t_\infty$}}
\hspace*{-1.275in}
\begin{subfigure}{\textwidth}
  \centering
  \includegraphics[width=0.55\linewidth]{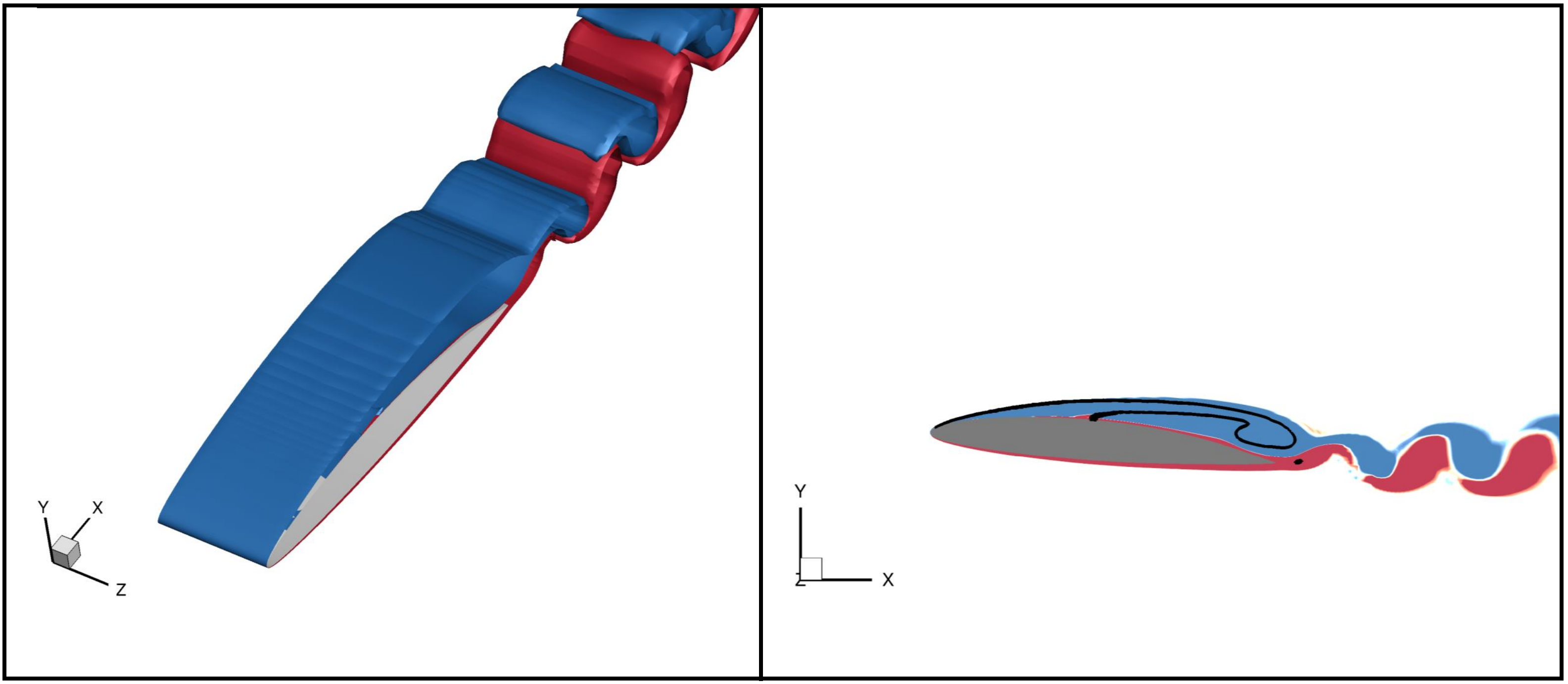}
\end{subfigure}
\hspace*{1.25in}\rotatebox[origin=c]{90}{\makebox[0.35in]{$t_1 = t_0 + 3.0/t_\infty$}}
\hspace*{-1.275in}
\begin{subfigure}{\textwidth}
  \centering
  \includegraphics[width=0.55\linewidth]{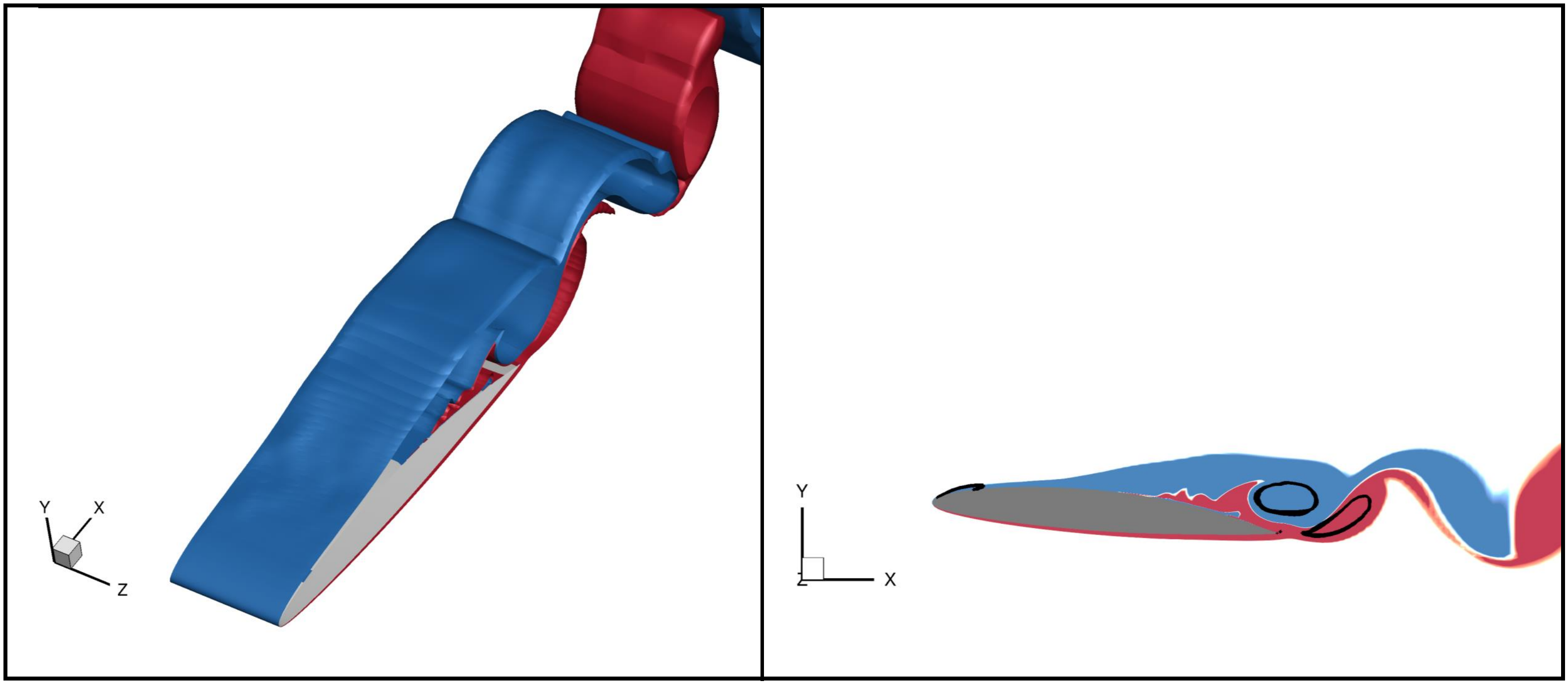}
\end{subfigure}
\hspace*{1.25in}\rotatebox[origin=c]{90}{\makebox[0.35in]{$t_1 = t_0 + 3.5/t_\infty$}}
\hspace*{-1.275in}
\begin{subfigure}{\textwidth}
  \centering
  \includegraphics[width=0.55\linewidth]{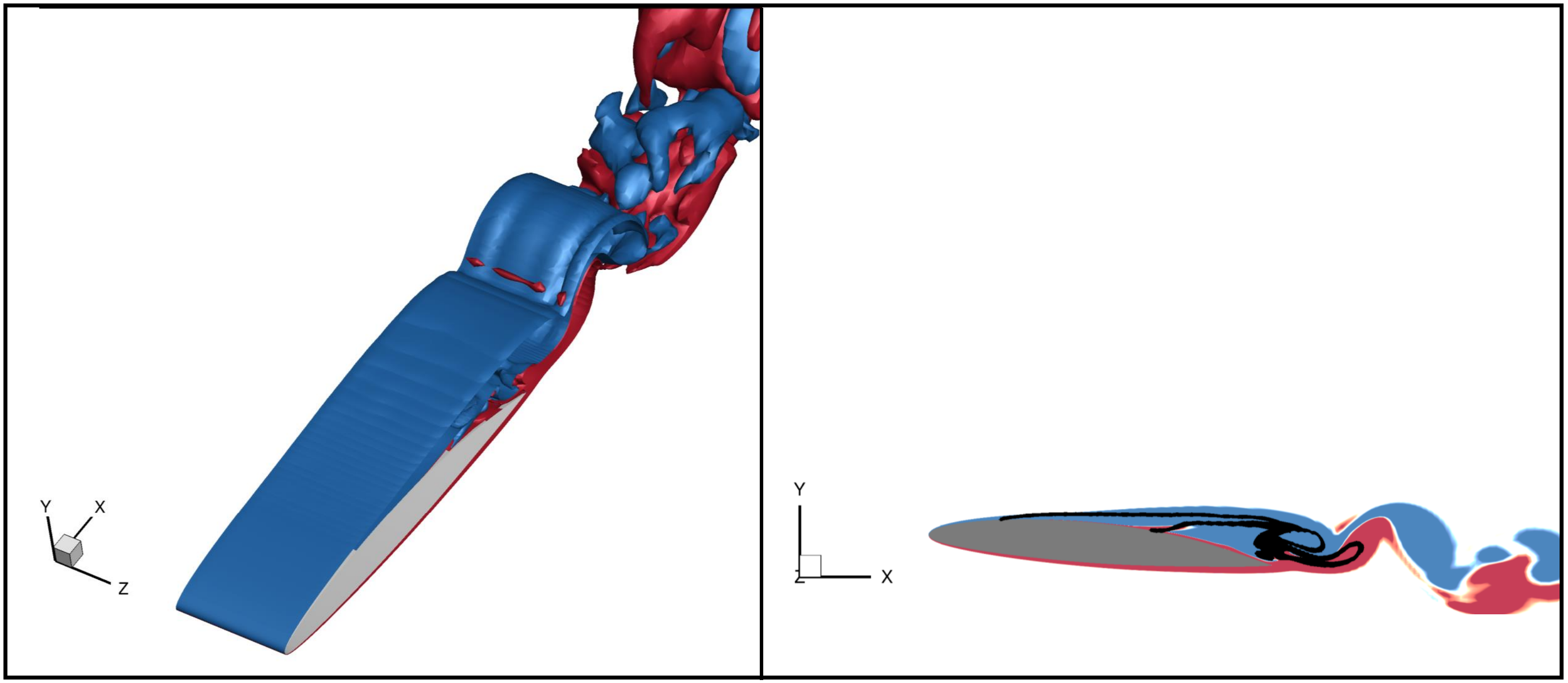}
\end{subfigure}
\caption{Key stages in the cavity growth-shedding cycle marked in terms of the mean flow time-scale $t_\infty = C/U_\infty$ for cavitating condition $\alpha = 5^{\circ}, \sigma = 0.52, U_r = 6$. (a) Iso-contours of $\boldsymbol{\omega_z} = \{-100, 100\}$ and (b) Contours of $\boldsymbol{\omega_z}$ with iso-contour of $\phi^{\mathrm{f}} = 0.95$ (in black) at mid-span of the hydrofoil.}
\label{fig:3Dcontours}
\end{figure}

Figure~\ref{fig:3Dcontours} shows some of the key stages in the cavity and corresponding vorticity cycles for the 3D computations. We find that the general observations made in Section~\ref{sec:cavVort} are consistent for the 3D cases. Thus we conclude that the 2D computational studies are sufficient to capture the salient flow features of interest for the parametric studies. 

\end{document}